\documentclass[abbrv,aps,prb,twocolumn,showpacs,preprintnumbers,amsmath,amssymb,
superscriptaddress,nofootinbib]{revtex4-1}

\usepackage{amsmath,amssymb,amsfonts}
\usepackage{bm,graphicx,color}
\usepackage{subfigure}

\newcommand{\laco}[1]{$\mathrm{LaCoO_3 }$}
\newcommand{\bH}{\mathcal{H}}
\newcommand{\rij}{\langle ij \rangle}

\newcommand{\bh}{\mathbf{h}}

\newcommand{\bS}{\mathbf{S}}
\newcommand{\bR}{\mathbf{R}}
\newcommand{\bd}{\mathbf{d}}
\newcommand{\bt}{\mathbf{t}}
\newcommand{\tj}{\mbox{$t$-$J\,$}}
\newcommand{\eps}{\epsilon}
\newcommand{\bph}{\boldsymbol{\phi}}
\newcommand{\btau}{\boldsymbol{\tau}}
\newcommand{\bk}{\mathbf{k}}
\newcommand{\bq}{\mathbf{q}}

\newcommand{\br}{\mathbf{r}}

\newcommand{\vac}{\text{vac}}
\newcommand{\bvac}{\emptyset}
\newcommand{\tef}{\text{eff}}

\def\lsim{~\rlap{$<$}{\lower 1.0ex\hbox{$\sim$}}}
\def\gsim{~\rlap{$>$}{\lower 1.0ex\hbox{$\sim$}}}

\begin{document}
\title{Excitonic condensation in systems of strongly correlated electrons}
\author{Jan Kune\v{s}} 
\affiliation{Institute of Physics, Academy of Sciences of the Czech republic, Na Slovance 2,
Praha 8, 182 21, Czech Republic}
\email{kunes@fzu.cz}
\date{\today}

\begin{abstract}
The idea of exciton condensation in solids was introduced in 1960's with the analogy to superconductivity in mind.
While exciton supercurrents have been realised only in artificial quantum-well structures so far, the application
of the concept of excitonic condensation to bulk solids leads to a rich spectrum of thermodynamic phases with diverse
physical properties. In this review we discuss recent developments in the theory of exciton condensation 
in systems described by Hubbard-type models. In particular, we focus on the connections to their various strong-coupling
limits that have been studied in other contexts, e.g., cold atoms physics. One of our goals is to provide a 'dictionary' 
which would allow the reader to efficiently combine results obtained in these different fields.
\end{abstract}
\maketitle

\tableofcontents
\section{Introduction}
The description of states of matter in terms of spontaneously broken symmetries~\cite{ginzburg50,nambu60} is one of the most fundamental
concepts in condensed matter theory. The text book examples are crystalline solids (broken translational symmetry) or magnets
(broken spin-rotational symmetry). Besides geometrical symmetries quantum-mechanical systems possess internal (gauge) symmetries 
associated with conserved charges. The most famous case of spontaneously broken internal symmetry is superconductivity.
Another example of broken internal symmetry is excitonic condensation.


\begin{figure}
\includegraphics[width=\columnwidth]{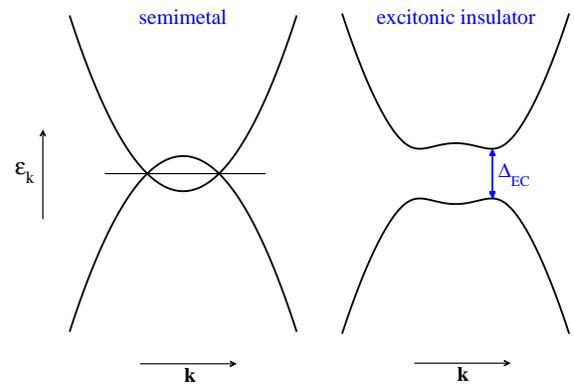}
\caption{\label{fig:cartoon} Schematic picture of excitonic condensation in a semimetal. Left: The band structure of a semimetal with overlapping bands
and no inter-band hybridisation. Right: The band structure of an excitonic insulator with a gap opened by the Weiss field $\Delta_{\text{EC}}$.}
\end{figure}
Excitonic condensation denotes spontaneous coherence between valence and conduction electrons that arises as a consequence of electron-electron interaction. 
As such, it is usually represented by orbital or band off-diagonal elements of one-particle density matrix, which possess finite expectation values
below the transition temperature $T_c$. It was originally proposed~\cite{mott61,knox63} to take place in the vicinity of semimetal-semiconductor transition depicted in Fig.~\ref{fig:cartoon}.
In an idealised system where the number of particles in the valence and conduction bands is separately conserved, the excitonic condensation 
breaks this symmetry and fixes the relative phase of valence and conduction electrons. 
Since inter-band hybridisation or pair- and correlated-hopping terms in the electron-electron interaction, which mix the valence and conduction bands, inevitably exist in real materials
the excitonic condensation in the above sense is always an approximation. The closest realisation of this idealised exciton condensate (EC) can be found in bi-layer systems.
In typical bulk structures, where the conduction-valence charge conservation does not hold and thus the relative phase of valence and conduction
electrons is not arbitrary, the excitonic condensation reduces to breaking of lattice symmetries, e.g., leading to electronic ferroelectricity~\cite{batista02}, and/or 
gives rise to exotic magnetic orders characterised by higher-order multipoles. The concept of EC allows unified description and understanding of these 
transitions.

The theory of excitonic condensation was developed in 1960's and 1970's with weakly correlated semimetals and semiconductors in mind. More recently the ideas of 
excitonic condensation were applied to strongly-correlated systems described by Hubbard-type models. In their strong-coupling limit, these models lead to spin or hard-core boson
problems that have been studied in other contexts. The purpose of this topical review is to summarise the recent work on the excitonic condensation in Hubbard-type models, discuss the 
corresponding phase diagram and describe connections to various other models such as the Blume-Emmery-Griffiths~\cite{beg}, bosonic \tj or bi-layer Heisenberg~\cite{sachdev90} models. 
In particular, we aim at providing a 'dictionary' for the names of mutually corresponding phases found in different models. We only briefly touch the active field of 
bi-layer systems in Sec.~\ref{sec:bilayer} and the spectacular transport phenomena observed there and refer the reader to specialised literature. We completely
omit another major direction of the exciton-polariton condensation~\cite{byrnes14}. 

\subsection{Brief history of the weak-coupling theory of excitonic condensation}
In 1961 Mott~\cite{mott61} considered metal-insulator transition in a divalent material (even number of electrons per unit cell) associated with
continuous opening of a gap, e.g., due to application of external pressure. He argued that the one-electron picture cannot be correct, that a semimetal
with small concentration of electrons and holes will be unstable if electron-hole interaction is taken into account. Knox~\cite{knox63} approached the transition from the insulator side and argued that if the excitonic binding energy overcomes the band gap the system becomes unstable. 
These proposals were put into a formal theory by Keldysh and Kopaev~\cite{keldysh65}, and des Cloizeaux~\cite{desc65} who developed 
a weak-coupling Hartree-Fock theory of excitonic condensation analogous to the BCS theory of superconductivity. 
The role of the pairing glue is played by the inter-band Coulomb interaction, which favours formation of bound electron-hole pairs, excitons.
The early theories employed so called dominant term approximation, which consists in keeping only the density-density
intra- and inter-band terms of the Coulomb interaction. This approximation appears well justified for weakly correlated metals or semiconductors
where the pair glues arise from the long-range part of the interaction. It leads to a large manifold
of degenerate excitonic states. It is due to this degeneracy that the small and so far neglected 
exchange and pair-hopping interaction terms may play an important role. Halperin and Rice~\cite{halperin68b} showed
that these terms select one out of the four following states as the lowest one: charge-density wave charge-current-density wave, spin-density-wave and spin-current-density wave. 

In 1970's Volkov and collaborators showed in a series of articles~\cite{volkov73,volkov75,volkov76} that excitonic condensation in slightly off-stoichiometric systems
leads to formation of a uniform ferromagnetic state although the normal state does not exhibit a magnetic instability. The idea of an excitonic ferromagnet was revived  
at the beginning of 2000's in the context of hexaborides, Sec.~\ref{sec:materials}.

\section{Spinless fermions}
\label{sec:spinless}
Before discussing the two band Hubbard model (\ref{eq:2bhm}), 
we look at its spinless version
\begin{equation}
\label{eq:efkm}
\begin{split}
   H_{\text{EFK}}=&\sum_{\rij} \left(t_aa_{i}^{\dagger}a^{\phantom\dagger}_{j}+
   t_bb_{i}^{\dagger}b^{\phantom\dagger}_{j}\right)+H.c.\\
   +&\sum_{\rij}\left(V_{ab}b_{i}^{\dagger}a^{\phantom\dagger}_{j} 
   +V_{ba}a_{i}^{\dagger}b^{\phantom\dagger}_{j}\right)+H.c. \\
   +&\frac{\Delta}{2}\sum_{i} \left(n^a_{i}-n^b_{i}\right)
   +\tilde{U} \sum_{i}n^a_{i}n^b_{i},
\end{split}
\end{equation}
which describes electrons of two flavours ($a$ and $b$) moving on a lattice and interacting via an on-site interaction. Here, 
$a_{i}^{\dagger}$ ($a^{\phantom\dagger}_{i}$) is an operator creating (annihilating) a fermion of flavour  $a$
on lattice site $i$, $n_{i}^a=a_{i}^{\dagger}a^{\phantom\dagger}_{i}$ is the corresponding local density operator, and analogically
for $b$ fermions. We will use $\psi_a(\br)$ and  $\psi_b(\br)$ to refer to the density distribution around  the lattice sites in orbitals $a$ and $b$, respectively,
when necessary.
The sum $\sum_{\rij}$ runs over the nearest neighbour (nn) bonds.
Model (\ref{eq:efkm}) is usually called extended Falicov-Kimball model (EFKM), where 'extended' implies that both $t_a,t_b\neq0$. 
In the original Falicov-Kimball model (FKM)~\cite{falicov69}, $t_b=V_{ab}=V_{ba}=0$, only the $a$-electrons can propagate while the
'heavy' $b$-electrons are immobile.~\footnote{More traditionally the mobile and heavy electrons are labelled $d$ and $f$, respectively.}

Three versions of EFKM have been discussed in the literature in the context of excitonic condensation:
(i) FKM with $t_b=V_{ab}=V_{ba}=0$, (ii) EFKM without cross-hopping $V_{ab}=V_{ba}=0$ and (iii) EFKM with cross-hopping $V_{ab},V_{ba}\neq0$.
Since the conservation of the charge per flavour plays a central role in excitonic condensation we compare models (i)-(iii) from that perspective. 
 Hamiltonian (\ref{eq:efkm}) conserves the total charge for any choice of parameters.
Since the corresponding $U(1)$ symmetry is not broken in any phase considered here, we will not mention this symmetry explicitly in the text. 
The FKM (i) conserves the number of the heavy $b$-electron on each site, which gives rise to a local $U(1)$ gauge symmetry.
Finite $t_b$ in (ii) removes the local gauge invariance, but preserves independently the number of $a$- and $b$-electrons, which gives
rise to a global $U(1)$ symmetry associated with the arbitrariness of the relative phase of the $a$- and $b$-electrons.
Finally, finite cross-hopping in (iii) removes this $U(1)$ symmetry. An important special case of (iii) is a system with 
a site symmetry that prohibits an on-site $a-b$ hybridisation. As an example, we may consider orbitals $a$ and $b$ of different
parity, which implies $V_{ab}=-V_{ba}$. In this case, the $U(1)$ symmetry is not removed completely 
but reduced to a discrete $Z_2$ symmetry reflecting the invariance of (\ref{eq:efkm}) under the transformation $a_{i}\rightarrow -a_{i}$, $V_{ab}\leftrightarrow V_{ba}$. 

On a bipartite lattice there are additional symmetries. 
Models (ii) with $t_at_b>0$ and $t_at_b<0$ can be mapped on each other by $a_{i}\rightarrow(-1)^{i}a_{i}$~\footnote{Changing sign on one sub lattice
of the bipartite lattice}. In
the EC phase the map turns ferro-EC order ($t_at_b<0$) into an antiferro-EC order ($t_at_b>0$).
This property will be important for understanding the nature of EC phase in FKM ($t_at_b=0$), which lies between the ferro- and 
antiferro- phases. There is also a symmetry with respect to the sign of $\Delta$ for $|t_a|=|t_b|$ consisting in the exchange $ a_{i}\leftrightarrow b_{i}$.

Model (ii) with symmetric bands $t_a=t_b$ is identical to a single band Hubbard model, where flavours $a$ and $b$ play the role of
the spin variable and $\Delta$ stands for an external magnetic field. 

Model (ii) with $t_a=-t_b$ is identical to the single band Hubbard model with attractive interaction. The corresponding mapping is achieved
by particle-hole transformation in one of the bands $b^{\phantom\dagger}_{i}\leftrightarrow b^{\dagger}_{i}$ and reversing the sign of
the interaction $\tilde{U}\rightarrow -\tilde{U}$.
This transformation maps an excitonic insulator onto an $s$-wave superconductor. Note that crystal-field $\Delta$ in the excitonic insulator
plays the role of chemical potential in the superconductor and the chemical potential in the excitonic insulator plays the role of magnetic field
in the superconductor. This shows that deviations from half-filling are detrimental for EC in the same 
way magnetic field is for spin-singlet superconductivity. The mapping between the superconductor and excitonic insulator
obviously breaks down in the presence of external electro-magnetic field.

\subsection{Falicov-Kimball model}
The interest in excitonic condensation in FKM started with the work of 
Portengen {\it et al.}~\cite{portengen96, portengen96b} who studied FKM model augmented with the hybridisation between the heavy $b$-electrons and light $a$-electrons $V_{ab}=-V_{ba}\neq0$. Using self-consistent mean-field (BCS-like) theory they found solutions with spontaneous on-site hybridisation -- excitonic condensate. They showed that in case that the $a$ and $b$ orbitals are of opposite parity the excitonic condensation gives rise to a ferroelectric polarisation. Importantly, their excitonic condensate existed 
also in the limiting case of $V_{ab}=V_{ba}=0$ suggesting a new ground state of the original FKM. 
This result was challenged by subsequent theoretical studies~\cite{czycholl99,farkasovsky99,farkasovsky02,brydon05} 
of FKM in one and infinite dimension, which found no spontaneous hybridisation. It was pointed out that the method of Portengen and collaborators
missed competing ordered states and the possibility of phase separation.
Freericks and Zlati\'{c}~\cite{freericks03} argued that spontaneous hybridisation in FKM 
breaks the local $U(1)$ gauge symmetry, associated with the phase of the heavy electron, and thus is prohibited by Elitzur's theorem~\cite{elitzur75}. 
The question is, however, quite subtle as can be illustrated, for example, by the excitonic susceptibility in $d=\infty$ FKM, which has a logarithmic singularity at $T=0$~\cite{zlatic01}.
The insight into the issue can be provided by going into EFKM. The mapping between $t_bt_a>0$ and $t_at_b<0$ models (on bipartite lattice) implies that if there is an (ferro-) ordered
state in the limit $t_b\rightarrow 0^-$ there must an antiferro- order in the limit $t_b\rightarrow 0^+$ and thus FKM is an unstable point between these two phases.
\begin{figure}
\includegraphics[width=0.7\columnwidth,clip]{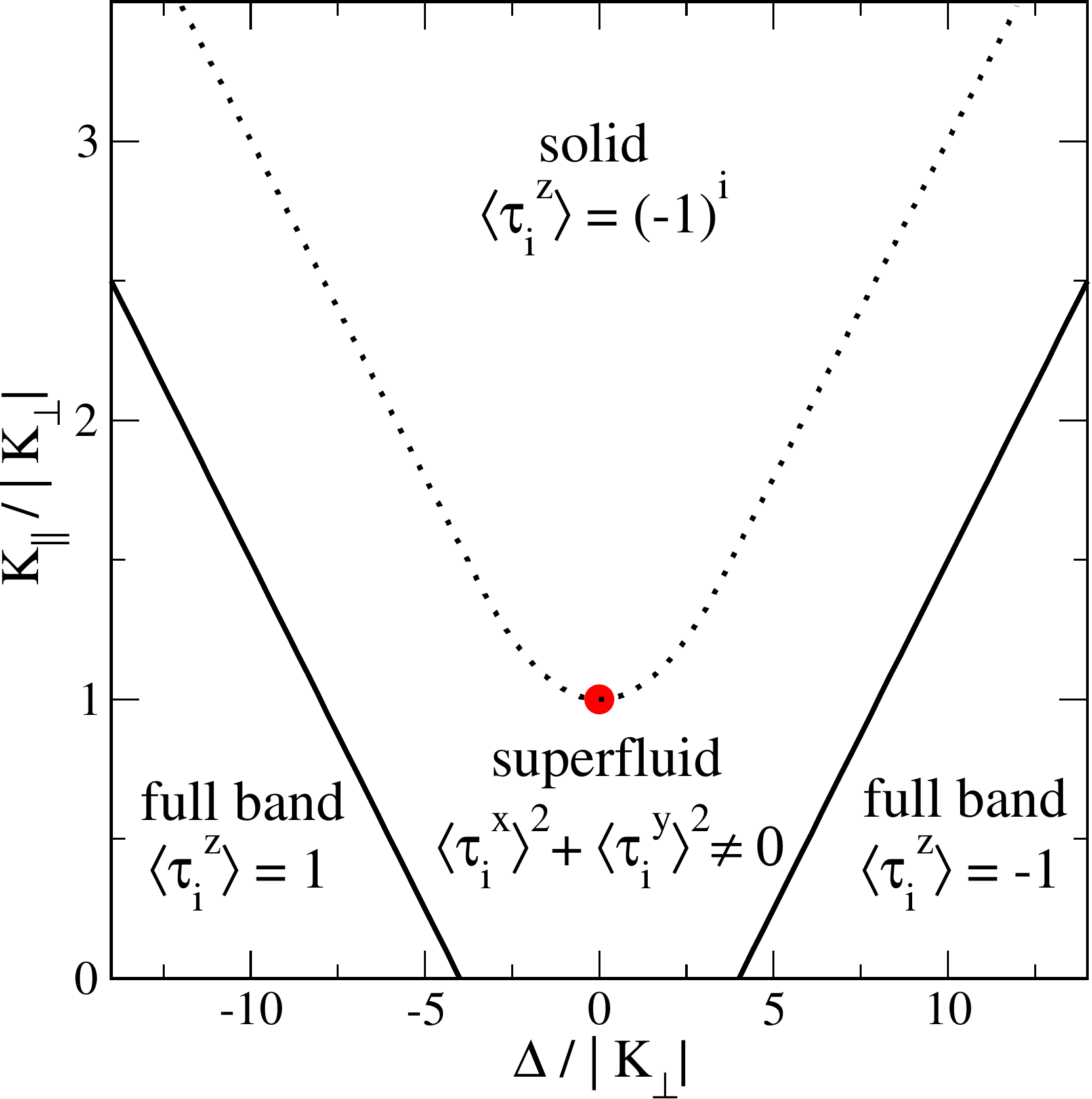}
\caption{\label{fig:xxz} The $T=0$ phase diagram of the $d=2$ $XXZ$ model constructed from data of Ref.~\onlinecite{schmid02}. 
The dotted line marks the first-order
solid-superfluid boundary, the full lines denote the continuous transitions between the superfluid and the fully polarised state.}
\end{figure}

\subsection{Extended Falicov-Kimball model}
\subsubsection{Strong-coupling limit}
\label{sec:efkm-strong}
An alternative way to EC order in FKM is to make both fermion species mobile, i.e., introducing EFKM.
Batista~\cite{batista02} studied EFKM at half-filling in the strong coupling limit $\tilde{U}\gg t_a,t_b$
as an alternative way to electronic ferroelectricity. Introducing pseudo spin variables
\begin{equation}
\label{eq:pseudospin}
\begin{aligned}
\tau_i^x &=a_i^{\dagger}b_i^{\phantom\dagger}+b_i^{\dagger}a_i^{\phantom\dagger}, \quad
\tau_i^y=i(b_i^{\dagger}a_i^{\phantom\dagger}-a_i^{\dagger}b_i^{\phantom\dagger}), \\
\tau_i^z &=a_i^{\dagger}a_i^{\phantom\dagger}-b_i^{\dagger}b_i^{\phantom\dagger},
\end{aligned}
\end{equation}
he arrived at a strong-coupling effective Hamiltonian, which in the case $V_{ab}=V_{ba}=0$ reads
\begin{equation}
\label{eq:strong_efkm}
\bH_{\text{eff}}=\frac{1}{2}\sum_{\rij} \left(K_{\parallel} \tau_i^z\tau_j^z+K_{\perp}\left(\tau_i^x\tau_j^x+\tau_i^y\tau_j^y\right)\right)+\frac{\Delta}{2}\sum_i\tau_i^z.
\end{equation}
$\bH_{\text{eff}}$ acts in the low-energy space built of states with singly occupied sites only. The coupling constants to the lowest order
in $t/\tilde{U}$ read $K_{\parallel}=\tfrac{t_a^2+t_b^2}{\tilde{U}}$ and $K_{\perp}=\tfrac{2t_at_b}{\tilde{U}}$. Labelling the two local states
$|\uparrow_i\rangle=a_i^{\dagger}|0\rangle$ and  $|\downarrow_i\rangle=b_i^{\dagger}|0\rangle$ the structure of the $S=1/2$ $XXZ$-model in 
an external field along the $z$-axis is apparent. Note that while $K_{\perp}$ can be both positive and negative, $K_{\parallel}$ is always positive. There is a well-known
exact mapping of (\ref{eq:strong_efkm}) onto the model of spinless hard-core bosons~\cite{matsubara56}
\begin{equation}
\label{eq:boson}
\tilde{\bH}_{\text{eff}}=\eps\sum_{i} n_i+K_{\perp}\sum_{\rij}\left(d^{\dagger}_{i}d^{\phantom\dagger}_{j}+H.c.\right)
               +2K_{\parallel}\sum_{\rij} n_in_j,
\end{equation}
with $\eps=\Delta-zK_{\parallel}$ ($z$ is the number of nearest neighbours). The operator $d^{\dagger}_{i}$ ($d^{\phantom\dagger}_{i}$)
creates (annihilates) a boson on site $i$, and $n_i=d^{\dagger}_{i}d^{\phantom\dagger}_{i}$ is the local density operator.
The physical states are constrained to those containing zero or one bosons per site.
In this language, the $|\downarrow_i\rangle$ is the local bosonic vacuum and $|\uparrow_i\rangle=d^{\dagger}_{i}|\downarrow_i\rangle$ 
is a state with one boson.  The transverse spin coupling translates into bosonic hopping, the longitudinal coupling into nearest-neighbour repulsion and
the magnetic field into bosonic chemical potential. Moving between the spin and boson formulation has been traditionally used to allow
convenient treatment of various models~\cite{holstein40,auerbach94}.

The model (\ref{eq:strong_efkm},\ref{eq:boson}) is much studied in the context of cold atoms on
optical lattices. Its $T=0$ phase diagram for a square lattice is shown in Fig.~\ref{fig:xxz}. Besides the trivial phases obtained for large $|\Delta|$, which correspond to orbitals of one flavour being filled and the other empty (saturated spin polarisation along the $z$-axis), there are two more phases. The solid phase ($z$-axis N\'eel antiferromagnet) favoured by $K_{\parallel}$, and the superfluid phase ($xy$ magnetic order) favoured by $K_{\perp}$. The solid phase is characterised by checker-board arrangement of sites with occupied $a$ and $b$ orbitals. The solid phase is connected to the ground state of half-filled FKM~\cite{brandt89}. The superfluid phase is characterised by a finite expectation value $\langle d_i \rangle$, i.e., can be described as a condensate 
of the $d$-bosons. In the language of the EFKM this means a finite $\langle a_i^{\dagger}b_i \rangle$ expectation value characterising the excitonic condensate. 
To understand the phase diagram in Fig.~\ref{fig:xxz} one may start form the familiar point of Heisenberg antiferromagnet, $\Delta=0$, $K_{\perp}=K_{\parallel}$. Upon application of a magnetic field
($-\Delta$) the ordered moments pick an arbitrary perpendicular orientation with a small tilt in the field direction (superfluid). For $|K_{\perp}|>K_{\parallel}$ the in-plane order is
obtained without the external field, while for $|K_{\perp}|<K_{\parallel}$ a finite field is needed to destroy the Ising (solid) order.

The solid and superfluid phases have quite different properties. The solid phase breaks the discrete translational symmetry. The superfluid phase breaks continuous
$U(1)$ symmetry associated with the phase of $d$-boson. For $K_{\perp}>0$ the superfluid also breaks the translational symmetry with the phase of $\langle d_i \rangle$ varying
between sublattices. Therefore in two dimensions only solid long-range order exists at finite temperature~\cite{mermin66}, while the superfluid phase has the form
of Kosterlitz-Thouless phase~\cite{berezinskii72,kosterlitz73}. The mismatch between the symmetries of solid and superfluid phases implies a first-oder transition between them 
or existence of an intermediate supersolid phase where both the orders are present. The stability of the supersolid  phase is a much studied question 
in the context of model (\ref{eq:boson}) and its generalisations. Investigations on cubic lattice in two~\cite{schmid02} and three~\cite{fisher74} dimensions 
found the first-order transition scenario to be realised. However, a robust supersolid phase was found on the triangular lattice~\cite{melko05,heidarian05,boninsegni05,wessel05}. 
Existence of a supersolid on triangular lattice is related to the fact that the solid and the superfluid adapt differently to the geometrical frustration.
For further reading on hard-core boson we refer the reader to specialised literature.


We conclude this section by considering the effect of cross-hopping. Non-zero cross-hopping $V_{ab}, V_{ba}$ in (\ref{eq:efkm}) breaks the $U(1)$ symmetry
of EFKM and generates additional terms in Hamiltonian (\ref{eq:boson})~\cite{batista02}. These are of two types. 
First, the 'correlated-hopping' terms of the form $(d_i^{\phantom\dagger}+d_i^{\dagger})(2n_j^{\phantom\dagger}-1)$ 
and of the order $(V_{ab}t_a-V_{ba}t_b)/\tilde{U}$. Summed over the nn sites their contribution can be split into a source field for $d$-bosons, proportional
to the mean boson density $\langle n \rangle$, and a coupling of $d_i^{\phantom\dagger}+d_i^{\dagger}$ to the density fluctuations. 
With a finite source field the situation is analogous to a ferromagnetic transition in an external
magnetic field, i.e., the sharp transition is smeared out and diminishes completely if the external field is comparable or larger than
the internal (Weiss) fields. In some symmetries the contributions of individual neighbours to the source fields add up to zero, e.g., in the case
of $a$- and $b$-orbitals in (\ref{eq:efkm}) of different parity~\cite{batista02}.
In this case, phase transition is still possible.
However, the second type of terms generated by cross-hopping of the form $d_i^{\phantom\dagger}d_j^{\phantom\dagger}$ and $d_i^{\dagger}d_j^{\dagger}$
and of the order $V_{ab}V_{ba}/\tilde{U}$ reduce the symmetry of the system to $Z_2$. 
Depending on the sign of this term the order parameter $\langle d_i^{\dagger} \rangle=\langle a_i^{\dagger}b_i\rangle$
is either real or imaginary. The meaning of the phase of the order parameter   $\langle a_i^{\dagger}b_i\rangle$ was discussed by Halperin and Rice~\cite{halperin68b}. 
In case or real orbitals $\psi_a(\mathbf{r})$ and $\psi_b(\mathbf{r})$, real  $\langle a_i^{\dagger}b_i\rangle$ gives rise to charge density modulation while imaginary $\langle a_i^{\dagger}b_i\rangle$
gives rise to periodic current pattern. In a model with inversion symmetry about the lattice sites, studied in Ref.~\onlinecite{batista02},
$V_{ab}=-V_{ba}$ leads to real  $\langle a_i^{\dagger}b_i\rangle$ implying an Ising-like ferroelectric transition. Solution with purely imaginary
$\langle a_i^{\dagger}b_i\rangle$ was reported in $d=1$ EFKM by Sarasua and Continentino~\cite{sarasua04} for somewhat artificial choice
of purely imaginary $V_{ab}=V_{ba}$.

\subsubsection{Intermediate and weak coupling}
Studies using various techniques and in different dimenssionalities, reviewed below, lead to the conclusion that the behaviour observed in the strong-coupling 
limit of EFKM extends to the intermediate coupling and connects to the weak-coupling regime.  
To be able to compare these results we have to understand how they are interpreted, in particular in $d=1,2$. Following the Mermin-Wagner theorem~\cite{mermin66}, 
the superfluid, which breaks continuous symmetry, is characterised by long-range order at finite $T$ in $d=3$, by long-range order at $T=0$ and algebraic correlations at finite $T$
in $d=2$, and by algebraic correlations at $T=0$ in $d=1$. In the Ising-like solid phase a long-range order exists at finite $T$ in $d=2,3$, while
in $d=1$ there is a long-range order at $T=0$.
The onset of algebraic correlations was used as a criterion to define the transition to the superfluid phase in the $d=1, T=0$ Monte-Carlo studies~\cite{batista04}.
Mean-field techniques on the other hand do not distinguish dimenssionalities and long-range order is obtained even if it does not exist in the exact solution. 
In these cases one compares the onset of algebraic correlations in more rigorous methods with the onset of mean-field long-range order.

\begin{figure}
\includegraphics[width=0.95\columnwidth,clip]{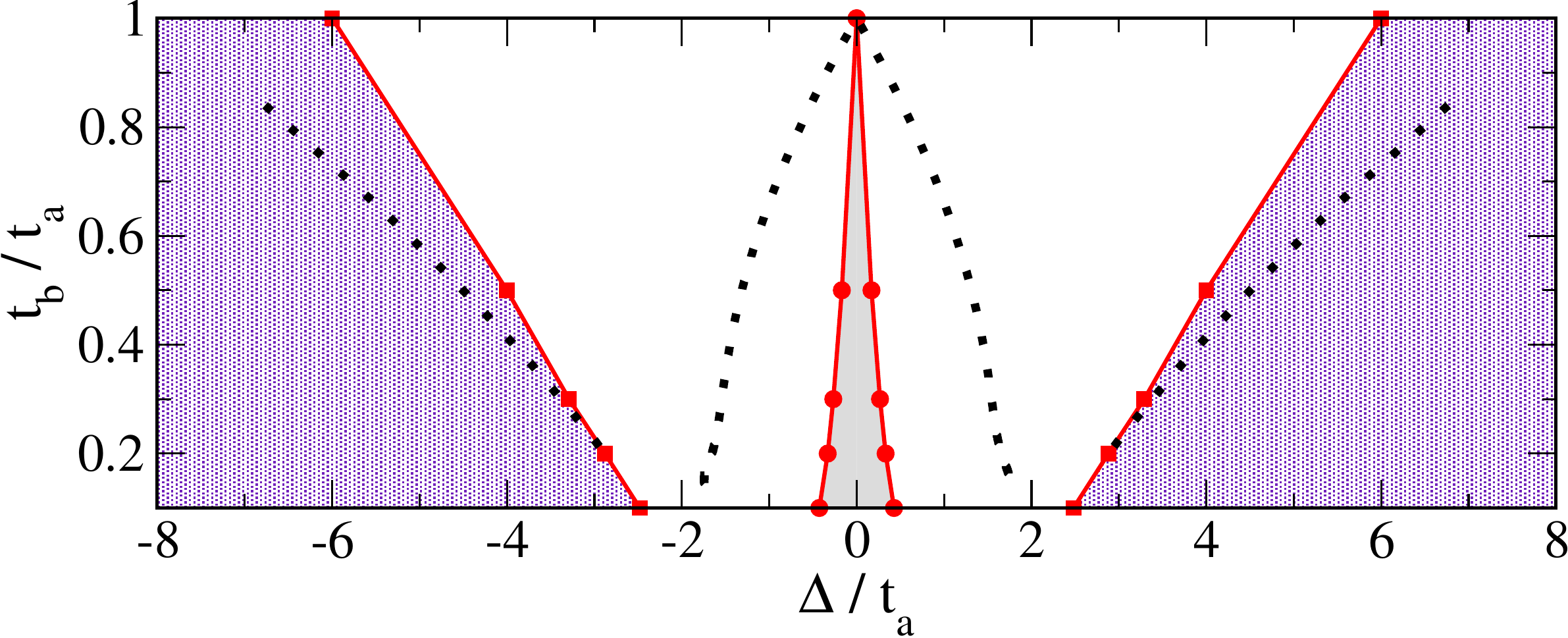}
\includegraphics[width=\columnwidth,clip]{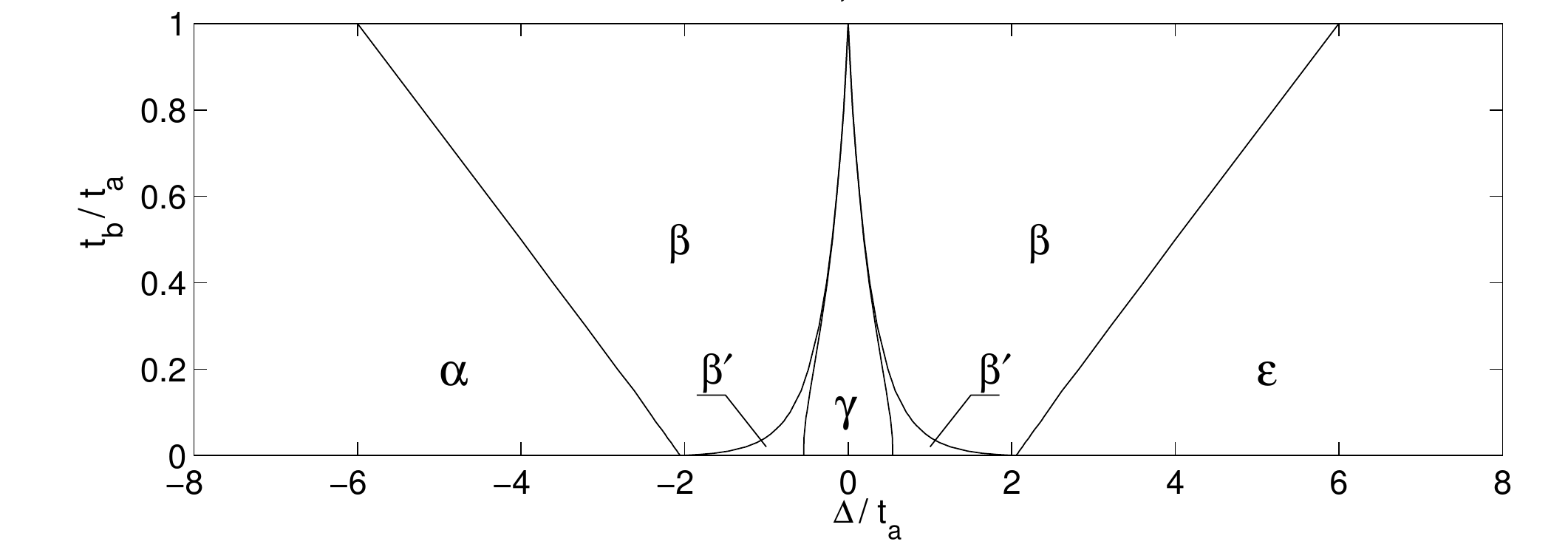}
\caption{\label{fig:efkm} The $T=0$ phase diagram of $d=2$ EFKM for $\tilde{U}/t_a=2$ obtained with constrained path Monte-Carlo~\cite{batista04} (top) and Hartree-Fock approximation~\cite{farkasovsky08} (bottom). The $\alpha$ and $\varepsilon$ (violet shaded) phases correspond to filled $b$ and $a$ band, respectively. The solid $\gamma$ phase (grey shaded) is surrounded by the excitonic (superfluid) phase $\beta$. The supersolid $\beta'$ was not found in the Monte-Carlo simulations on square lattice. The dotted lines in the top panel represent the strong-coupling phase boundaries of Fig.~\ref{fig:xxz}. The bottom panel was adapted with permission of author from Ref.~\onlinecite{farkasovsky08}. 
Copyrighted by the American Physical Society.}   
\end{figure}

Using the constrained path Monte-Carlo method Batista {\it et al.}~\cite{batista04} obtained the $T=0$ phase diagram of the
$d=1,2$ model shown in Fig.~\ref{fig:efkm}. In fact, Farka\v{s}ovsk\'y~\cite{farkasovsky08} showed that self-consistent Hartree-Fock 
method reproduces the Monte-Carlo phase diagram in $d=2$ remarkably well (see Fig.~\ref{fig:efkm}) and extended the study to $d=3$ and $t_b\ll 1$ .
A similar HF phase diagram was obtained by Schneider and Czycholl~\cite{schneider08} for semielliptic  densities of states 
($d=\infty$ Bethe lattice). In both studies, the excitonic phase persists in the limit $|t_b|\rightarrow 0$ for certain range of $\Delta$. Due to the symmetry 
of EFKM under $t_b\leftrightarrow -t_b$, $t_b=0$ is an unstable fixed point between the ferro and antiferro excitonic phases.   
Interestingly, for $t_b\sim 0$ Farka\v{s}ovsk\'y~\cite{farkasovsky08} finds also a phase where both the solid and excitonic orders are present simultaneously
\footnote{The $\beta'$ phase in Ref.~\onlinecite{farkasovsky08}} - a supersolid phase. This result has not been confirmed by other studies
yet and the supersolid phase was shown to be unstable in the strong-coupling limit~\cite{schmid02}. Moreover, existence of a phase with finite $\langle a^{\dagger}b\rangle$ at $t_b=0$, which smoothly connects to both $t_b>0$ and $t_b<0$  violates the Elitzur's theorem~\cite{elitzur75}. 

\subsubsection{BCS-BEC crossover}
The similarity of the strong- and weak-coupling phase diagrams opens an interesting question of how the physics described by the
the hard-core bosons (\ref{eq:strong_efkm}) evolves into the Hartree-Fock physics of fermions in (\ref{eq:efkm}). Similar question 
is known as the BCS-BEC crossover in the context of superconductivity or crossover between Slater and Heisenberg antiferromagnet.
The issue of BCS-BEC crossover in the excitonic phase of half-filled EFKM was studied by several authors using random phase approximation (RPA)~\cite{zenker12},
slave-boson~\cite{zenker10,zenker11}, projective renormalisation~\cite{phan10}, variational cluster~\cite{seki11}, exact diagonalisation~\cite{kaneko13}, and
density-matrix renormalisation group~\cite{ejima14} techniques. The different methods provide quite consistent picture of the phase
diagram with the provision that long-wavelength fluctuations of the order parameters are ignored by some of the mean-field methods, 
which therefore describe critical phase in $d=1, T=0$ and the Kosterlitz-Thouless phase in $d=2$ as phases with true long-range order.
\begin{figure}
\includegraphics[width=0.9\columnwidth,clip]{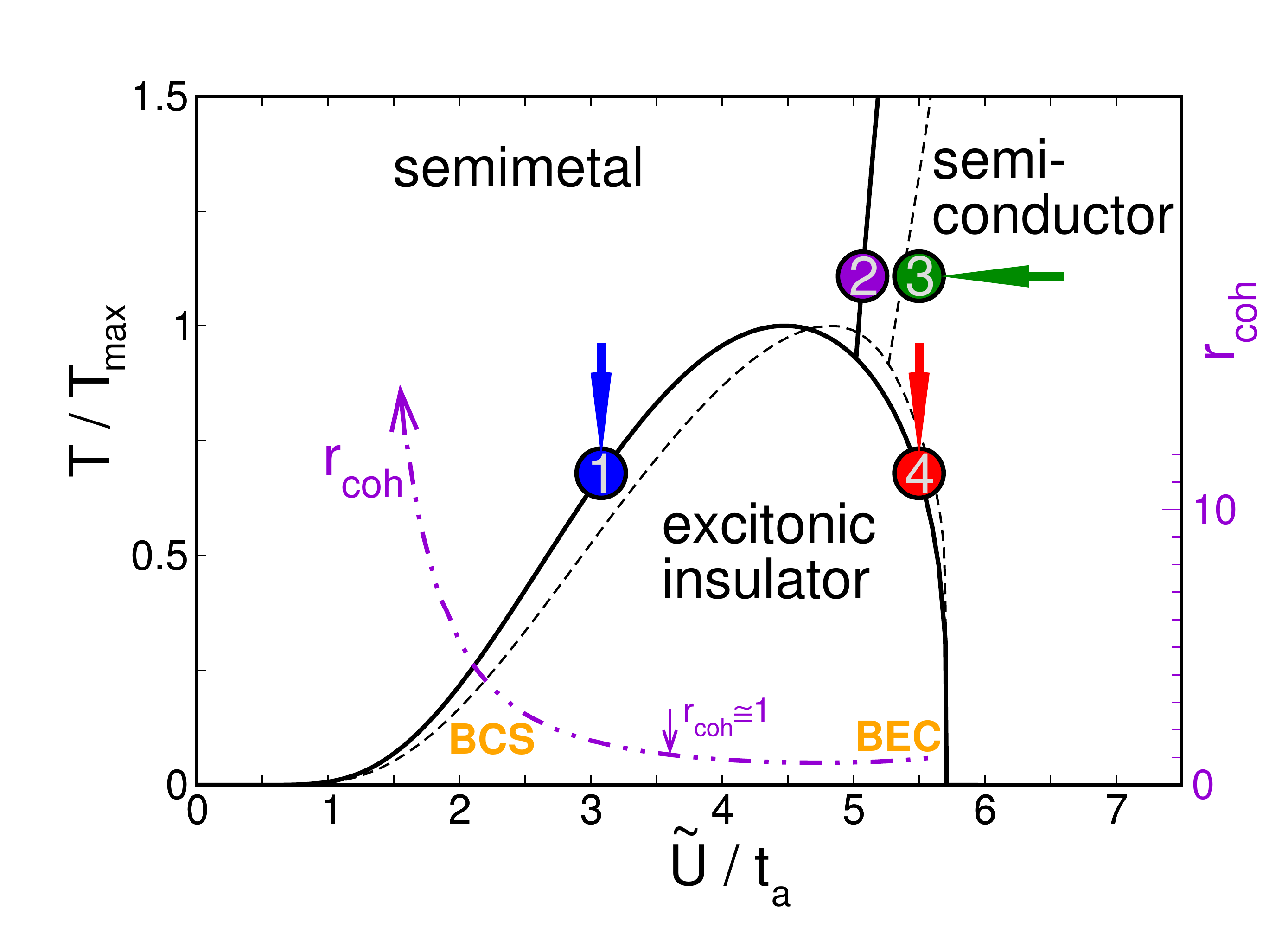}
\caption{\label{fig:fehske1} The phase diagram of $d=2$ half-filled EFKM for $\Delta=2.4$, $t_a=1$, $t_b=-0.8$ obtained with RPA (solid lines)
and slave bosons (dashed lines). Note that the temperature is scaled to the maximum critical temperature $T_{\text{max}}$, where $T_{\text{max}}=0.361$ (RPA) and $T_{\text{max}}=0.256$ (slave bosons). The points marked 1-4 are discussed in the text.
Adapted with permission of authors from Ref.~\onlinecite{zenker12}. Copyrighted by the American Physical Society.}
\end{figure} 

\begin{figure}[t]
\subfigure[ $\tilde{U}=0.50$,\newline $T/T_{\text{max}}=1.108$]{
\includegraphics[height=0.25\linewidth]{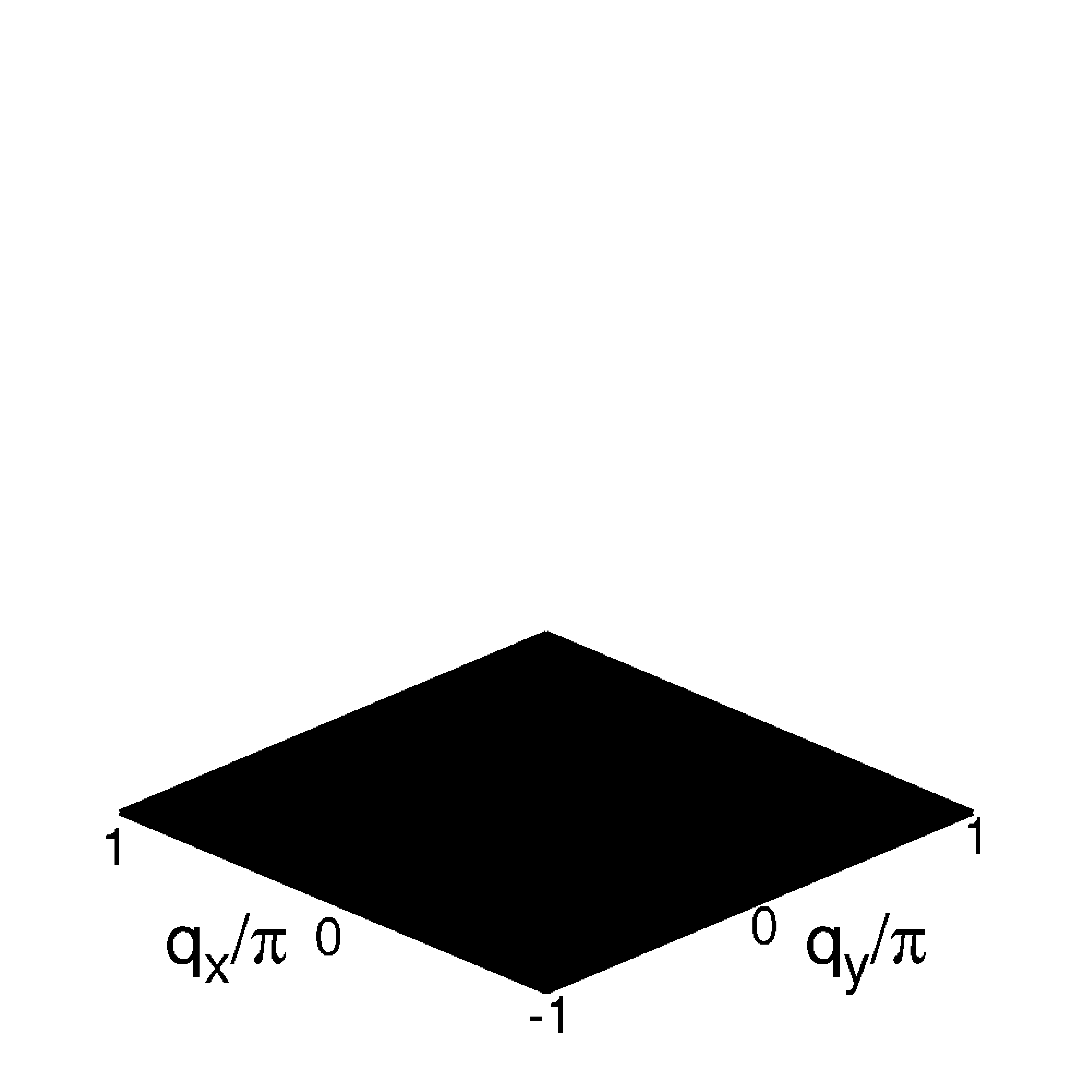}
}
\subfigure[ $\tilde{U}=3.08$,\newline $T/T_{\text{max}}=0.679$]{
\includegraphics[height=0.25\linewidth]{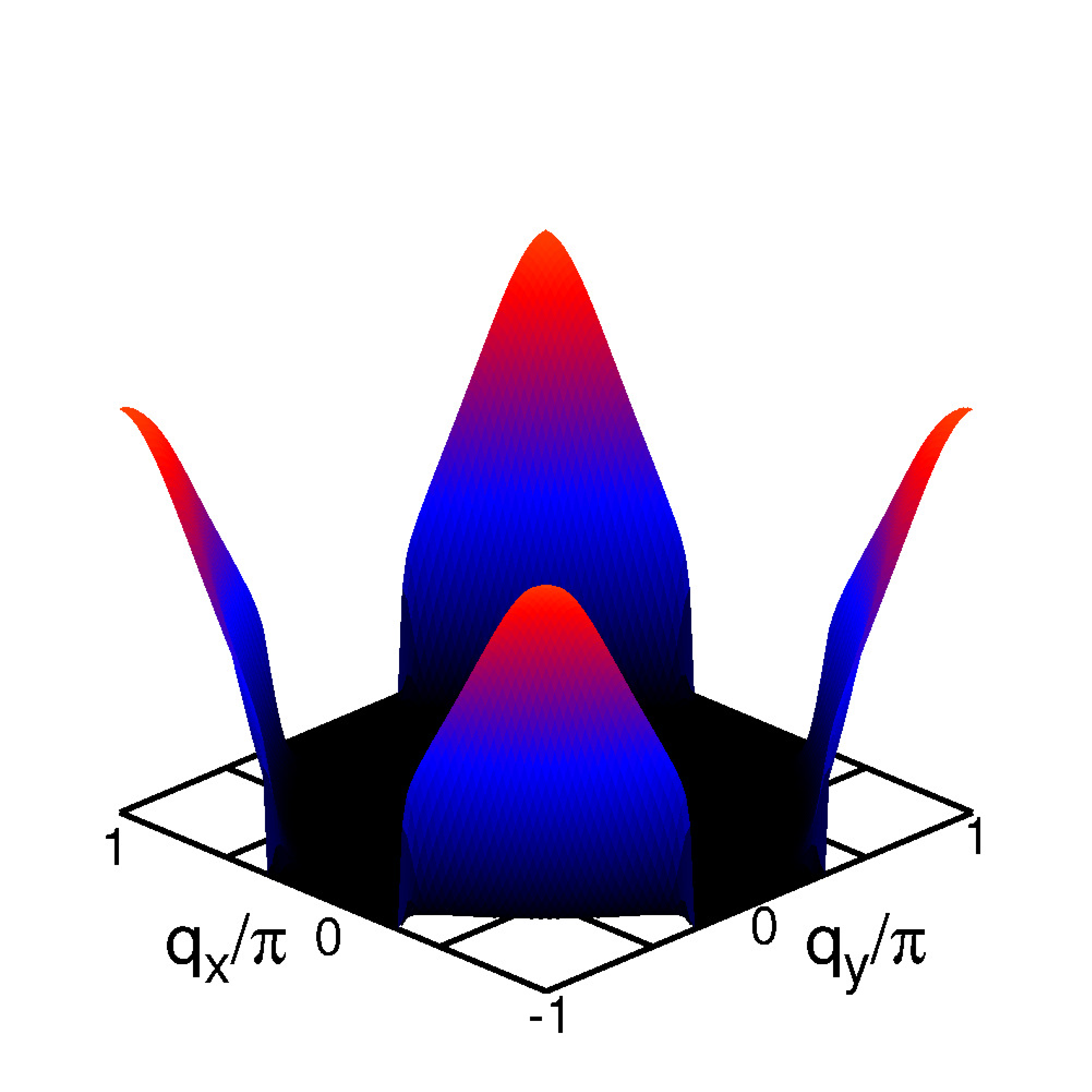}
}
\subfigure[ $\tilde{U}=5.07$,\newline $T/T_{\text{max}}=1.108$]{
\includegraphics[height=0.25\linewidth]{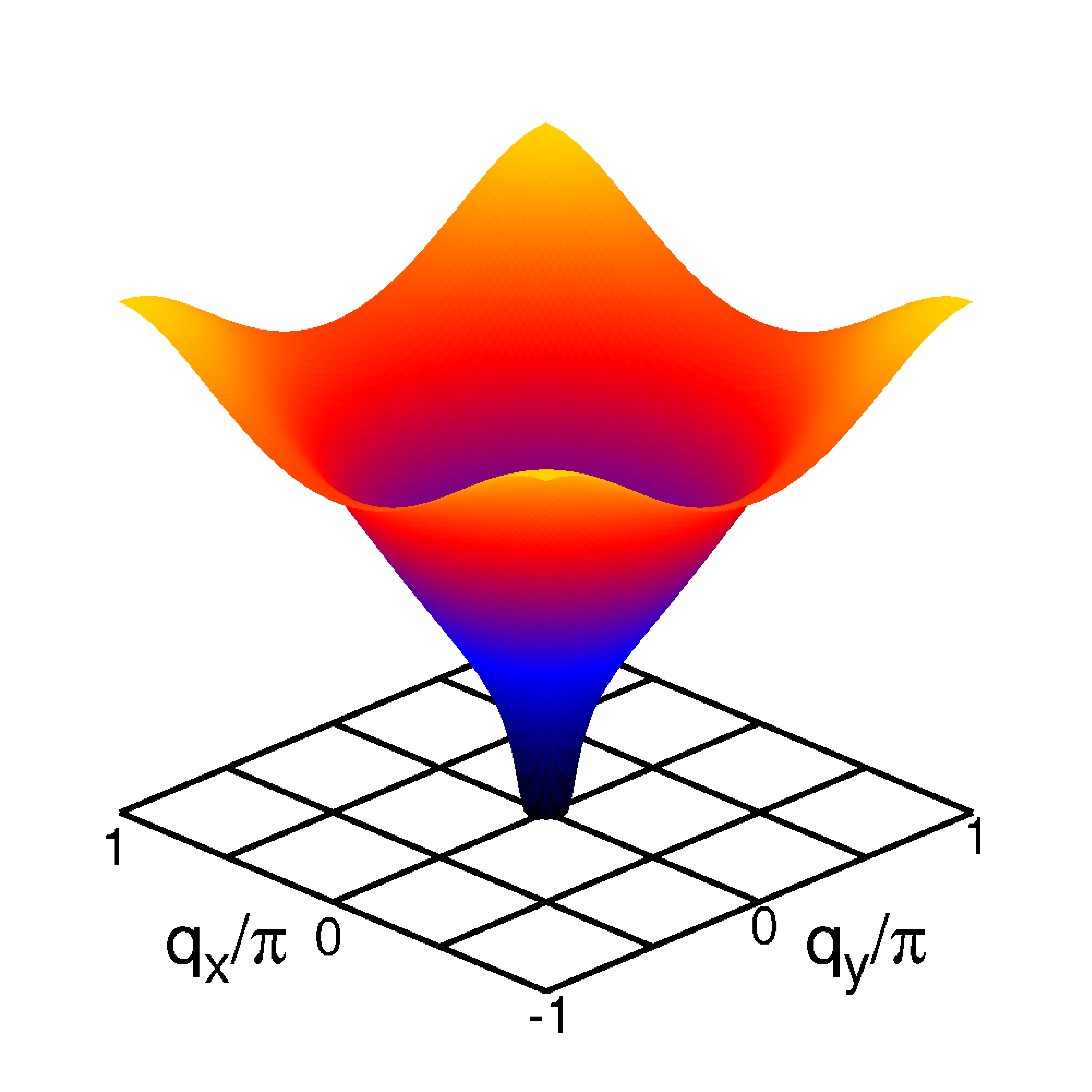}
}
\subfigure{
\includegraphics[height=0.25\linewidth]{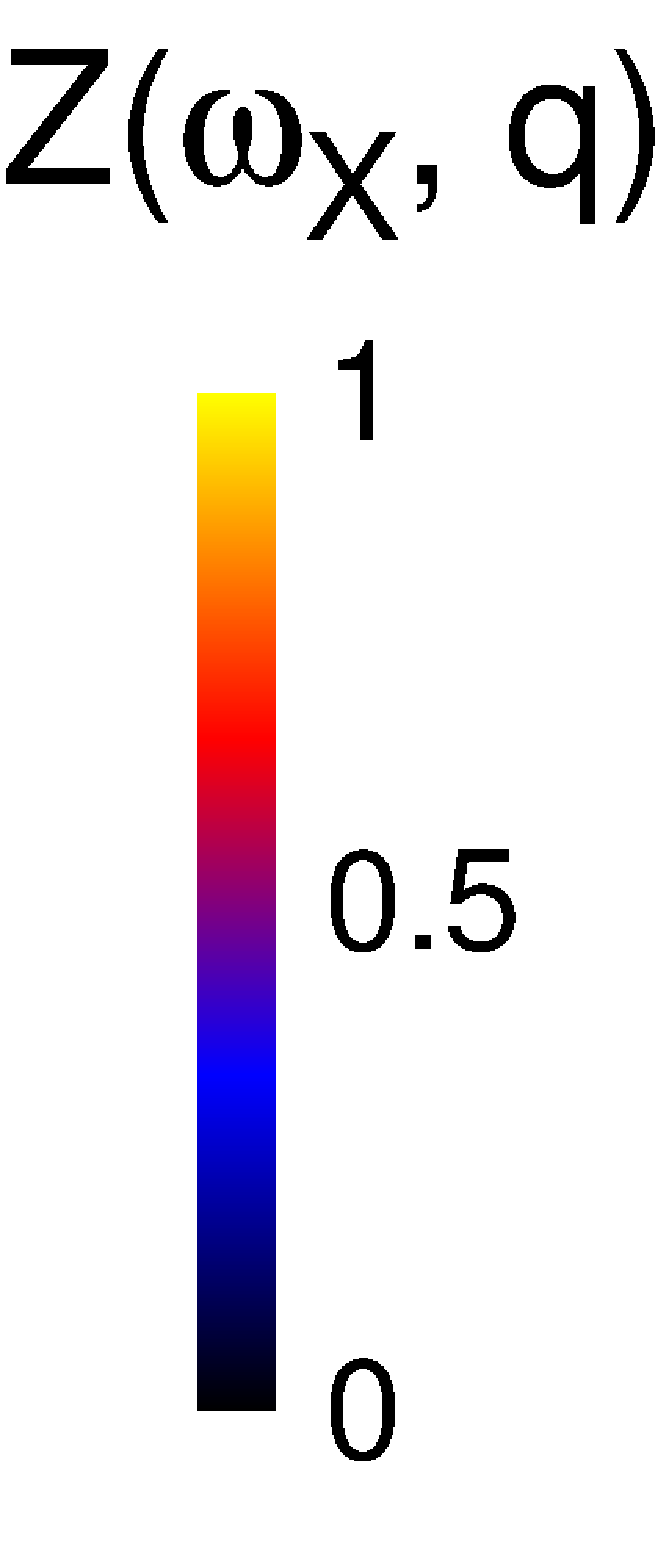}
}

\subfigure[$\tilde{U}=5.50$,\newline$T/T_{\text{max}}=1.108$]{
\includegraphics[height=0.25\linewidth]{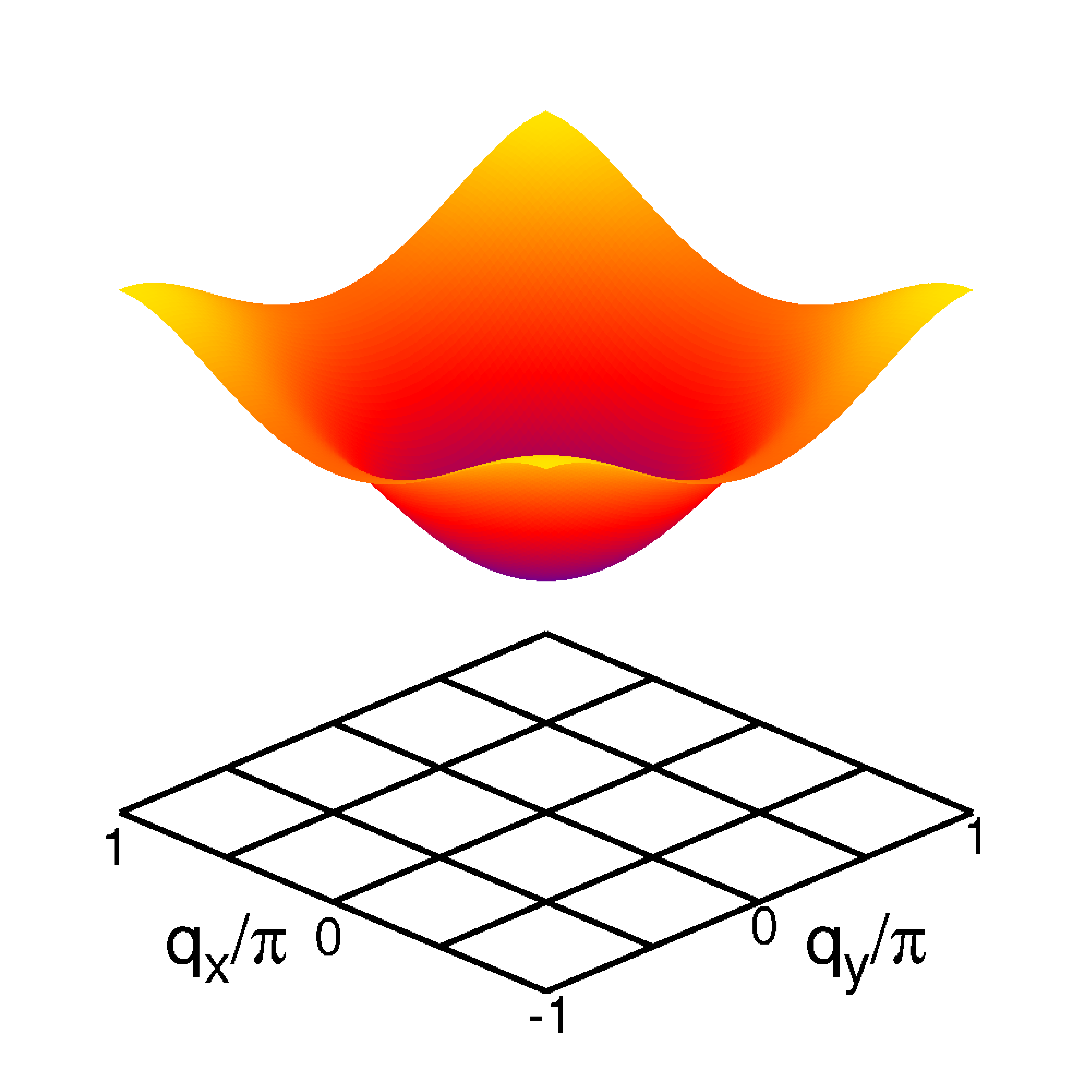}
}
\subfigure[$\tilde{U}=5.50$,\newline $T/T_{\text{max}}=0.679$]{
\includegraphics[height=0.25\linewidth]{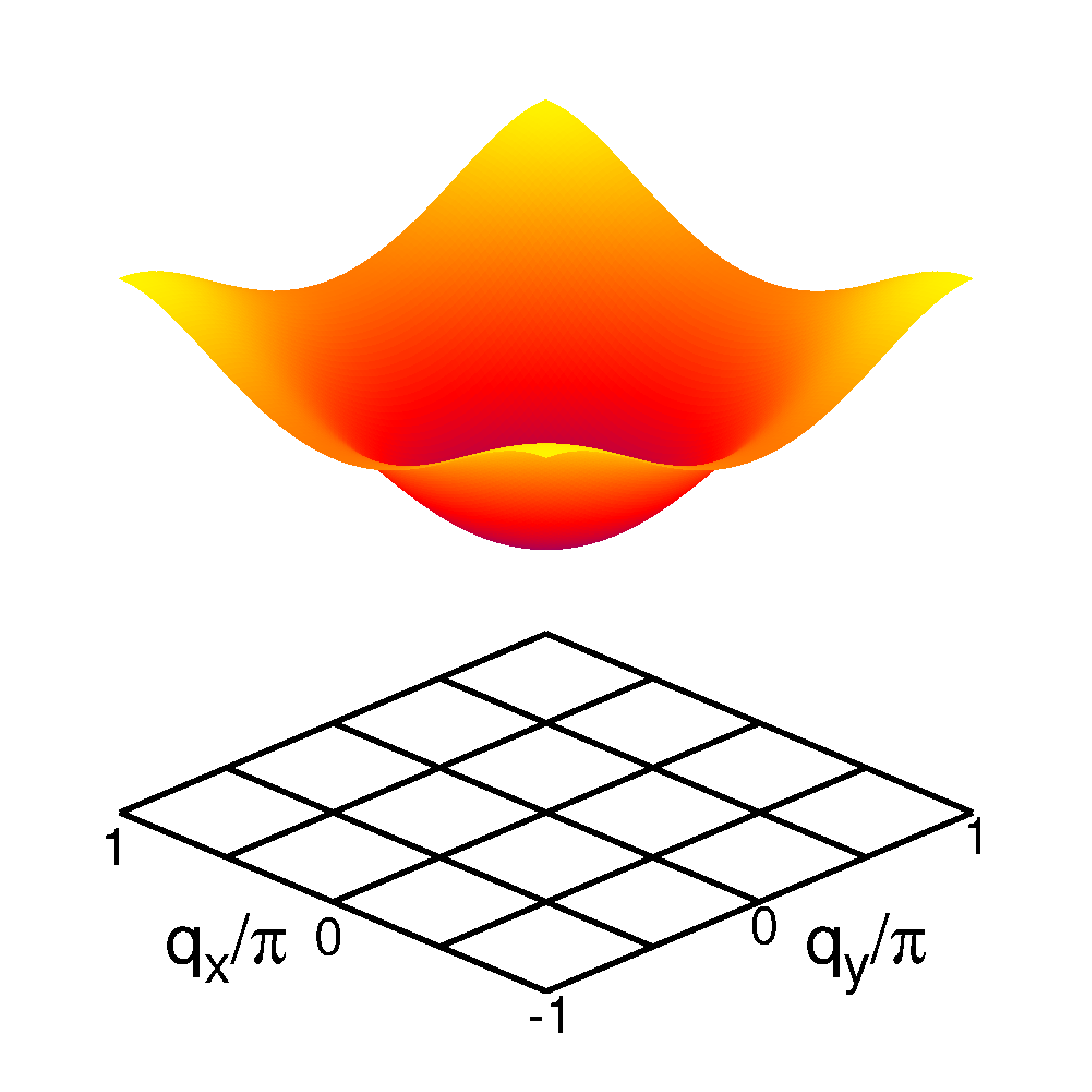}
}
\subfigure[$\tilde{U}=50.0$,\newline $T/T_{\text{max}}=1.108$]{
\includegraphics[height=0.25\linewidth]{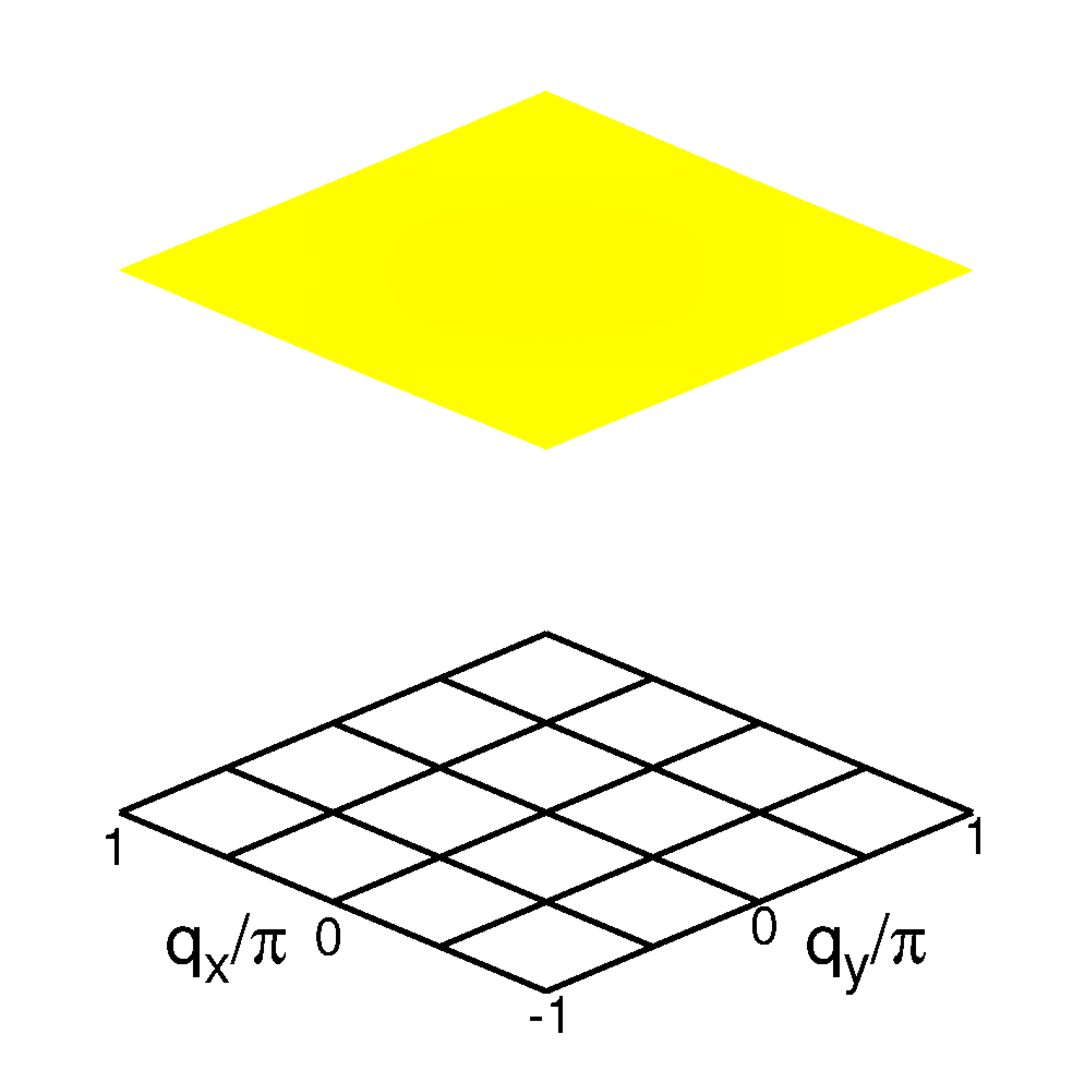}
}
\subfigure{
\includegraphics[height=0.25\linewidth]{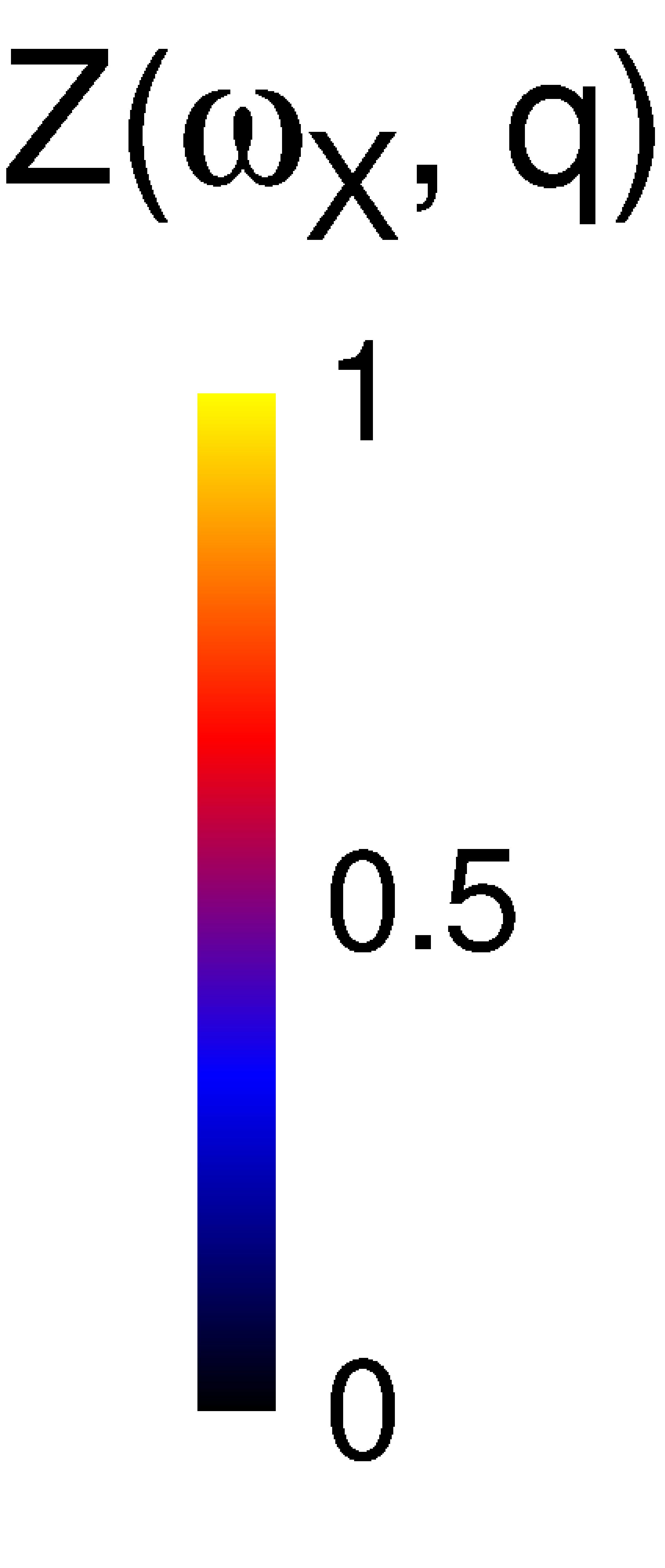}
}
\caption{\label{fig:zx} Evolution of the exciton quasiparticle weight $Z_X(\bq)$, obtained with RPA, across the semimetal--semiconductor crossover.
On the metallic side (a-c) non-zero $Z_X(\bq)$ exists only in a part of the Brillouin zone. On the semiconductor side (e-g), $Z_X(\bq)$ is finite everywhere in the
Brillouin zone and approaches $Z_X(\bq)=1$ in the strong-coupling limit (g). Panel (a) corresponds to a weakly coupled semimetal with no excitonic quasi-particle
residues anywhere in the Brillouin zone. Panels (b)-(f) correspond to the points 1-4 of Fig.~\ref{fig:fehske1}. Adapted with permission of authors from Ref.~\onlinecite{zenker12}. Copyrighted by the American Physical Society.}
\end{figure}

\begin{figure}
\includegraphics[width=0.4\columnwidth,clip]{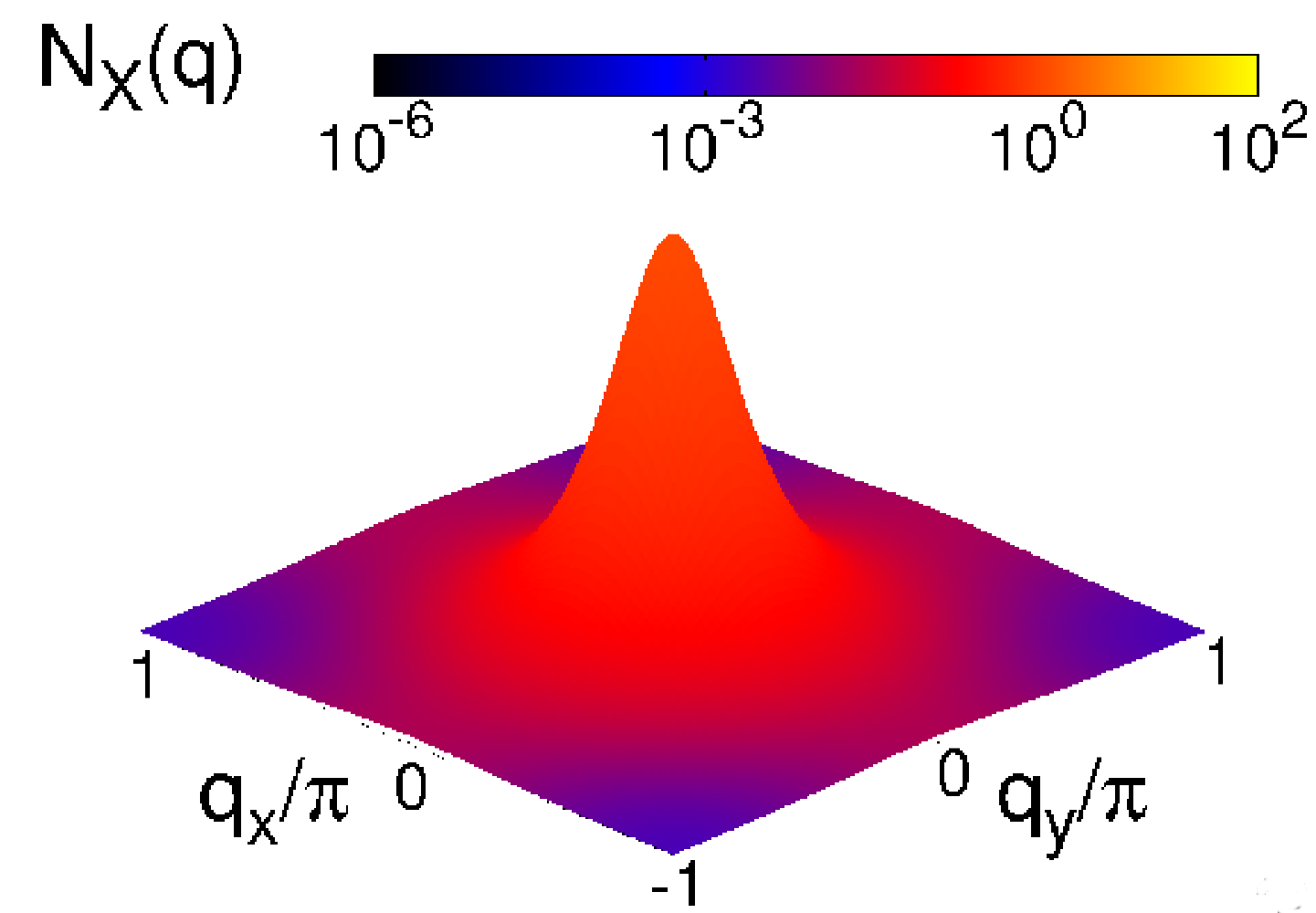}
\includegraphics[width=0.4\columnwidth,clip]{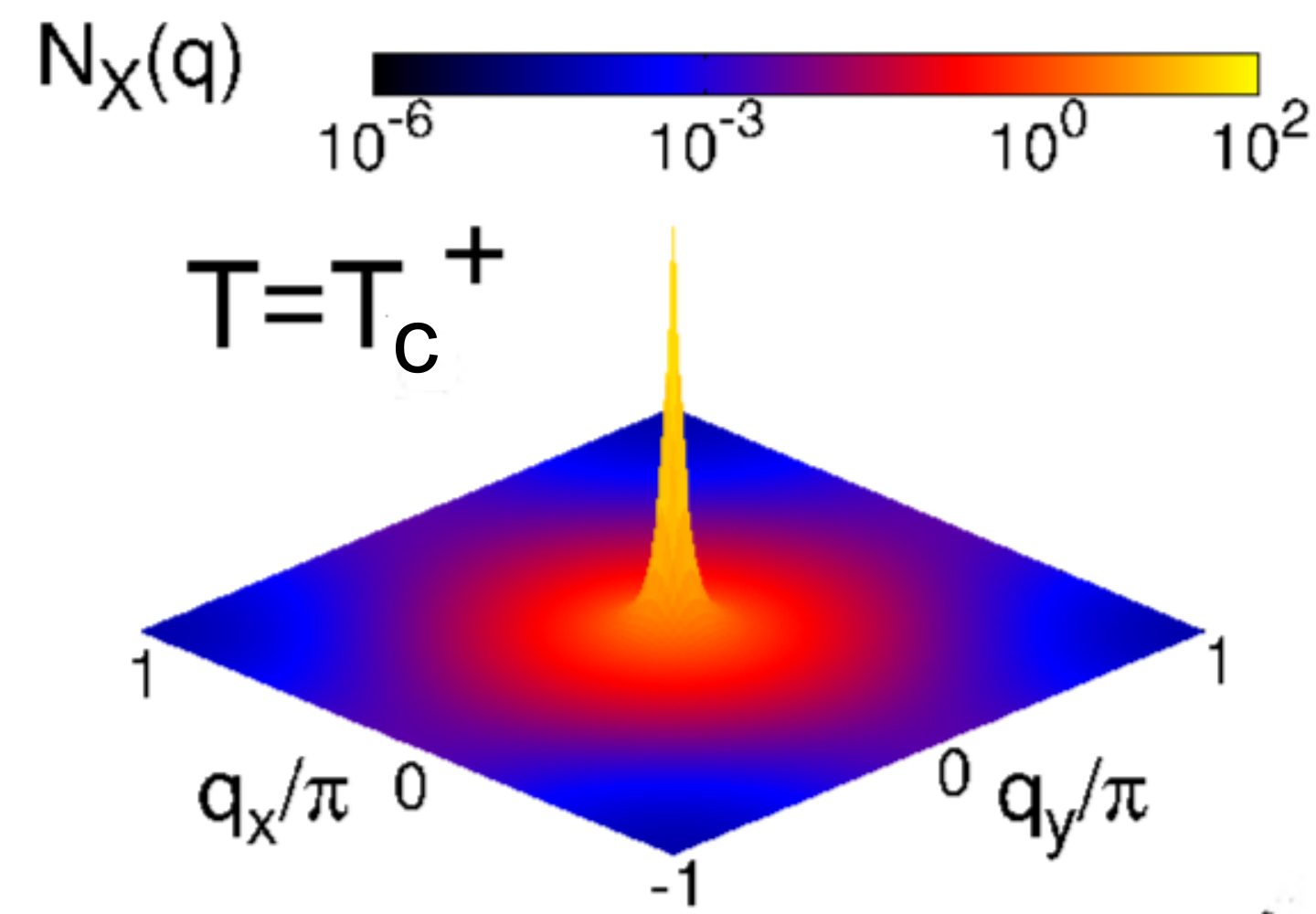}
\caption{\label{fig:nx} The occupancy of the excitonic states $N_X(\bq)$ at points 3 (left panel) and 4 (right panel) of Fig.~\ref{fig:fehske1} on the semiconductor side of the phase diagram. The diverging occupancy of the lowest mode, $\bq=(0,0)$, at $T_c$ is clearly visible. Adapted with permission of authors from Ref.~\onlinecite{zenker12}. Copyrighted by the American Physical Society.}
\end{figure}

The BCS vs BEC question can be formulated as: Are there long-lived excitons above $T_c$ that condense
at the transition? The typical $T-U$ phase diagram of half-filled EFKM with exciton condensate is
shown in Fig.~\ref{fig:fehske1}.
Following Ref.~\onlinecite{zenker12} we introduce a boson creation operator 
$d_{\bq}^{\dagger}=\tfrac{1}{\sqrt{N}}\sum_{\bk}a_{\bk+\bq}^{\dagger}b^{\phantom\dagger}_{\bk}$ and the corresponding propagator $\chi^{ab}_{\bq}(\omega)=\langle\langle d^{\phantom\dagger}_{\bq};d^{\dagger}_{\bq}\rangle\rangle_{\omega}$. The excitonic transition proceeds differently on the semi-metal and semi-conductor sides. 
In semimetal, the particle-hole continuum extends to zero energy in a finite $\bq$-region of the Brillouin zone. While long-lived excitons may exist 
in some other parts of the Brillouin zone (Fig.~\ref{fig:zx}b), they are not the lowest energy excitations and the excitonic transition follows the BCS scenario.
In semiconductor, the excitonic band in the two-particle spectrum extends throughout the whole Brillouin zone (Fig.~\ref{fig:zx}e-g) and lies below the particle-hole continuum. $B_{\bq}(\omega)$ , the coherent part of $\operatorname{Im}\chi^{ab}_{\bq}(\omega)$, can be approximated as
\begin{equation}
\label{eq:zx}
B_{\bq}(\omega)=-\pi Z_X(\bq)\delta(\omega-\omega_X(\bq)),
\end{equation}
where $Z_X(\bq)$ is the weight of the excitonic quasiparticle and $\omega_X(\bq)$ is its dispersion. Note that both $Z_X(\bq)$ and $\omega_X(\bq)$ depend on temperature. The number of excitons
with crystal-momentum $\bq$ is given by
\begin{equation}
\label{eq:nx}
N_X(\bq)=Z_X(\bq)n_{\text{BE}}(\omega_X(\bq)),
\end{equation}
where $n_{\text{BE}}(\omega)$ is the Bose-Einstein (BE) function. The excitonic transition in a semiconductor is connected
with the minimum of $\omega_X(\bq)$ going to zero at $T_c$ and the resulting divergence of $N_X(\bq)$, Fig.~\ref{fig:nx}.

The semimetal and semiconductor side of the phase diagram appear to be qualitatively different.
It is therefore instructive to see that the evolution from a semimetal to a semiconductor with increasing $\tilde{U}$ is in fact
smooth although they differ by the presence of the excitonic band. The key observation is that the parts of the exciton band that split off
the particle-hole continuum carry little spectral weight $Z_X(\bq)$. Moreover, the notion of infinitely sharp $B_{\bq}(\omega)$ is an idealisation,
which holds only for excitonic peak well separated from the particle-hole continuum.
Fig.~\ref{fig:zx} shows the evolution of $Z_X(\bq)$. Starting from a semimetal with no exciton band, through a semimetal with exciton band
in a part of the Brillouin zone, the region of $Z_X(\bq)=0$ shrinks to a point at the semimetal/semiconductor crossover. On the semiconductor
side, $Z_X(\bq)$ remains finite over the entire Brillouin zone. Vanishing of the excitonic insulator for large $\tilde{U}$ is the consequence of $\omega_X(\bq)>0$
being finite for all $\bq$ down to $T=0$.

Another question worth understanding is the relationship of the present picture to the strong-coupling limit $t_{a,b}\ll \tilde{U}$. We point out that to reach this limit
one cannot keep $\Delta$ fixed but have to scale it as $1/\tilde{U}$ (see also Fig.~\ref{fig:xxz}). Although the bosonic excitons exist in a semiconductor away from the 
strong-coupling limit, there is no separation of the bosonic and fermionic energy scales. This is reflected in the $Z_X(\bq)$ substantially different from
1. The dynamics of the excitons therefore cannot be described by purely bosonic Hamiltonian. One has the option to keep the fermions in the model, use 
a more general action description of the exciton dynamics or resort to an effective bosonic Hamiltonian describing only the vicinity of the minimum of the exciton band. While the particle-hole gap grows linearly with increasing $\tilde{U}$, the exciton band remains located around $\Delta$ with its width approaching 
the $t_a t_b/\tilde{U}$ scaling. If the particle-hole gap is sufficiently large, the excitons are no more dressed with the particle-hole excitations
and thus $Z_X(\bq)\approx 1$ everywhere in the Brillouin zone. The exciton dynamics in this limit is described by purely bosonic Hamiltonian (\ref{eq:boson}).


\section{Two-band Hubbard model (S=1/2 fermions)}
\label{sec:2bhm}
The two-band Hubbard model (2BHM) generalises EFKM (\ref{eq:efkm}) to include electron spin. In absence of
external magnetic field and spin-orbit coupling the one-particle part $H_{\text{t}}$ consists of two identical 
copies of the corresponding terms in EFKM. The interaction part $H_{\text{int}}$ is richer.
\begin{equation}
\begin{aligned}
\label{eq:2bhm}
H_{\text{2BH}}=&H_{\text{t}}+H_{\text{int}}\\
H_{\text{t}}=&\sum_{\rij,\sigma} \left(
  t_aa_{i\sigma}^{\dagger}a^{\phantom\dagger}_{j\sigma}+
   t_bb_{i\sigma}^{\dagger}b^{\phantom\dagger}_{j\sigma}
   \right)+H.c.\\
   +&\sum_{\rij,\sigma} \left(
   V_{ab}b_{i\sigma}^{\dagger}a^{\phantom\dagger}_{j\sigma}+
   V_{ba}a_{i\sigma}^{\dagger}b^{\phantom\dagger}_{\j\sigma}
   \right)+H.c.\\
   +&\frac{\Delta}{2}\sum_{i,\sigma} \left(n^a_{i\sigma}-n^b_{i\sigma}\right)\\
H_{\text{int}}=&U\sum_{i} \left(n^a_{i\uparrow}n^a_{i\downarrow}+n^b_{i\uparrow}n^b_{i\downarrow}\right)+
  U'\sum_{i,\sigma\sigma'} n^a_{i\sigma}n^b_{i\sigma'}\\
-&J\sum_{i\sigma} \left(n^a_{i\sigma}n^b_{i\sigma} 
+a_{i\sigma}^{\dagger}a_{i\bar{\sigma}}^{\phantom\dagger}b_{i\bar{\sigma}}^{\dagger}b_{i\sigma}^{\phantom\dagger}\right)\\
+&J'\sum_{i} \left(a_{i\uparrow}^{\dagger}a_{i\downarrow}^{\dagger}b_{i\downarrow}^{\phantom\dagger}
b_{i\uparrow}^{\phantom\dagger}\right)+H.c.,
\end{aligned}
\end{equation}
with the notation $\bar{\sigma}=-\sigma$.
There are two general set-ups in which EC in 2BHM have been studied: a lattice of two-orbital atoms and a bi-layer Hubbard model
system.  

In the bi-layer model the orbitals $a_i$ and $b_i$ are assumed to be spatially well separated and 
the exchange-interaction is vanishingly small $J',J\approx0$. This is equivalent to the dominant-term approximation
used extensively for long-range interaction~\cite{halperin68b}. Similarly the on-site and inter-site inter-layer tunnelling, 
$V_{ab},V_{ba}\approx 0$, is negligible due to the spatial separation of
layers.

In two-orbital atoms, the overlap of (orthogonal) $a$ and $b$ orbitals gives rise to a sizeable ferromagnetic Hund's exchange $J$
and  pair hopping $J'$. The on-site $a$-$b$ hopping vanishes either by symmetry or can be eliminated by a basis transformation.
The cross-hopping $V_{ab}$, $V_{ba}$ is in general non-zero, however, in materials with high symmetry it may vanish as well.~\footnote{For example
for $xy$ and $x^2-y^2$ orbitals on a square lattice}.

\subsection{Normal state}
\begin{figure}
\includegraphics[width=0.33\columnwidth]{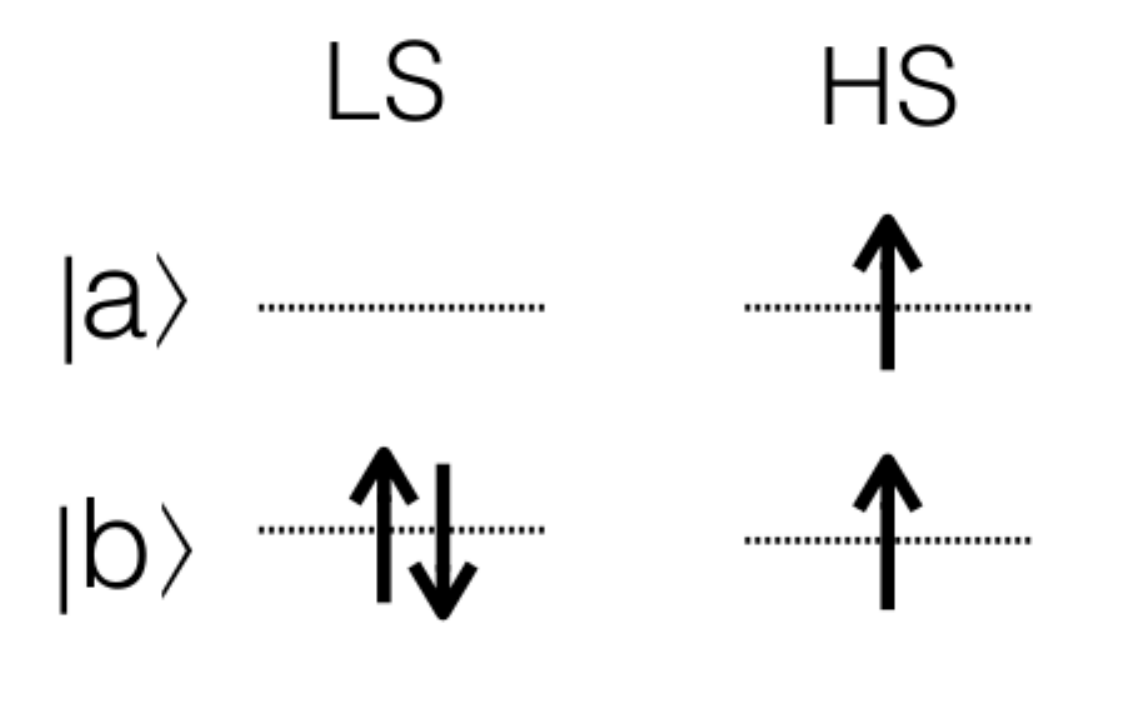}
\includegraphics[width=0.65\columnwidth,clip]{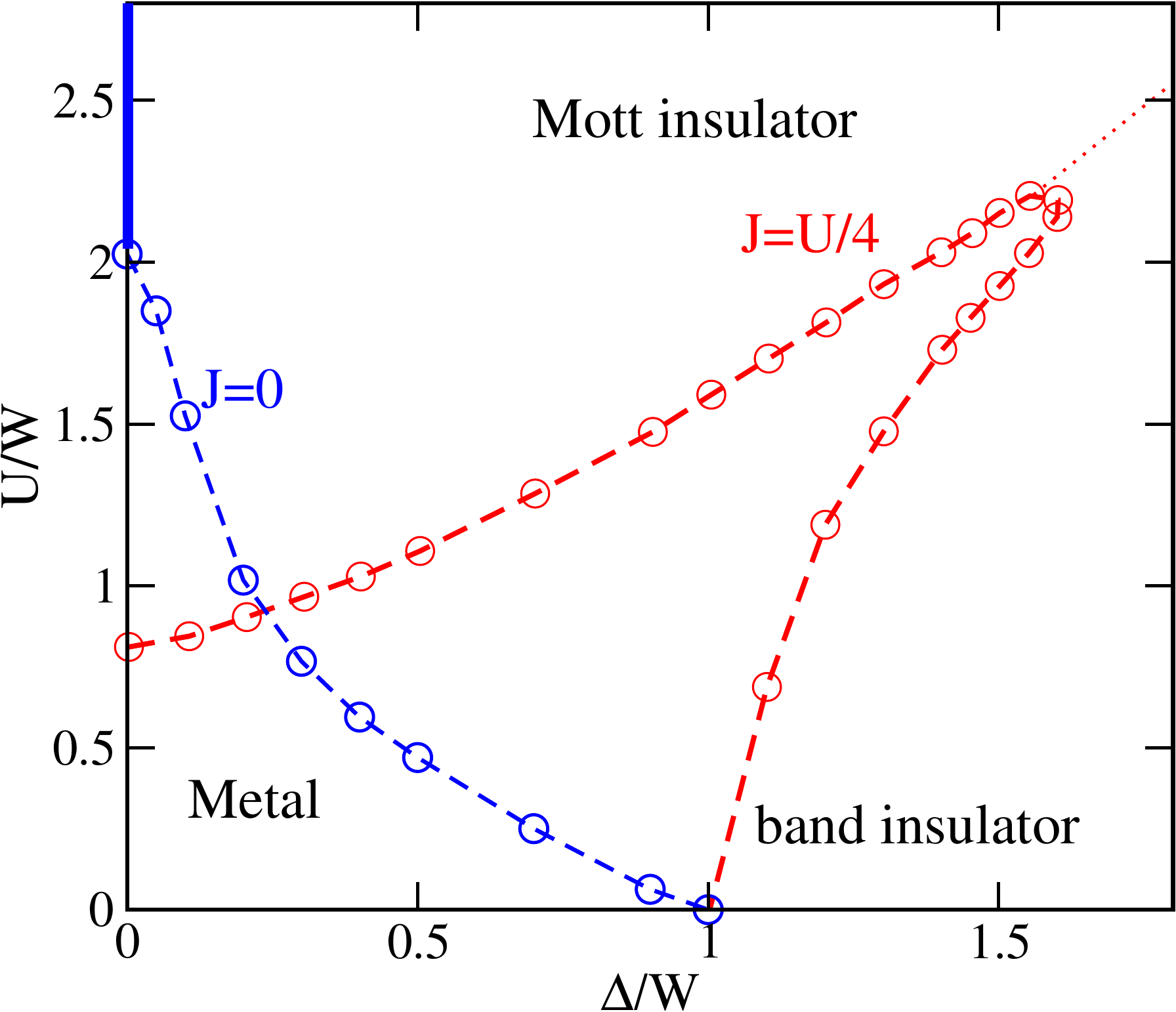}
\caption{\label{fig:werner} Left: The cartoon representation of the atomic low-spin (LS) and high-spin (HS) states. Right:
The phase diagram of $d=\infty$ 2BHM with symmetric $t_a=t_b$ bands (semi-elliptic density of states)
for $J'=J=U/4$; $U'=U-2J$ (red) and $J'=J=0$; $U'=U$ (blue)  obtained by Werner and Millis~\cite{werner07}.
The dotted red line marks  the HS-LS crossover. The vertical blue line for $J=0$ corresponds to a model with 6-fold degenerate atomic ground state. 
$U$ and $\Delta$ are expressed in the units of bandwidth $W$.}
\end{figure}

Before discussing the ordered phases we briefly summarise the basic properties of half-filled 2BHM without broken symmetry. 
Its physics at strong and intermediate coupling is controlled by competition between the Hund's coupling $J$ and the crystal-field splitting
$\Delta$~\cite{werner07,suzuki09}. Large $\Delta$ favours the singlet low-spin (LS) state, while large $J$ favours the triplet high-spin (HS) state,
see Fig.~\ref{fig:werner}. For finite $J$ the phase diagram, shown in Fig.~\ref{fig:werner}, contains three regions:
HS Mott insulator connected to the limit $U\gg W$, with $W$ standing for the bare bandwidth, and $J\gg\Delta$, LS band insulator connected to the limit $\Delta\gg W,U$ and a metal
connected to the non-interacting limit and $\Delta<W$. The low-energy physics deep in the Mott phase is described by $S=1$ Heisenberg model with anti-ferromagnetic interaction. 
The band insulator far away from the phase boundaries is a global singlet separated by large gap from the excited states. In the vicinity 
of the HS-LS crossover both LS and HS states have to be taken into account. The physics arising in this parameter region is the subject of subsequent 
sections. 

The physics of the $J,J'=0$ model is different. This setting corresponds to a bi-layer system where one would typically choose $U>U'$. In this case, 
the low-energy physics for $W\ll U$ and $\Delta\ll (U-U')$ is described by $S=1/2$ bi-layer Heisenberg model, while for $\Delta\gg (U-U')$ one obtains
a band insulator. The region $\Delta\approx (U-U')$ is described by exciton-\tj model, discussed later. The choice $U=U'$, shown in Fig.~\ref{fig:werner},
is anomalous in the sense that the bi-layer Heisenberg region is absent and the region $\Delta\approx0$ corresponds to exciton-\tj model with two exciton
flavours.

The low-energy physics of the metallic phase is sensitive to Fermi surface nesting, in particular for weak coupling. Nesting plays an important role in some popular models, e.g.,
the Fermi surfaces derived from $a$ and $b$ bands are perfectly nested on cubic lattice both in $d=2$ and $d=3$.

\subsection{Excitonic order parameter}
Excitonic condensation is characterised by a spontaneous coherence between the $a$- and $b$-electrons in (\ref{eq:2bhm}), i.e., 
appearance of matrix elements of the form 
\begin{equation}
F^{\sigma\sigma'}_{ii'}=\langle a^{\dagger}_{i\sigma}b^{\phantom\dagger}_{i'\sigma'}\rangle
\quad\text{and}\quad
\mathcal{F}^{\sigma\sigma'}_{\bk\bk'}=\langle a^{\dagger}_{\bk\sigma}b^{\phantom\dagger}_{\bk'\sigma'}\rangle.
\end{equation}
These can be in general complicated objects characterised by translational symmetry, internal structure and spin symmetry. 

We will consider only ordered phases with \mbox{single-$\bq$} translational symmetry, ${\mathcal{F}_{\bk\bk'}\sim \delta_{\bk',\bk+\bq}}$. 
On the square lattice we will encounter only ferro-EC, $\bq=0$, and antiferro-EC, $\bq=(\pi,\pi)$ states,
which correspond to uniform ${F_{ii}}$ and staggered
${F_{ii}\sim(-1)^i}$, respectively.

The internal structure describes the behaviour of the above matrix elements as a function of the reciprocal vector $\bk$ or the relative position $\bR_i-\bR_{i'}$. 
The internal structure reflects the symmetry of the pairing interaction. 
Isotropic pairing leads to isotropic $\mathcal{F}_{\bk,\bk+\bq}$ and, in particular, to a finite local element $F_{ii}$. 
Since this is the case for local Hubbard interaction (\ref{eq:2bhm}) we can use the local element $F_{ii}$ as an order parameter
for the exciton condensation.
The spatial decay of the $a-b$ coherence can be quantified by a correlation length $r_{\text{coh}}$ defined
as 
\begin{equation}
\label{eq:rcoh}
r_{\text{coh}}=\sqrt{\frac{\sum_{\bR} |\bR|^2 |F_{\bR_i,\bR_i+\bR}|^2}{\sum_{\bR} |F_{\bR_i,\bR_i+\bR}|^2}}=
\sqrt{\frac{\sum_{\bk} |\nabla_{\bk}\mathcal{F}_{\bk\bk}|^2}{\sum_{\bk} |\mathcal{F}_{\bk\bk}|^2}}
\end{equation}
was studied by several authors for EFKM~\cite{seki11,kaneko13,ejima14} as well as 2BHM~\cite{kaneko14}. Since the largest contribution
to the pairing interaction in these models is on-site the correlation length reflects the relative size of the 
pairing field to the bandwidth. Weak pairing yields sizeable $\mathcal{F}_{\bk\bk}$ only in the vicinity of the 
Fermi level and one ends with $r_{\text{coh}}$ over many unit cells (BCS limit). Strong pairing leads to 
almost constant $\mathcal{F}_{\bk\bk}$ and $r_{\text{coh}}$ limited to a few sites (BEC limit).

Finally, we discuss the spin structure of the EC order parameter, which we divide into the spin-singlet and
spin-tripels parts
\begin{equation}
F^{\sigma\sigma'}_{ii}=\frac{1}{2}\left(\phi_{i}^s\delta_{\sigma\sigma'}+\bph^t_i\cdot\btau^*_{\sigma\sigma'}\right),
\end{equation}
i.e., singlet $\phi^s$ and triplet $\bph^t$ components are defined by
\begin{equation}
\label{eq:opar}
\begin{split}
\phi^s_{i}&=\sum_{\sigma}\langle a^{\dagger}_{i\sigma}b^{\phantom\dagger}_{i\sigma}\rangle\\
\bph^t_{i}&=\sum_{\sigma\sigma'}\langle a^{\dagger}_{i\sigma}b^{\phantom\dagger}_{i\sigma'}\rangle\btau_{\sigma\sigma'},
\end{split}
\end{equation}
where $\btau$ are the Pauli matrices and $^*$ denotes the complex conjugation. Tensorial character of $\{\phi^s,\bph^t\}$ and the fact that their elements are complex
numbers allow numerous distinct phases as will be discussed in what follows. 
Some of the phases lead to a uniform spin polarisation while others have no ordered spin moments.

\subsection{Strong-coupling limit}
At half-filling and large $U,U'\gg t_a, t_b,V_{ab},V_{ba},J,J'$ the charge fluctuations are strongly suppressed. Similar to the strong-coupling treatment of 
EFKM, the low-energy physics can be described by an effective model built on states containing only doubly occupied sites. 
The virtual excitations to states with singly and triply occupied sites provide inter-site couplings in the low-energy model. 
Using Schrieffer-Wollf transformation to the second order one arrives at expressions of the type $t^2/U$, known from the analogous
transformation from single-band Hubbard to Heisenberg model.
By construction, the low-energy model does not capture one-particle or charge excitations, which take place only at energies of the order $U$.
The general strong-coupling expressions for the case of symmetric bands $t_a=t_b$ were derived by Balents~\cite{balents00b}. In the following 
we discuss several special parameter choices of the strong-coupling model, which are studied in the literature.

\subsubsection{Large $J$ - general formulation}
\label{sec:Jgf}
\begin{figure}
\includegraphics[width=0.9\columnwidth]{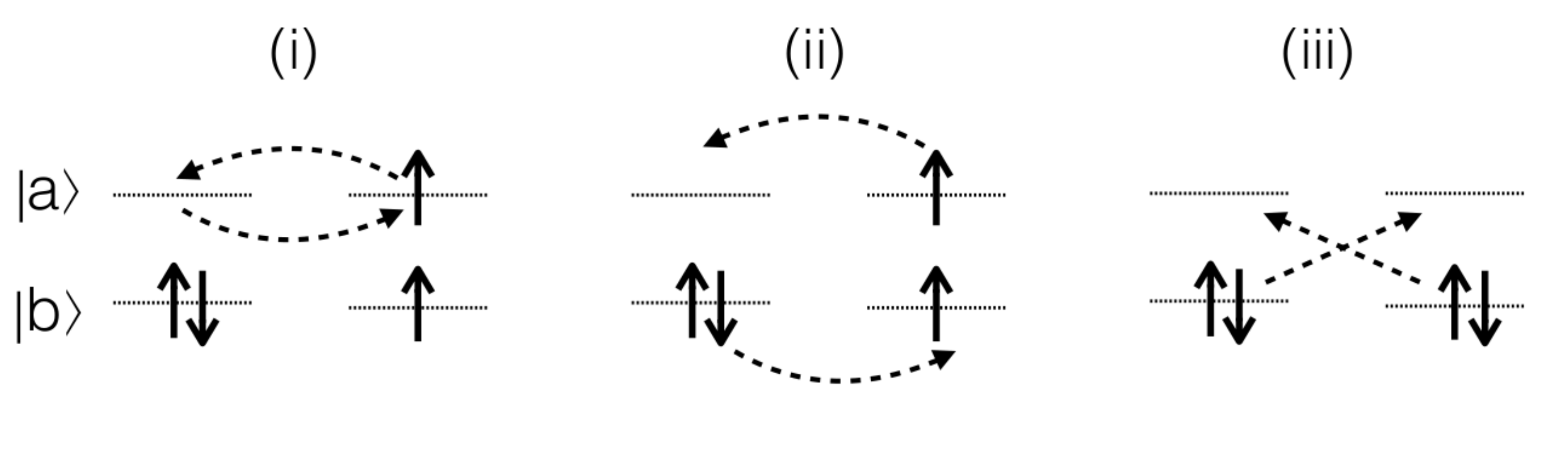}
\caption{\label{fig:hop} Typical nn hopping processes that give rise to couplings in the effective Hamiltonian (\ref{eq:hamT}, \ref{eq:boson1}, \ref{eq:boson2}).}
\end{figure}

For a sufficiently large Hund's coupling $J$, the low-energy Hilbert space can be constructed from the atomic HS and LS states:
${|1\rangle=a_{\uparrow}^{\dagger}b^{\dagger}_{\uparrow}|\vac\rangle}$, 
${|0\rangle=\tfrac{1}{\sqrt{2}}(a_{\uparrow}^{\dagger}b^{\dagger}_{\downarrow}+
a_{\downarrow}^{\dagger}b^{\dagger}_{\uparrow})|\vac\rangle}$, 
${|-1\rangle=a_{\downarrow}^{\dagger}b^{\dagger}_{\downarrow}|\vac\rangle}$, 
and ${|\bvac\rangle=b_{\uparrow}^{\dagger}b^{\dagger}_{\downarrow}|\vac\rangle}$, where $|\vac\rangle$ is the fermionic vacuum.
The local Hilbert space is further reduced if we assume easy axis anisotropy and can drop the $|0\rangle$ state. In this case
only the density-density part of the Hund's interaction contributes.~\footnote{The density-density approximation is often used
in numerical simulation with the quantum Monte-Carlo method for Anderson impurity model.} In the following, we will derive the strong-coupling model
for this case to demonstrate the principle. The generalisations are straightforward and can be found  in the literature~\cite{balents00b,kunes14a}.
Similar to (\ref{eq:pseudospin}), the low-energy Hamiltonian can be formulated in terms of pseudospin variables.
Following Ref.~\onlinecite{balents00b}, we introduce on-site standard-basis operators~\cite{haley72} $T_i^{mn}$ with matrix elements in the local basis
\begin{equation}
\label{eq:T-op}
\langle m'|T^{mn}|n'\rangle=\delta_{mm'}\delta_{nn'}.
\end{equation}
The effective Hamiltonian reads
\begin{equation}
\begin{split}
\label{eq:hamT}
\bH^{(1)}_{\tef}=&\eps\sum_{i,s}T_i^{ss}+K_{\perp}\sum_{\langle ij\rangle,s}\left(T_i^{s\bvac}T_j^{\bvac s}+i\leftrightarrow j\right)\\
+&\sum_{\langle ij \rangle,s,s'}\left(K_{\parallel}+(-1)^{s+s'}K_{0}\right)T_i^{ss}T_j^{s's'}\\
+&K_1\sum_{\langle ij \rangle,s}\left(T_i^{s\bvac}T_j^{\bar{s}\bvac}+T_i^{\bvac s}T_j^{\bvac\bar{s}}\right),
\end{split}
\end{equation}
where $s=\pm 1$ and $\bar{s}=-s$. Typical hopping processes contributing to $\bH^{(1)}_{\tef}$ are shown in Fig.~\ref{fig:hop}.
The process (i) lowers the energy of HS-LS pair on nn bond relative to the LS-LS pair. Therefore it lowers the 
energy $\eps$ of a single HS site on otherwise LS lattice relative to the single atom value of $E_{HS}-E_{LS}$,
and contributes to a nn repulsion between HS states $K_{\parallel}$. A similar process between two HS states with
opposite $s$ gives rise to the exchange term $K_0$. The process (ii) exchanges the HS and LS states in a nn bond and
introduces quantum fluctuations in the model. Neglecting the cross-hopping contributions (see Ref.~\onlinecite{kunes14a}
for the general expressions including cross-hopping corrections) the coupling constants read: ${\eps=\Delta-3J-z\tfrac{t_a^2+t_b^2}{U'}}$, 
${K_{\perp}=\frac{2t_at_b}{U'}}$, ${K_{\parallel}=(t_a^2+t_b^2)\tfrac{2U-U'+2J}{U'(U+J)}}$, and
${K_0=\tfrac{t_a^2+t_b^2}{U+J}}$, where $z$ is the number of nearest neighbours. Finally, the process (iii) converts
HS-HS pairs with zero total moment into LS-LS pairs and vice versa. This process is possible only with finite cross-hopping,
${K_1=-2V_{ab}V_{ba}\tfrac{U'}{(U+J-\Delta)(2U'-U-J+\Delta)}}$.

Hamiltonian (\ref{eq:hamT}) can be formulated in terms of hard-core bosons. In this picture, $|\bvac\rangle$ is identified with
bosonic vacuum and $|s\rangle$ with a state containing one boson of flavour $s$. However, practical implementation of the hard-core 
constraint prohibiting more than one boson per site is complicated. The standard way to treat the constraint is to introduce a 
new vacuum state $|\Omega\rangle$ and Schwinger-like bosons: $|s\rangle=d_s^{\dagger}|\Omega\rangle$, $|\bvac\rangle=h^{\dagger}|\Omega\rangle$. 
The physical states are required to obey the local constraint
\begin{equation}
\label{eq:constr}
h_i^{\dagger}h_i^{\phantom\dagger}+\sum_s d_{i,s}^{\dagger}d_{i,s}^{\phantom\dagger}=1.
\end{equation}
Rewriting (\ref{eq:hamT}) in terms of $d$- and $h$-boson, using the replacement
$T_i^{s\bvac}=d_{i,s}^{\dagger}h_i^{\phantom\dagger}$ and introducing the
$d$-number operator $n_i=\sum_sd_{i,s}^{\dagger}d_{i,s}^{\phantom\dagger}=\sum_s T_i^{ss}$ and the spin operator 
$S_i^z=\sum_ssd_{i,s}^{\dagger}d_{i,s}^{\phantom\dagger}=\sum_s sT_i^{ss}$ one arrives at
\begin{equation}
\label{eq:boson1}
\begin{split}
\tilde{\bH}^{(1)}_{\tef}&=\eps\sum_{i} n_i+K_{\perp}\sum_{\langle ij\rangle,s}\left(d^{\dagger}_{i,s}d^{\phantom\dagger}_{j,s}h_j^{\dagger}h_i^{\phantom\dagger}+H.c.\right)\\
               &+K_{\parallel}\sum_{\langle ij \rangle}n_in_j
               + K_0\sum_{\langle ij \rangle}S^z_i S^z_j \\
&-K_1\sum_{\langle ij \rangle,s} \left(
d^{\dagger}_{i,s}d^{\dagger}_{j,\bar{s}}h_j^{\phantom\dagger}h_i^{\phantom\dagger}+H.c.
\right).
\end{split}
\end{equation}
This allows us to interpret the $K_{\perp}$ term as nn hopping of $d$-bosons, $K_{\parallel}$ as nn repulsion between the $d$-bosons, $K_0$ as nn spin-spin
interaction and $K_1$ as pairvise creation and annihilation of $d$-bosons on nn sites. 
Besides technical advantages, the introduction of $h$-boson for $|\bvac\rangle$ treats the LS and HS states on equal footing and thus is well suited 
for $\eps<0$, where $|\bvac\rangle$ is not the atomic ground state and thus cannot be viewed as the vacuum state. 

The effective Hamiltonian for the case with $SU(2)$ spin symmetry has a similar structure, but differs by the presence of a third bosonic flavour $|0\rangle=d^{\dagger}_0|\Omega\rangle$. 
Introducing a cartesian vector $\bd^{\dagger}_i$ with components
\begin{equation}
\label{eq:dcart}
\begin{pmatrix} d^{\dagger}_{x} \\ d^{\dagger}_{y}  \\ d^{\dagger}_{z}
\end{pmatrix}=
\frac{1}{\sqrt{2}}
\begin{pmatrix} d^{\dagger}_{-1}-d^{\dagger}_{1} \\ i(d^{\dagger}_{-1}+d^{\dagger}_{1}) \\ \sqrt{2} d^{\dagger}_{0}
\end{pmatrix},
\end{equation}
the effective Hamiltonian can be written in a compact form
\begin{equation}
\label{eq:boson2}
\begin{split}
\tilde{\bH}^{(2)}_{\tef}&=\eps\sum_{i} n_i+K_{\perp}\sum_{\langle ij\rangle}\left(\bd^{\dagger}_{i}\cdot \bd^{\phantom\dagger}_{j}h_j^{\dagger}h_i^{\phantom\dagger}
    + H.c.\right)\\
               &+K_{\parallel}\sum_{\langle ij \rangle}n_in_j
               + K_0\sum_{\langle ij \rangle}\bS_i \cdot \bS_j \\
&+K_1\sum_{\langle ij \rangle} \left(
\bd^{\dagger}_{i}\cdot\bd^{\dagger}_{j}h_j^{\phantom\dagger}h_i^{\phantom\dagger}+H.c.
\right)\\
&+K_2\sum_{\langle ij\rangle}\left(\left(\bd_i^{\phantom\dagger}h_i^{\dagger}+\bd_i^{\dagger}h_i^{\phantom\dagger}\right)\cdot\bS_j+i\leftrightarrow j\right).
\end{split}
\end{equation}
Here, $\bS_i$ are spin $S=1$ operators with $S^x_i=\tfrac{1}{\sqrt{2}}(T_i^{10}+T_i^{0-1}+T_i^{01}+T_i^{-10})$, $S^y_i=\tfrac{i}{\sqrt{2}}(T_i^{01}+T_i^{-10}-T_i^{10}-T_i^{0-1})$.
Expressed in vector notation, $\bS_i$ takes the form of a cross product
\begin{equation}
\label{eq:s}
\bS_i=-i\bd_i^{\dagger}\wedge\bd_i.
\end{equation}
Unlike the Ising case (\ref{eq:boson1}) where cross-hopping is necessary to generate $K_1$, in the $SU(2)$ symmetric case (\ref{eq:boson2}) $K_1$ appears also without cross-hopping if 
the pair-hopping term is finite, $J'\neq0$.
The last term with coupling constant $K_2\sim V_{ba}t_a+V_{ab}t_b$, which couples the d-operators to the spin operators, does not have an analogy in the Ising case.
The full expressions for the coupling constants can be found in Ref.~\onlinecite{kunes14a}.

Exciton condensation in (\ref{eq:boson2}) is characterised by a finite expectation value $\langle \bd^{\dagger}_ih_i\rangle$,
 related to the spin-triplet order parameter (\ref{eq:opar}) of 2BHM by
\begin{equation}
\label{eq:scop}
 \langle \bd^{\dagger}_ih_i\rangle=\bph^t_i/\sqrt{2}.
 \end{equation}

Models (\ref{eq:boson1}) and (\ref{eq:boson2}) for special choices of the coupling constants are known under their own names. We discuss these cases below.

\subsubsection{Blume-Emmery-Griffiths model}
\begin{figure}
\includegraphics[width=0.7\columnwidth]{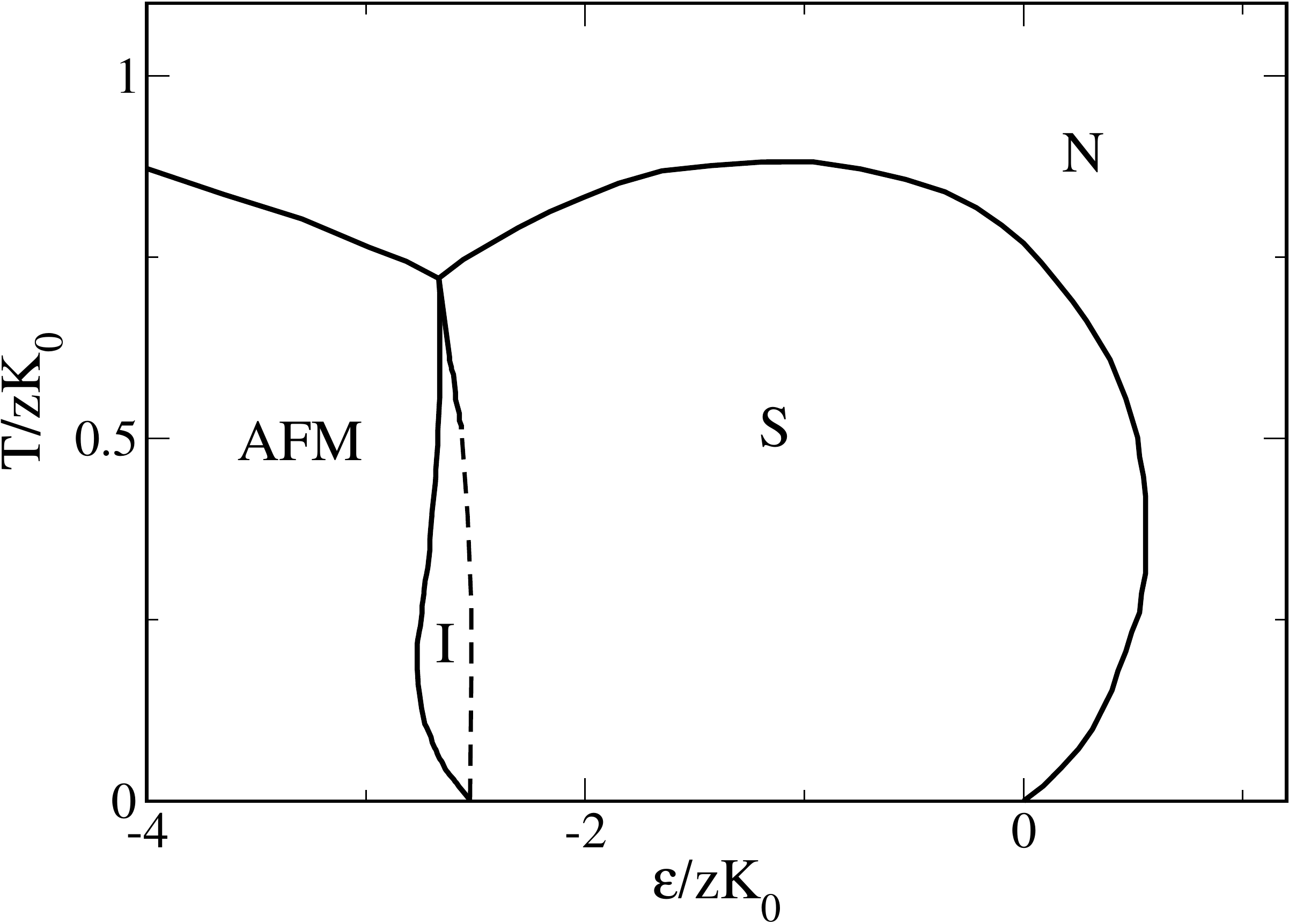} 
\caption{\label{fig:beg} The phase diagram of BEG model for ${K_{\parallel}/K_0=3.5}$. Besides the normal phase (N), it contains the solid phase (S), antiferromagnetic phase (AFM) and
ferrimagnetic phase (I). The data were taken from Ref.~\onlinecite{hoston91}.
}
\end{figure}

Without cross-hopping and one fermionic species immobile, $t_b,V_{ab},V_{ba}=0$, 
(\ref{eq:boson1})  becomes purely classical, with $K_{\perp},K_1=0$.
Known as the Blume-Emmery-Griffiths (BEG) model~\cite{beg}, it was originally introduced to describe mixtures
of $^3$He and $^4$He. In the standard formulation of the BEG model discrete index $s$ is used to describe the local
state $s=\pm 1$ corresponding to $|\pm 1\rangle$ and $s=0$ is assigned to $|\bvac\rangle$. The model is then written as
\begin{equation}
H_{\text{BEG}}=\eps \sum_i s_i^2 + K_{\parallel}\sum_{\langle ij \rangle} s_i^2s_j^2+K_0\sum_{\langle ij \rangle}s_i s_j.
\end{equation}
The BEG model found its use in many areas of statistical physics. Despite its simplicity it has a rich phase 
diagram~\cite{hoston91}. The phase diagram of the model on bipartite lattice does not depend on the sign of $K_0$ as the ferromagnetic $K_0<0$ and antiferromagnetic $K_0>0$
models can be mapped on each other by the  transformation $s_i\rightarrow (-1)^is_i$.
For $K_{\parallel}<|K_0|$ there are only magnetically ordered (AFM)  and normal (N) phases. At $T=0$ the N phase corresponds to
empty lattice $\langle n_i \rangle=0$, while in the AFM phase there is one boson on each site $\langle n_i \rangle=1$. 
A typical phase diagram for $K_{\parallel}>|K_0|$ is shown in Fig.~\ref{fig:beg}. 
Besides the magnetically ordered (AFM) phase there is the solid (S) phase (called antiquadrupolar in Ref.~\onlinecite{hoston91}) 
characterised at $T=0$ by bosons occupying one sublattice, the other being empty, $\langle n_i \rangle=\tfrac{1}{2}(1-(-1)^i)$.
Note that this phase has a residual spin degeneracy on the occupied sites. This is a consequence of the exchange interaction beyond nn being strictly zero. 
In a more realistic model one expects the occupied sites to order magnetically at a sufficiently low temperature.
Coexistence of the magnetic and solid order (I phase in Fig.~\ref{fig:beg} ) is found at finite temperature in a narrow range 
of $\epsilon$ separating the magnetic and solid phases. For positive $\epsilon$ at $T=0$, the system is in the vacuum (empty lattice). 
Interestingly, for moderate $\epsilon>0$ the solid phase is found at elevated temperature. This reentrant behaviour of the solid phase was found also
for finite $t_b$ in Monte-Carlo simulations of the spinless bosons~\cite{schmid02} as well as in DMFT simulations of 2BHM~\cite{kunes11b}.
Generalisation of BEG model to the case with spin-rotational symmetry (\ref{eq:boson2}) is straightforward. On the mean-field level it leads to quantitative modification of the 
phase diagram.

\subsubsection{Bosonic \tj model}
Next, we discuss models (\ref{eq:boson1}) and (\ref{eq:boson2}) in the parameter range where they describe conserved bosons.
We assume that both $a$ and $b$ electrons are mobile generating a finite hopping $K_{\perp}$ in (\ref{eq:boson1}) and (\ref{eq:boson2})
of the $d$-bosons. In addition, we require that $K_1,K_2=0$, which is the case for $V_{ab},V_{ba},J'=0$. 
Hamiltonians (\ref{eq:boson1}) and (\ref{eq:boson2}) describe interacting bosons carrying Ising and Heisenberg spin $S=1$, respectively. 
For suitable parameters, the system may undergo a BE condensation characterised by a complex vector order parameter $\bph_i^t$ (\ref{eq:scop}).
In case of the Ising interaction (\ref{eq:boson1}) the vector $\bph^t_i$ is confined to the $xy$ plane. The vector character of the order parameter
adds additional structure to the condensation of spinless bosons discussed in Sec.~\ref{sec:efkm-strong}.
\vspace{0.2cm}

{\it The role of $K_0$.}
The $SU(2)$ symmetric Hamiltonian (\ref{eq:boson2}) describes spinful $S=1$ bosons with infinite on-site repulsion, nn repulsion $K_{\parallel}$
and nn spin-exchange $K_0$. Hamiltonians of this kind have been studied both theoretically and experimentally for cold atoms
in optical traps. Interestingly, in a genuine system of bosons with spin-independent interactions the structure of the exact ground state prohibits
the spin-exchange from appearing in any low-energy effective Hamiltonian~\cite{eisenberg02}. As pointed out in Ref.~\onlinecite{eisenberg02}, it can only 
arise as a low-energy effective description in fermionic system - as in the present case. 

BE condensation in the continuum version of (\ref{eq:boson2}) was studied in Refs.~\cite{ho98,ohmi98}.  
The sign of the exchange was shown to play a crucial role for the properties of the superfluid phase as it determines
the residual symmetry of the BE condensate. Antiferromagnetic exchange ($K_0>0$)
selects the so called polar state characterised by ${(\bph_i^t)^*\wedge\bph^t_i=0}$, while ferromagnetic exchange ($K_0<0$) selects ferromagnetic state, where
${|(\bph_i^t)^*\wedge\bph^t_i|}$ is maximised. Different residual symmetries of these states result in different low-energy dynamics and qualitatively
different behaviour of topological defects (vortices)~\cite{ho98,ohmi98}. 
The polar and ferromagnetic phases are found for Ising spins (\ref{eq:boson1})~\cite{kunes14c}, although the topological aspects that determine
the low-energy excitations and possible topological defects are different from Heisenberg spins (\ref{eq:boson2}).
\vspace{0.2cm}

\begin{figure}
\includegraphics[width=0.8\columnwidth,clip]{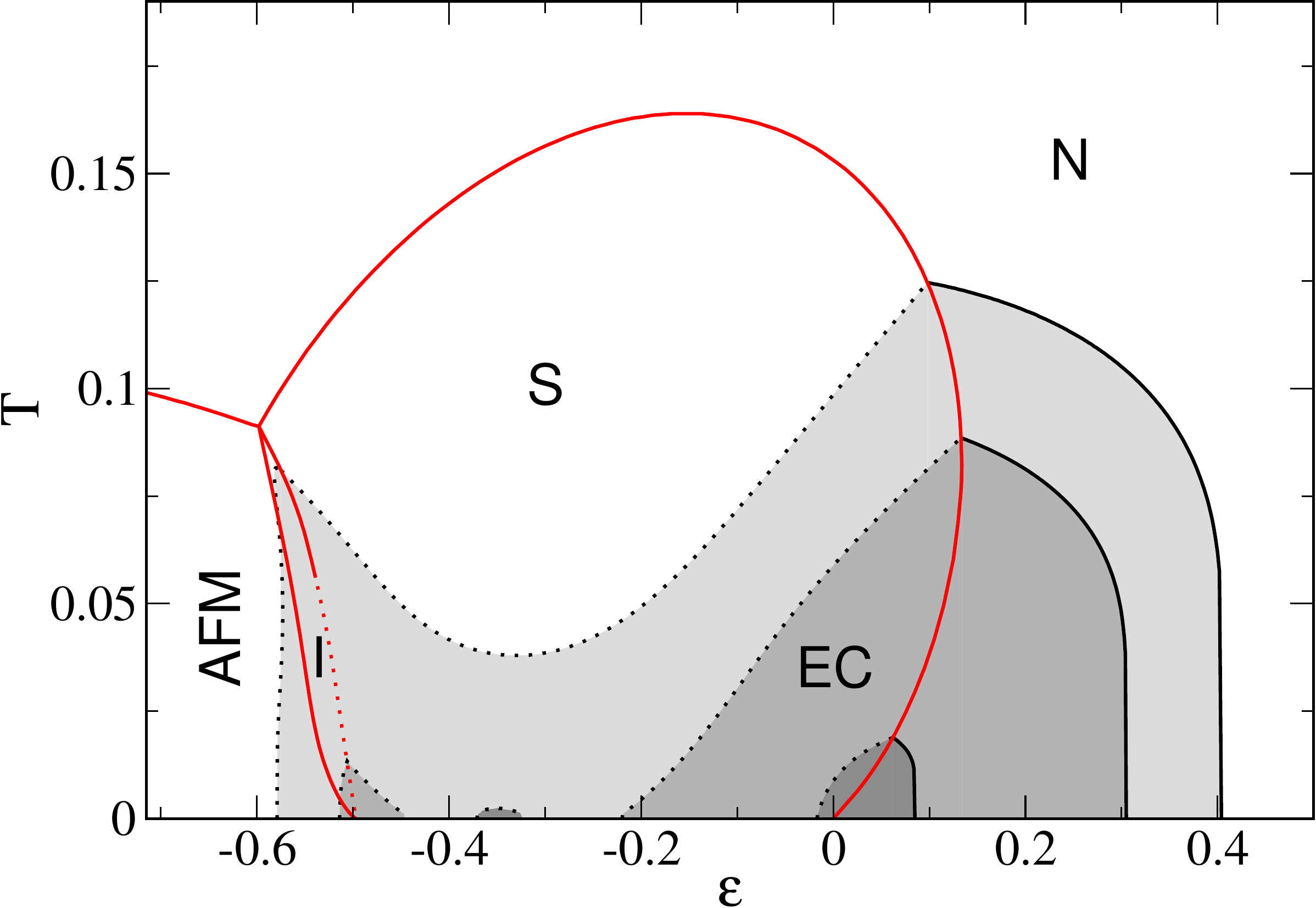}
\caption{\label{fig:btj_mf} Mean-field $\eps$-$T$ phase diagram~\cite{august14} of the bosonic \tj model (\ref{eq:boson2})
for the parameters: $K_0=0.041$, $K_{\parallel}=0.164$, and $K_{\perp}$=\{-0.103, -0.076, -0.023\}. Solid lines mark continuous
transitions, while dotted lines mark first-order transitions. The red lines correspond to the phase boundaries of the
$S=1$ BEG model with its phases described in Fig.~\ref{fig:beg}. The shaded regions correspond to the EC phase for the three choices
of $K_{\perp}$ (with the brightest colour for largest $|K_{\perp}|$). Red lines inside the EC phase have no meaning and are shown 
only to put several phase diagrams in one figure.}
\end{figure}

{\it Mean-field phase diagram.} 
The continuum model may be viewed as an effective description of the lattice model at low boson concentrations. At
higher boson concentrations other phases, e.g., these present in the BEG phase diagram Fig.~\ref{fig:beg}, exist on a lattice and 
compete with the superfluid. In Fig.~\ref{fig:btj_mf} we show the phase diagram of (\ref{eq:boson2}) on a square lattice ($z=4$) obtained 
with a mean-field decoupling of the pseudospin variables (\ref{eq:hamT}) $T_iT_j\approx \langle T_i\rangle T_j+ T_i\langle T_j\rangle- \langle T_i\rangle \langle T_j \rangle$. 
In absence of condensed bosons, this mean-field theory reduces to that of a slightly modified BEG model
(Heisenberg instead of Ising spin $S=1$) with phase boundaries marked with the red lines.
The grey area marks the superfluid (exciton condensate) phase -- the different shades of grey correspond to different values of $K_{\perp}$. 
The transition between the normal and the EC phase is continuous, while the transitions to the other ordered phases are of the first-order as dictated by 
symmetry. The intermediate phases, such as supersolid (with S and EC orders) or AFM-supersolid with (AFM and EC orders), which would allow
continuous transitions, do not exist or are thermodynamically unstable. 

At large $\eps$ the ground state is an empty lattice, a state that is connected to normal Bose gas at elevated temperatures. 
Upon reduction of $\eps$ the system undergoes a continuous transition to the superfluid EC phase with $T_c$ determined by $|K_{\perp}|$ and $\eps$. The sign of $K_{\perp}$
determines the periodicity of the order parameter: for $K_{\perp}<0$ the system goes to uniform ferro-EC state (not to be confused with ferromagnetic EC state) with 
$\bph^t_i=\bph$, while for $K_{\perp}>0$ the system goes into antiferro-EC state with $\bph^t_i=(-1)^i\bph$, the models with $\pm |K_{\perp}|$ are connected by 
the gauge transformation $d_i\rightarrow (-1)^i d_i$. The BEG phase boundaries are not affected by $K_{\perp}$.

Another parameter that determines the nature of the EC phase is the exchange coupling $K_0$. The phase diagram in Fig.~\ref{fig:btj_mf} was obtained for
antiferromagnetic $K_0>0$, but some of its features remain unchanged when the sign of $K_0$ is flipped. In particular, the BEG phase boundaries are unchanged.
This is so because the on-site and interaction terms in the Hamiltonian are invariant under the exchange of spin species on one of the sublattices ($d_{is}\rightarrow d_{i\bar{s}}$),
which maps ferromagnetic BEG to antiferromagnetic BEG model. This transformation works only for Ising spins, but in the mean-field treatment the difference
between Ising and Heisenberg spins disappears. The transformation changes the hopping term in (\ref{eq:boson1},\ref{eq:boson2}) and thus the argument
cannot be used for the EC phase. Nevertheless, the mean-field phase boundary between the normal and EC phase does not depend on $K_0$ at all, 
because it contributes to the free energy in the order $\phi^4$.  On the other hand, the first-order phase boundaries depend $K_0$ and its sign.
\vspace{0.2cm}

{\it Ferromagnetic EC state.}
It is instructive to see where exactly the difference between the $\pm|{K_0}|$ models comes from. First, we point out that  
an order parameter of the form $\bph^t_j=e^{i\mathbf{q}\cdot\mathbf{R}_j}\bph$ implies a uniform magnetisation  
$\langle\bS_j\rangle\sim 
i\bph^{*}\wedge\bph$ irrespective of its periodicity $\mathbf{q}$.
Second, let us consider the free energy of the FMEC and polar EC states with the same magnitude $|\bph|$ of the order parameter.
The on-site, hopping and interaction terms contribute the same for the two states. 
However, the exchange energy in the polar EC state is zero while the FMEC state has finite exchange energy.
Therefore, like in the continuum models, $K_0>0$ leads to a non-magnetic polar EC state, while $K_0<0$ leads to FMEC with a finite uniform magnetisation. 
For the same $|K_0|$, FMEC/S boundary for $K_0<0$ is shifted in favour of the EC phase compared to the polar-EC/S boundary for $K_0>0$
because the FMEC energy is lower than the corresponding polar EC energy.

Continuous transition between the FM and FMEC phases is allowed by symmetry.
While we are not aware of an explicit calculation of the FM/FMEC transition for $K_0<0$, one can gain insight from the $T=0$ mean-field
wave function, which has the product form~\cite{balents00b}
\begin{equation}
|\Psi\rangle=\Pi_i \left( \tilde{h}h^{\dagger}+\tilde{\bd}\cdot \bd^{\dagger}_i\right)|\Omega\rangle,
\end{equation}
with the coefficients fulfilling
${|\tilde{h}|^2+\sum_s |\tilde{d}^{\phantom\dagger}_{s}|^2=1}$. The magnetisation  (\ref{eq:s}) 
in terms of the EC order parameter  (\ref{eq:scop}) ${\bph=\sqrt{2}\tilde{\bd}^*\tilde{h}}$ is given by
\begin{equation}
\langle \Psi|\bS_i|\Psi \rangle=-i\tilde{\bd}^*\wedge \tilde{\bd}=\frac{i}{2|\tilde{h}|^2}\bph^*\wedge\bph.
\end{equation}
This shows that at a continuous FMEC/FM transition $\tilde{h}$ goes to zero, while the magnetisation reaches smoothly its saturation value.
\vspace{0.2cm}

{\it Two flavour models.} We are of not aware of specific theoretical studies of spinful hard-core bosons (\ref{eq:boson1},\ref{eq:boson2}) on a lattice. However, models
(\ref{eq:boson1},\ref{eq:boson2}) can be viewed as special cases of more general multi-component boson systems.
Bosonic \tj model describing mixtures of two boson species with hard-core constraint have been investigated in several studies. 
This set-up is similar to the model (\ref{eq:boson1}) with Ising spin and $K_1=0$, which conserves both bosonic species separately. 
In the following, we discuss briefly how and which of the results of these studies can be used to describe (\ref{eq:boson1}).

The bosonic \tj model is usually formulated in terms of $I=1/2$ pseudospin:  $I_z=d^{\dagger}_{1}d^{\phantom\dagger}_{1}-d^{\dagger}_{-1}d^{\phantom\dagger}_{-1}$, $I^+=d^{\dagger}_{1}d^{\phantom\dagger}_{-1}$, and $I^-=d^{\dagger}_{-1}d^{\phantom\dagger}_{1}$, and anisotropic $XXZ$ inter-site exchange interaction.
While the definition of spin operator $S_z$ in  (\ref{eq:boson1}) coincides with that of $I_z$, the transverse components $I^+$ and $I^-$ have no spin counterpart in 
(\ref{eq:boson1}). Therefore a great care is required when interpreting the results of bosonic \tj model in the language of model (\ref{eq:boson1}). In particular,
the $z$-axis pseudospin order (often referred simply as an anti-ferromagnetic order) corresponds to a true spin order in (\ref{eq:boson1}).
However, the $xy$ pseudospin order (e.g. $xy$-ferromagnet of Refs.~\cite{altman03, kuklov03}) does not correspond to a spin order in (\ref{eq:boson1}) or
its $SU(2)$-symmetric generalisation (\ref{eq:boson2}). 

Another difference between the bosonic \tj model and (\ref{eq:boson1}) concerns the hard-core constraint. In most 
studies on the bosonic \tj model simultaneous presence of two bosons of different flavour on the same site is allowed either explicitly (The hard-core constraint applies only to
particles of the same species.) or implicitly (Double occupancy is allowed in the high-energy model from which the \tj model is derived.), giving rise to the $xy$ pseudospin
exchange. On the other hand, the total hard-core constraint and the absence of the $xy$ exchange in (\ref{eq:boson1}) are viewed as kinematic constraints~\footnote{The $xy$ exchange
processes in \ref{eq:boson1}) arise in fourth order expansion in $t_a,t_b$ of 2BHM.} . This calls for care when using the phase diagrams of bosonic \tj model since existence of some phases, e.g., the counter-superfluid of Ref.~\cite{kuklov03}, depends crucially on the $xy$ exchange and so such phases are not present in (\ref{eq:boson1}).

In the following we discuss two types of studies on the bosonic \tj model, which can be translated to provide information about (\ref{eq:boson1}). The first concerns
the transition between AFM and SF phases and the possibility of intermediate AFM-supersolid phase.
Boninsegni~\cite{boninsegni01} and Boninsegni and Prokof\'ev~\cite{boninsegni08} studied the bosonic t-J$_z$ model, 
corresponding to (\ref{eq:boson1}) for $K_{\parallel}=-K_0<0$ using Monte-Carlo simulations. Since the solid phase is absent for $K_{\parallel}<|K_0|$~\cite{hoston91}, 
one may expect AFM-SF transition for moderate $K_0$ (For strong $K_0$ the BEG model
predicts first order transition between the Mott AFM and vacuum states at low temperatures.) The Monte-Carlo calculations find a first-order AFM-SF
transition with no indication of AFM-supersolid characterised by coexistence of the AFM and SF order parameters.

Numerical calculations with repulsive $K_{\parallel}$ ($K_0=K_{\parallel}>0$), motivated by observation
of the supersolid in the spinless case, have been reported on a triangular lattice recently~\cite{trousselet14,lv14}. 
Although double occupancy by different bosonic flavours was allowed and the lattice differs from the square lattice, 
the basic features common to Fig.~\ref{fig:btj_mf} were found in the phase diagram for large on-site (inter-species) repulsion.
In particular, the sequence of $T=0$ phases 'empty lattice-SF-S-SF'-AFM' with decreasing $\eps$ can be expected.

 \subsubsection{Bi-layer Heisenberg model}
\begin{figure}
\includegraphics[width=0.8\columnwidth]{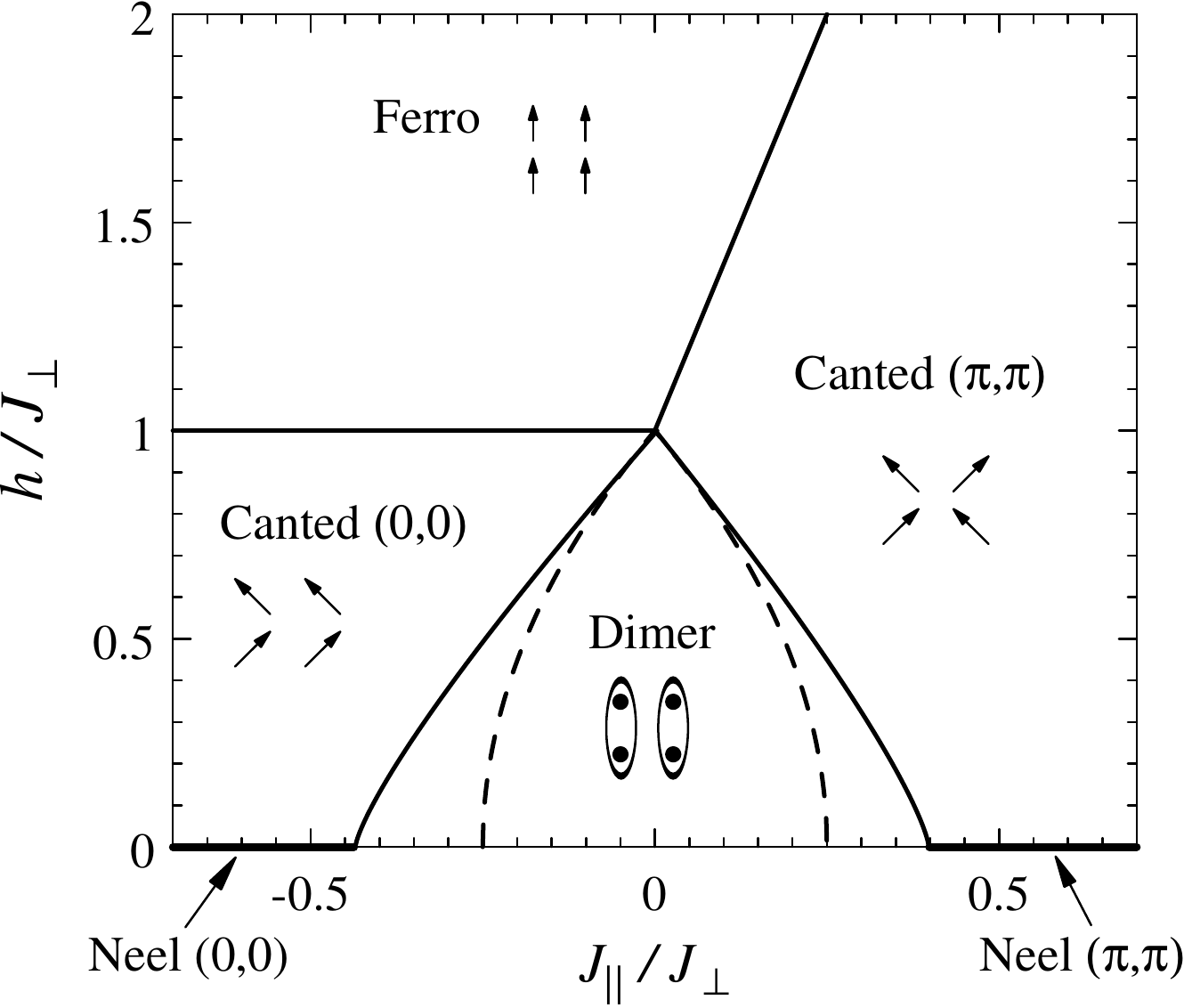}
\caption{\label{fig:sommer} $T$=0 phase diagram of the bi-layer Heisenberg obtained by
Sommer, Vojta and Becker. Reproduced from Ref.~\onlinecite{sommer01} 
with kind permission of The European Physical Journal (EPJ).}
\end{figure}

Another special case of the $SU(2)$-symmetric model (\ref{eq:boson2}) studied in the literature corresponds to  
bi-layer Heisenberg model
\begin{equation}
\label{eq:bi-heisenberg}
H_{\text{bi-H}}=J_{\perp}\sum_i\mathbf{S}_{ia}\cdot\mathbf{S}_{ib}+J_{\parallel}\sum_{\rij,m}\mathbf{S}_{im}\cdot\mathbf{S}_{jm}
-\bh\cdot\sum_{im}\bS_{im},
\end{equation}
with $S=1/2$ spin operators $\bS_{im}$, $i$ being the site index and $m=a,b$ a layer index. The model arises as the large
$U$ limit of the half-filled bi-layer Hubbard model (\ref{eq:2bhm}), which describes two identical layers, indexed by the orbital index $m$, coupled
on inter-layer rungs. This situation corresponds to the parameter set $U\gg |t_a|=|t_b|$, $\{\Delta,U',J'\}=0$, $J<0$ in (\ref{eq:2bhm}), where
the anti-ferromagnetic coupling on the rung $J_{\perp}=J$ may arise from inter-layer tunnelling. Integrating out the 
charge fluctuations one arrives at (\ref{eq:bi-heisenberg}) with $J_{\parallel}=4t_a^2/U$.
Model (\ref{eq:boson2}) is obtained by going to the singlet-triplet basis, which diagonalises the rung exchange (local)  part  of (\ref{eq:bi-heisenberg}).
Introducing a map
\begin{equation}
\bS_{ia,b}=\frac{1}{2}\left(\pm \bt^{\dagger}_is_i^{\phantom\dagger} \pm \bt^{\phantom\dagger}_is_i^{\dagger}-i\bt^{\dagger}_i \wedge \bt^{\phantom\dagger}_i\right)
\end{equation}
one arrives at~\cite{sommer01,sachdev90}
\begin{equation}
\label{eq:bi_h}
\begin{split}
H_{\text{bi-H}}&=\frac{J_{\perp}}{4} \sum_i  \left( \bt_i^{\dagger}\cdot\bt_i{\phantom\dagger}-3 s_i^{\dagger}s_i{\phantom\dagger} \right)\\
&+\frac{J_{\parallel}}{2}\sum_{\langle ij\rangle}\left(\bt^{\dagger}_{i}\cdot \bt^{\phantom\dagger}_{j}s_j^{\dagger}s_i^{\phantom\dagger}+ H.c.\right)\\
&+\frac{J_{\parallel}}{2}\sum_{\langle ij \rangle}(-i)^2\left(\bt^{\dagger}_{i}\wedge\bt^{\phantom\dagger}_{i}\right)\cdot\left(\bt^{\dagger}_{j}\wedge\bt^{\phantom\dagger}_{j}\right)\\
&+\frac{J_{\parallel}}{2}\sum_{\langle ij \rangle} \left(\bt^{\dagger}_{i}\cdot\bt^{\dagger}_{j}s_j^{\phantom\dagger}s_i^{\phantom\dagger}+H.c.\right)\\
&+\sum_ii\bh\cdot\left(\bt^{\dagger}_{i}\wedge\bt^{\phantom\dagger}_{i}\right),
\end{split}
\end{equation}
where $s_i$ and $\bt_i=(t_{i,x},t_{i,y},t_{i,z})$ are bosonic operators and the physical subspace fulfils the local constraint ${s_i^{\dagger}s_i^{\phantom\dagger}+
\bt_i^{\dagger}\cdot\bt_i^{\phantom\dagger}=1}$. Hamiltonian (\ref{eq:bi_h}) is equivalent to (\ref{eq:boson2}) with parameters $K_{\parallel}=K_2=0$, 
$K_{\perp}=K_0=K_1=J_{\parallel}/2$ and $\epsilon=J_{\perp}$. 
The main difference of (\ref{eq:bi-heisenberg}) to the bosonic \tj model consist in the presence of non-zero $K_1$,
which leads to the number of $d$-bosons being a non-conserved quantity. While the system undergoes
transitions~\cite{sommer01} to phases characterised by non-zero $\bph_i=\langle \bt_i^{\phantom\dagger}s_i^{\dagger}\rangle$, the phase of 
$\bph_i$ is not arbitrary and thus the transitions cannot be viewed as Bose-Einstein condensation
of spinful bosons. In zero magnetic field $\bh=0$ the ordered phases are characterised by real $\bph_i$ which breaks the $SU(2)$ symmetry of (\ref{eq:bi-heisenberg}) and corresponds to a N\'eel state with ordered moments $\langle\bS_{ia,b}\rangle=\pm\bph_i$, see Fig.~\ref{fig:sommer}. Non-zero magnetic field $\bh\neq0$
reduces the symmetry of (\ref{eq:bi-heisenberg}) to $U(1)$. The vector $\bph_i$ gets oriented perpendicular to $\bh$ and acquires an
imaginary part perpendicular to the real one. The real part of $\bph_i$ describes the AFM order, while $i\bph^*_i\wedge\bph_i$ describes
the FM component along $\bh$. Breaking of the $U(1)$ symmetry can be described as BE condensation of spinless bosons. 
This is particularly easy to see for $J_{\perp}\gg J_{\parallel}$, because close to the transition $h\approx J_{\parallel}$ and model (\ref{eq:boson2}) 
reduces to the Hamiltonian of spinless bosons for the flavour polarised along the field direction.

BE condensation in quantum magnets has been an active area of research and the reader is referred to the recent review~\cite{zapf14} for 
further reading and references. In general, the basic difference between BE condensation in spin systems and excitonic condensation discussed here
consists in the microscopic origin of the $U(1)$ symmetry broken by the condensate. In spin models, it is the spin rotation in the $xy$-plane of systems
with uniaxial anisotropy and condensation refers to some kind of in-plane magnetic order. The $U(1)$ symmetry in excitonic systems
is more abstract and refers to the arbitrariness of the relative phase between $a$ and $b$ orbitals in (\ref{eq:2bhm}). 

\subsubsection{Exciton \tj model}
The strong-coupling limit of 2BHM with anti-ferromagnetic Hund's exchange was studied by Rademaker and collaborators~\cite{rademaker12a, rademaker12b, rademaker13a,rademaker13b,rademaker13c}. They started from the antiferromagnetic bi-layer Heisenberg model~(\ref{eq:bi_h}) and considered doping one layer by holes and 
the other by the same amount  of electrons. This is equivalent to introducing the crystal-field $\Delta$ in (\ref{eq:2bhm}) while keeping the total electron concentration
at half-filling. The anti-ferromagnetic Hund's coupling $J<0$ (corresponding to inter-layer exchange) was assumed to originate from small, but finite 
inter-layer tunnelling. Starting from the undoped Heisenberg limit with on-site states ${|1m\rangle=t_m^{\dagger}|\Omega\rangle}$, and 
${|00\rangle=s^{\dagger}|\Omega\rangle=\tfrac{1}{\sqrt{2}}(a_{\uparrow}^{\dagger}b^{\dagger}_{\downarrow}-a_{\downarrow}^{\dagger}b^{\dagger}_{\uparrow})|\vac\rangle}$, the
authors of Refs.~\onlinecite{rademaker12a, rademaker12b, rademaker13a,rademaker13b,rademaker13c} introduce a bound exciton state 
$|\bvac\rangle=h^{\dagger}|\Omega\rangle$ to the system. The new Hamiltonian, which is a special case of that discussed in Ref.~\onlinecite{balents00b},
reads
\begin{equation}
\label{eq:etj}
\begin{split}
H&=H_{\text{bi-H}}+\mu \sum_i h_i^{\dagger}h_i^{\phantom\dagger}+V\sum_{\rij}h_i^{\dagger}h_i^{\phantom\dagger}h_j^{\dagger}h_j^{\phantom\dagger}\\
&+t_{\text{ex}}\sum_{\langle ij\rangle}\left(\bt^{\dagger}_{i}\cdot \bt^{\phantom\dagger}_{j}h_j^{\dagger}h_i^{\phantom\dagger}+ 
s^{\dagger}_{i}s^{\phantom\dagger}_{j}h_j^{\dagger}h_i^{\phantom\dagger}+H.c.\right),
\end{split}
\end{equation}
with the hard-core constraint ${h_i^{\dagger}h_i^{\phantom\dagger}+s_i^{\dagger}s_i^{\phantom\dagger}+
\bt_i^{\dagger}\cdot\bt_i^{\phantom\dagger}=1}$. The exciton site energy $\mu$ controls the exciton concentration $\rho$. The spin-singlet bosons
$h$ and $s$ appear symmetrically in (\ref{eq:etj}), except for the term 
${t^{\dagger}t^{\dagger}s^{\phantom\dagger}s^{\phantom\dagger}}$ which does not have a ${t^{\dagger}t^{\dagger}h^{\phantom\dagger}h^{\phantom\dagger}}$
counterpart. The exciton hopping $t_{\text{ex}}$ and repulsion $V$ play roles analogous to $K_{\perp}$ and $K_{\parallel}$, respectively, in (\ref{eq:boson1}, \ref{eq:boson2}).

Note that the notion of exciton and vacuum is exchanged with respect to that introduced in Sec.~\ref{sec:Jgf}. 
Unlike the unambiguous vacuum state of real bosons, the meaning of vacuum for hard-core bosons is ambiguous, similar to the notion of
electron and hole for fermions. The condensation of hard-core boson generally means that the system is in a quantum-mechanical superposition of the
vacuum and one-particle states, whatever their definition is. 

In Refs.~\onlinecite{rademaker12a, rademaker12b} Rademaker{\it et al.} studied the propagation of a single $|\bvac\rangle$ exciton in the model
with antiferromagnetic coupling $J_{\parallel},J_{\perp}>0$. The propagation of exciton strongly 
depends on the ratio of the Hund's (inter-layer) and intra-layer exchange $J_{\perp}/J_{\parallel}$. For $J_{\perp}\gg J_{\parallel}$ the ground state of undoped system is a product of 
local singlets $|00\rangle$ and the exciton can propagate as a free particle forming a band with the width $2zt_{\text{ex}}$. 
In the opposite limit, $J_{\perp}\ll J_{\parallel}$ the system consists of weakly coupled AFM layers. It is well know from the fermionic $t-J$ model that the motion
of a hole in the AFM background is strongly inhibited since a moving hole disturbs the AFM order. This physics is also reflected in the motion of an exciton. 
For $t_{\text{ex}}\ll J_{\parallel}$, exciton propagation is severely limited and the exciton bandwidth is reduced to the order $t^2_{\text{ex}}/J_{\parallel}$.
In the opposite limit $t_{\text{ex}}\gg J_{\parallel}$, the exciton can explore neighbouring sites over larger distances 
which gives rise to the typical incoherent string-state spectrum~\cite{rademaker12a, rademaker12b,rademaker13c}.

\begin{figure}
\includegraphics[width=0.9\columnwidth]{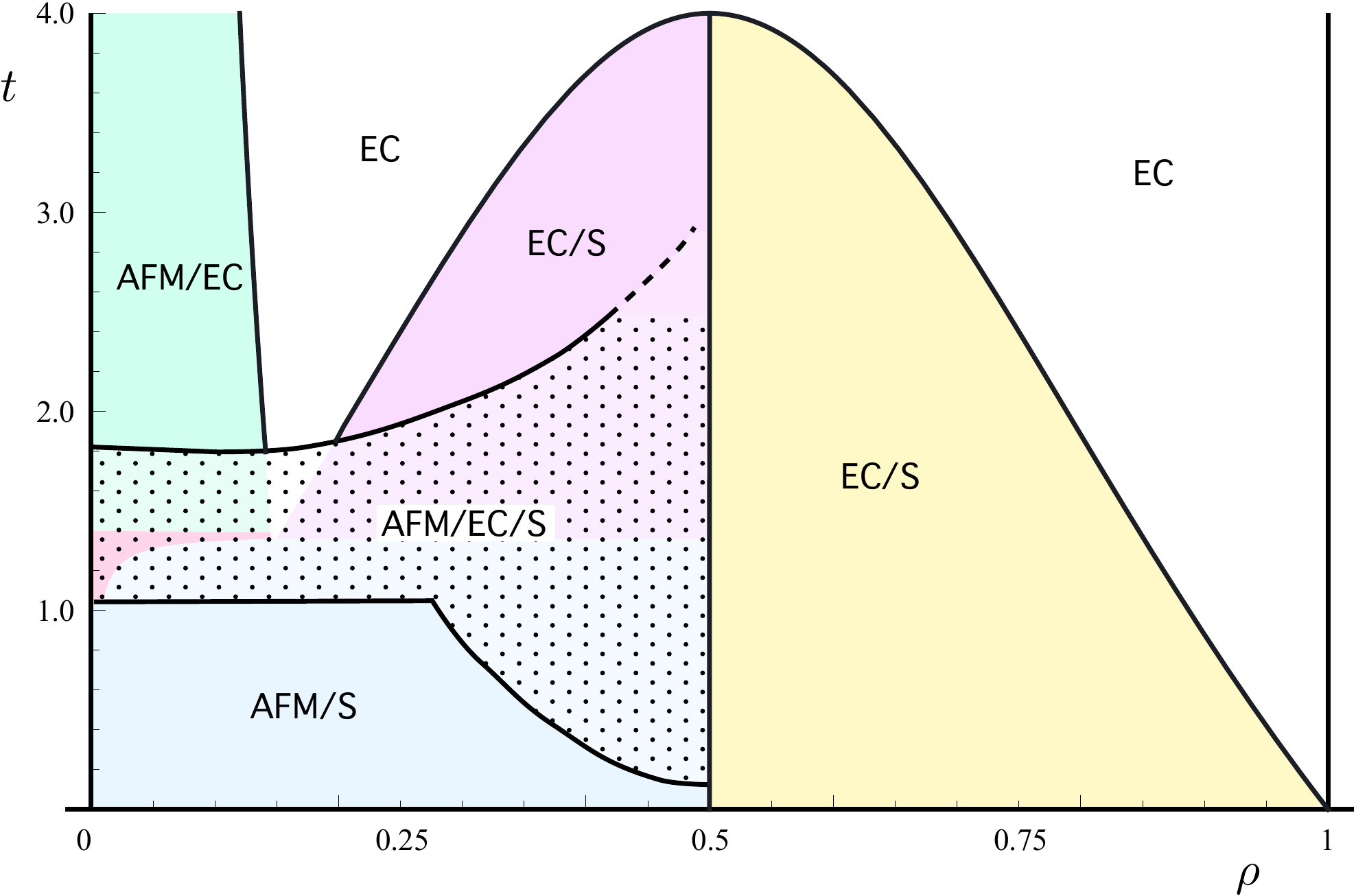}
\caption{\label{fig:rademaker} The $T=0$ phase diagram of the exciton \tj model for $J_{\perp}=0.02$, $J_{\parallel}=0.125$ and $V=2$. The parameters
$t$ and $\rho$ denote the exciton hopping amplitude and concentration, respectively. 
The names of phases are the same as in Fig.~\ref{fig:btj_mf}: antiferromagnet (AFM), exciton condensate (EC) and solid (S). 
The slash, e.g., 'EC/S', denotes phase separation.
Adapted with permission of authors from Ref.~\onlinecite{rademaker13c}. Copyrighted by the American Physical Society.}
\end{figure}

Exciton condensation in the excitonic \tj model was studied in the Refs.~\onlinecite{rademaker13a,rademaker13c}. The general features of the mean-field 
phase diagram, shown in Fig.~\ref{fig:rademaker}, are similar to the case of antiferromagnetic Hund's exchange in Fig.~\ref{fig:btj_mf} and can be traced back to 
their 'common ancestor' in the $XXZ$ model. In particular, there are three basic phases: AFM, solid and the EC superfluid and first-order AFM/solid, 
solid/superfluid and AFM/superfluid transitions.
We can compare these Figs. \ref{fig:rademaker} and  \ref{fig:btj_mf} keeping in mind that increasing $\eps$ in Fig.~\ref{fig:btj_mf} corresponds
to increasing $\rho$ in Fig.~\ref{fig:rademaker}. 
In both models the $T=0$ solid (S) phase is absent for large exciton hopping $K_{\perp}$ ($t_{\text{ex}}$). At intermediate
hopping $K_{\perp}$ ($t_{\text{ex}}$), the AFM-EC-S-EC sequence of phases is found in both models, while at small hopping
there is a direct transition between the AFM and S phases. 
The small positive $J_{\perp}$ selects the spin-singlet condensate which is characterised by spontaneous coherence between the $|00\rangle$ and $|\bvac\rangle$ states. 
The mean-field ground state can be written as ${|\Psi\rangle=\Pi_i(\sqrt{\rho}h_i^{\dagger}+\sqrt{1-\rho}s_i^{\dagger})|\Omega\rangle}$. An interesting consequence of the 
singlet condensation is an enhanced propagation of spin excitations (triplons). The authors  of Refs.~\onlinecite{rademaker13a,rademaker13c} observed that the triplet excitations in the condensate of mobile excitons $t_{\text{ex}}\gg J_{\parallel}$ propagate faster that in the quantum paramagnet. Surprisingly, they found the effective
triplon hopping scales with the density of the condensate $\rho_{\text{SF}}=\sqrt{\rho(1-\rho)}$ rather than the exciton density $\rho$.


\subsection{Weak coupling}
\label{sec:weakc}
In the weak-coupling (BCS) limit the formation of excitons and their condensation take place at the same temperature. The physics of the system in this limit
can be described by approaches such as Hartree-Fock approximation and RPA. The key feature of the weak-coupling theories is that
EC instability is driven by nesting between the Fermi surface sheets formed by the valence and conduction bands. The lack of perfect nesting in real materials 
is likely one of the reasons why EC is rarely found in nature. The weak-coupling theory of the excitonic condensation in systems
with equal concentration of holes and electrons was developed in 1960's~\cite{keldysh65,desc65} and summarised in the review articles of Halperin and Rice~\cite{halperin68a,halperin68b}. The pairing glue considered in the  weakly-coupled semi-metals or semiconductors comes from the long-range part of the 
Coulomb interaction. The exchange part of the long-range Coulomb interaction is small, i.e. similar to the choice $J,J'\approx0$ in (\ref{eq:2bhm}), and
one has to consider both spin-singlet and spin-triplet pairing. Halperin and Rice~\cite{halperin68b} classified the possible excitonic condensates in systems
with single Fermi surface sheet per band into the charge-density wave (real singlet), charge-current-density wave (imaginary singlet), 
spin-density wave (real triplet), spin-current-density wave (imaginary triplet) type~\footnote{This assignment assumes that the underlying orbitals are real Wannier functions.}.
In case that there are multiple Fermi surface sheets related by point-group symmetries, e.g., as in hexaborides, a more complex symmetry classification is necessary~\cite{balents00a,murakami02}. The classification of Ref.~\onlinecite{halperin68b} includes only the non-magnetic polar EC states. In mid 1970's Volkov and collaborators showed that in a doped material with unequal number of electrons and holes ferromagnetic EC state develops characterised by simultaneous presence of finite singlet and triplet components
~\cite{volkov73,volkov75,volkov76}. 
\begin{figure}
\includegraphics[width=\columnwidth]{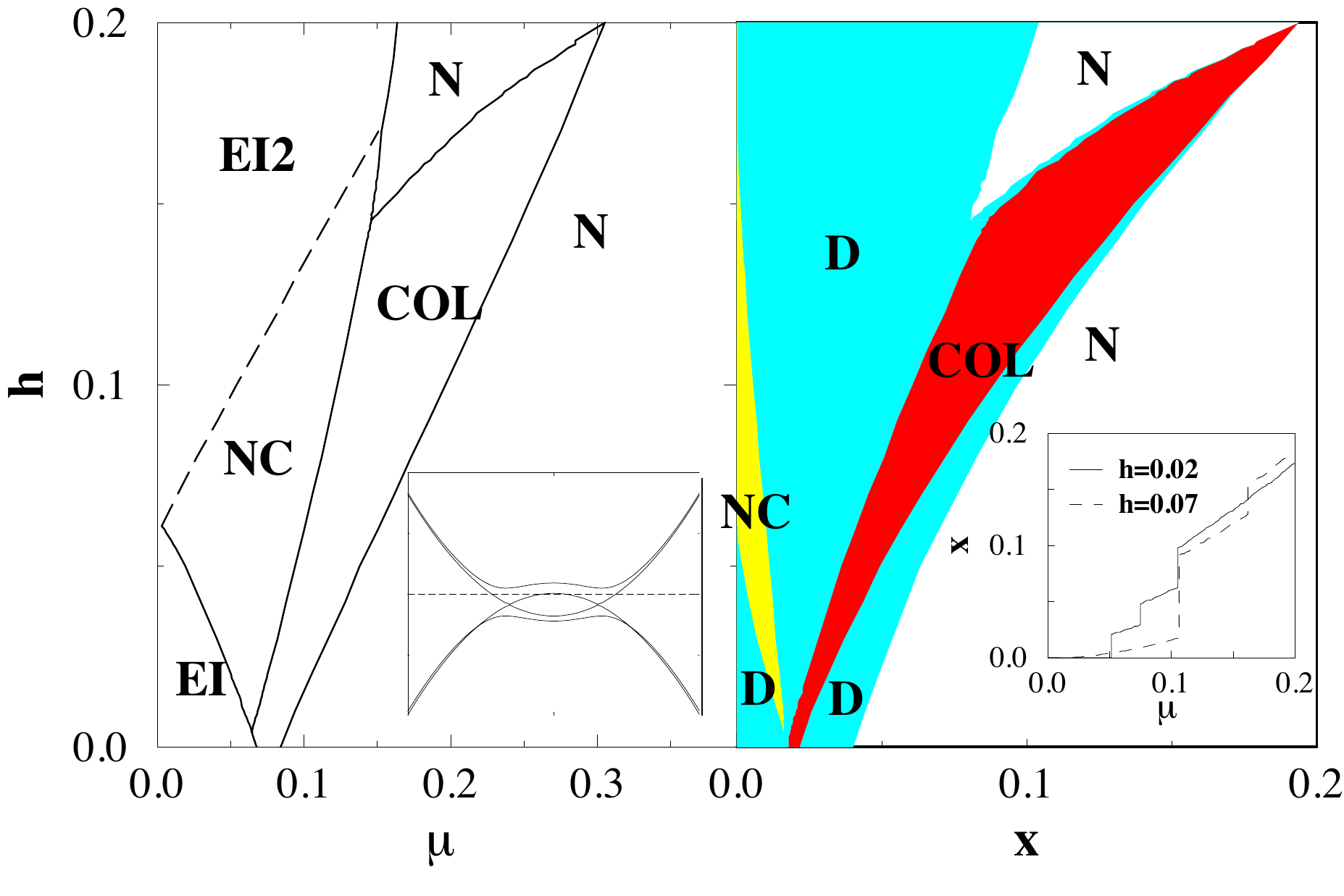}
\caption{\label{fig:bascones} The mean-field phase diagram of doped 2BHM as a function of external field $\bh$ and the chemical potential $\mu$ (left) and as a function of doping $x$ (right). The solid lines in the left panel mark first-order transitions. The light blue areas (marked D) in the right panel represent phase coexistence regions. The meaning of the different phases is described in the text. Adapted with permission of authors from Ref.~\onlinecite{bascones02}. Copyrighted by the American Physical Society.}
\end{figure}

We briefly review the mean-field (Hartree-Fock) theory of (\ref{eq:2bhm}) and consider the simplest case of uniform EC order ($t_at_b<0$).
The mean-field Hamiltonian that allows polar as well as ferromagnetic EC order reads
\begin{equation}
\label{eq:H_mf}
\begin{split}
H&=\sum_{\bk,\sigma\sigma'}\left(\eps_{\bk a}\delta_{\sigma\sigma'}-\bh_a\cdot\btau_{\sigma\sigma'}\right)a^{\dagger}_{\bk\sigma}a^{\phantom\dagger}_{\bk\sigma'}\\
&+\sum_{\bk,\sigma\sigma'}\left(\eps_{\bk b}\delta_{\sigma\sigma'}-\bh_b\cdot\btau_{\sigma\sigma'}\right)b^{\dagger}_{\bk\sigma}b^{\phantom\dagger}_{\bk\sigma'}\\
&-\sum_{\bk,\sigma\sigma'}\left(\left(\Delta_s\delta_{\sigma\sigma'}+\boldsymbol{\Delta}_t\cdot\btau_{\sigma\sigma'}\right) b^{\dagger}_{\bk\sigma}a^{\phantom\dagger}_{\bk\sigma'}+H.c.\right).
\end{split}
\end{equation}
Here, $a^{\phantom\dagger}_{\bk\sigma}, b^{\phantom\dagger}_{\bk\sigma}$ are Fourier transforms of $a^{\phantom\dagger}_{i\sigma}, b^{\phantom\dagger}_{i\sigma}$,
respectively. The crystal-field splitting $\Delta$ as well as the spin-independent part of the self-energy are absorbed in band dispersions $\eps_{\bk a}, \eps_{\bk b}$. 
In the following discussion we will assume $a$ to form the conduction band and $b$ to form the valence band.
The Weiss fields $\bh_a,\bh_b,\Delta_s,\boldsymbol{\Delta}_t$ are given by
\begin{equation}
\label{eq:weiss}
\begin{split}
\bh_a=\bh+\frac{U}{2N}\sum_{\bk,\sigma\sigma'}\langle a^{\dagger}_{\bk\sigma}a^{\phantom\dagger}_{\bk\sigma'} \rangle \btau_{\sigma\sigma'}\\
\Delta_s=\frac{U'}{2N}\sum_{\bk,\sigma\sigma'}\langle a^{\dagger}_{\bk\sigma}b^{\phantom\dagger}_{\bk\sigma'} \rangle \delta_{\sigma\sigma'}\\
\boldsymbol{\Delta}_t=\frac{U'}{2N}\sum_{\bk,\sigma\sigma'}\langle a^{\dagger}_{\bk\sigma}b^{\phantom\dagger}_{\bk\sigma'} \rangle \btau_{\sigma\sigma'},\\
\end{split}
\end{equation}
where $\bh$ is the external magnetic field (acting on spin only). The field $\bh_b$ is defined as $\bh_a$ with the orbital flavour replaced. Note that for model (\ref{eq:2bhm})
with local interactions ${(\Delta_s,\boldsymbol{\Delta}_t)=U'(\phi_s,\bph_t)}$, where $\phi_s$, $\bph_t$ are the local order parameters. The generalisation to models with non-local interaction, which leads to $\bk$-dependent Weiss fields, is straightforward and can be found in the literature. 

Bascones {\it et al.}~\cite{bascones02} studied model (\ref{eq:H_mf}) as a function of doping and external field $\bh$ at $T=0$. The phase diagram, shown in Fig.~\ref{fig:bascones}, contains four phases: the normal phase (N), polar excitonic insulator (EI) phase and two metallic FMEC phases called NC and COL in Ref.~\onlinecite{bascones02}.
To illuminate the nature of these phases it is helpful to introduce $\Delta_{\sigma\sigma'}=\tfrac{1}{2}(\Delta_s\delta_{\sigma\sigma'}+\boldsymbol{\Delta}_t\cdot\btau_{\sigma\sigma'})$.

In the zero-field EI phase, the singlet and triplet EC orders are degenerate. A finite field $\bh$, assuming $\bh=h_z \hat{\mathbf{z}}$ and $h_z>0$ in the following, lifts the degeneracy. The undoped system selects a triplet state with $\Delta_{\uparrow\uparrow}=\Delta_{\downarrow\downarrow}=0$ and $|\Delta_{\uparrow\downarrow}|>|\Delta_{\downarrow\uparrow}|$. For sufficiently large $h_z$ the system enters the EI2 phase of Fig.~\ref{fig:bascones} with $\Delta_{\downarrow\uparrow}=0$. Upon doping two distinct FMEC phases are found. 
In the NC the only non-zero element of the oder parameter is $\Delta_{\uparrow\downarrow}\neq 0$. The EI2 and NC phases are distinguished by presence of a $\bh$-dependent
gap between the uncondensed $a_{\downarrow}$ and $b_{\uparrow}$ bands. The order parameter in the NC phase is purely spin-triplet and 
the phase coincides with the FMEC phase of $S=1$ bosons in the strong-coupling limit.
In the COL phase, the only non-zero element of $\Delta_{\sigma\sigma'}$ is $\Delta_{\downarrow\downarrow}\neq 0$. In this phase, predicted by Volkov {\it et al.}~\cite{volkov75} the singlet and triplet order parameters mix with equal weight ($\Delta_s=-\Delta^z_t$).

The $\bh=0$ spin-wave spectrum of the COL state for a general chemical potential $\mu$ is characterised by two gapless modes with a quadratic dispersion at small $|\mathbf{q}|$ and an additional soft but gapped mode~\cite{bascones02}. At a special value of $\mu$ the soft mode becomes gapless and the spectrum has one quadratic and two linear modes.  
The mean-field phase diagram and magnetic excitations of the polar EI phase in the undoped model were studied by Brydon and Timm~\cite{brydon09b}
and Zocher {\it et at.}~\cite{zocher11} using RPA. They observed acoustic-like modes with linear dispersion at small $|\mathbf{q}|$ predicted by
Kozlov and Maksimov~\cite{kozlov66} and J\'erome {\it et al.}~\cite{jerome67}.


Moving away from the strong-coupling limit the separation of energy scales of exciton formation and exciton condensation is progressively less well defined and eventually
these scales are not separated at all. As in the spinless case of EFKM, EC exists also at weak-coupling and one can follow the BEC-BCS crossover as the interaction 
strength is lowered. The weak-coupling methods, such as Hartree-Fock approximation and RPA, were applied to study the EC in its early days
and are summarised in the review article of Halperin and Rice~\cite{halperin68a,halperin68b}.

\subsection{Intermediate  coupling}
Investigations of systems with intermediate-coupling strength are notoriously difficult due to the lack of small parameters. The general approaches to this problem
include numerical simulations of finite systems such as exact diagonalisation or quantum Monte-Carlo (QMC) methods, large-N expansions, and embedded impurity or imbedded cluster methods such as dynamical mean-field theory~\cite{dmft} (DMFT), variational cluster approximation~\cite{vca}, dynamical cluster approximation or cluster 
DMFT~\cite{dca}. A major obstacle in simulation of ordering phenomena with finite-system methods is the necessity of scaling analysis, i.e., the cluster size must be large
enough to show the 'diverging' correlation length. Monte-Carlo simulations on large clusters are available for many bosonic and spin systems, but usually not for fermions. 
Rademaker {\it et al}~\cite{rademaker13b} applied the determinant Monte-Carlo method to bi-layer Hubbard model and
were able to demonstrate an enhanced response to the excitonic pairing field, but could not reach temperatures below $T_c$. 
Besides the Green's function QMC methods for calculation of correlation functions, wave function QMC approaches can be applied to variational search for ground states. 
While we are not aware of variational QMC studies of excitonic condensation in Hubbard-type lattice models, 
variational QMC has been used to study the corresponding continuum problem~\cite{zhu96}. 

The DMFT methods have been very valuable for investigation of Hubbard model and its multi-band generalisations in the past two decades. However, most applications of DMFT so far focused on one-particle quantities and normal (paramagnetic) phase. With an exception of a multi-band study of Ref.~\onlinecite{park11}, linear response calculations needed to identify instabilities have been so far limited to simple one-~\cite{jarrell92} and two-band models~\cite{jarrell93,ulmke95,kunes14a}. Recent DMFT investigations of ordered phases in models as simple as 2BHM
uncovered a rich physics. Capone {\it et al.}~\cite{capone02,capone04}, Koga and Werner~\cite{koga15} and Vanhala {\it at al.}~\cite{vanhala15} found various forms of
superconductivity in 2BHM at half-filling. Recently, dynamical cluster approximation was also applied to study excitonic condensation.~\cite{vanhala15} 

\subsubsection{Half filling}
Kune\v{s} and collaborators studied 2BHM (\ref{eq:2bhm}) in the vicinity of spin-state transition, see also Fig.~\ref{fig:werner}, using DMFT~\cite{kunes11b,kunes14a,kunes14b,kunes14c}
for 2BHM (\ref{eq:2bhm}) with $V_{ab},V_{ba},J'=0$ and density-density interaction. 
In Ref.~\cite{kunes11b}, they reported observation of the solid phase for the case of strongly asymmetric bands $|t_a|\gg |t_b|$. 
The normal/solid phase boundary qualitatively agrees with strong-coupling BEG model. In particular, the re-entrant transition as a function of temperature for large $\Delta$ has been found. 
In Ref.~\onlinecite{kunes14a}, an unbiased linear response DMFT approach~\cite{kunes11a} was used to probe stability of the normal phase. Two types of instabilities were found, an instability
towards the solid and an instability towards the excitonic condensate. This situation resembles EFKM. Indeed, the physics behind formation of the solid and the superfluid
phases in the strong-coupling limits of EFKM and 2BHM is similar. As in EFKM, the excitonic instability in 2BHM extends to the weak-coupling limit.
This is not so for the solid phase. In the weak-coupling limit, the instability towards solid either does not exist at all or is weaker than the antiferromagnetic instability.
\begin{figure}
\includegraphics[width=0.46\columnwidth]{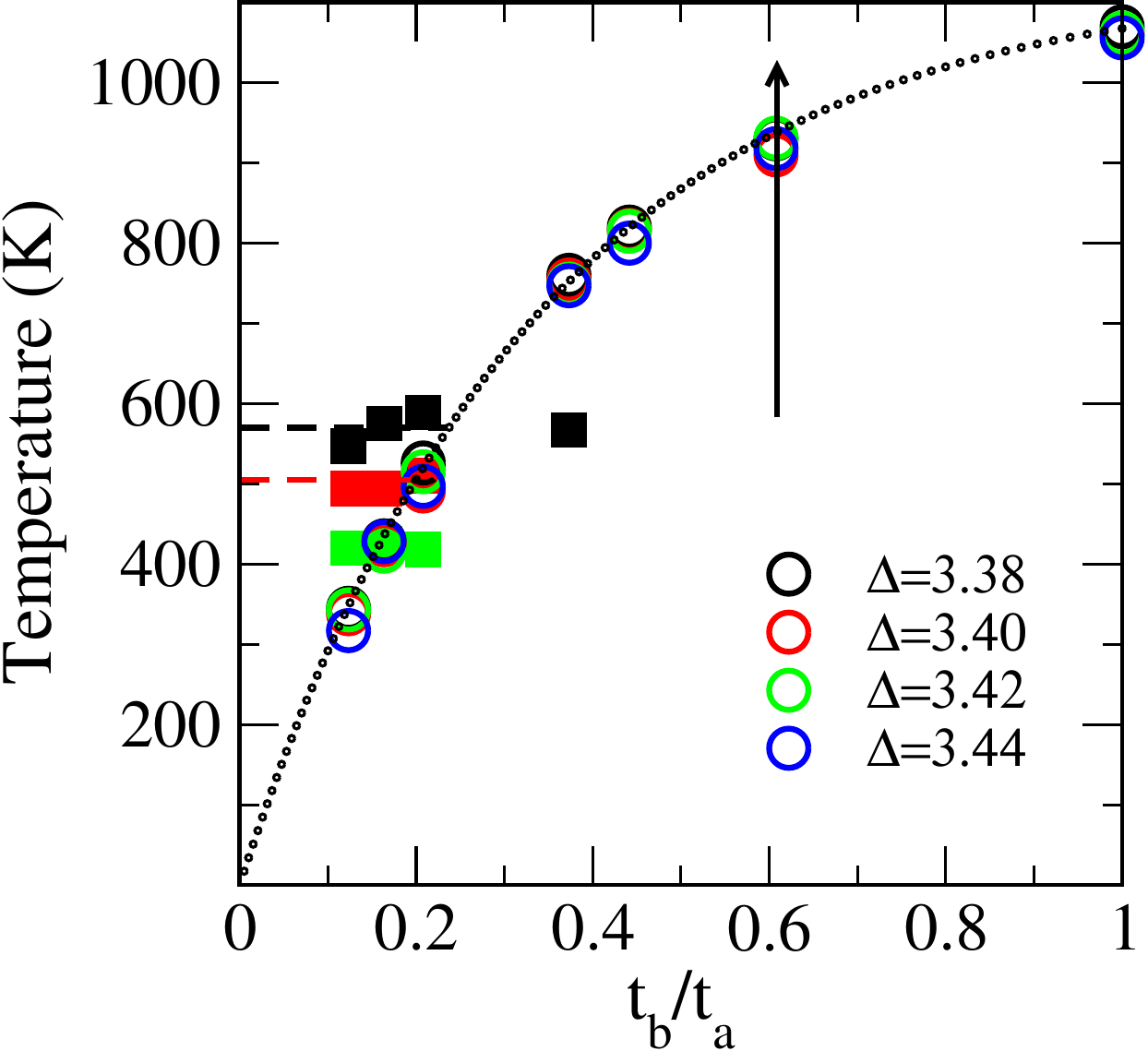}
\includegraphics[width=0.44\columnwidth]{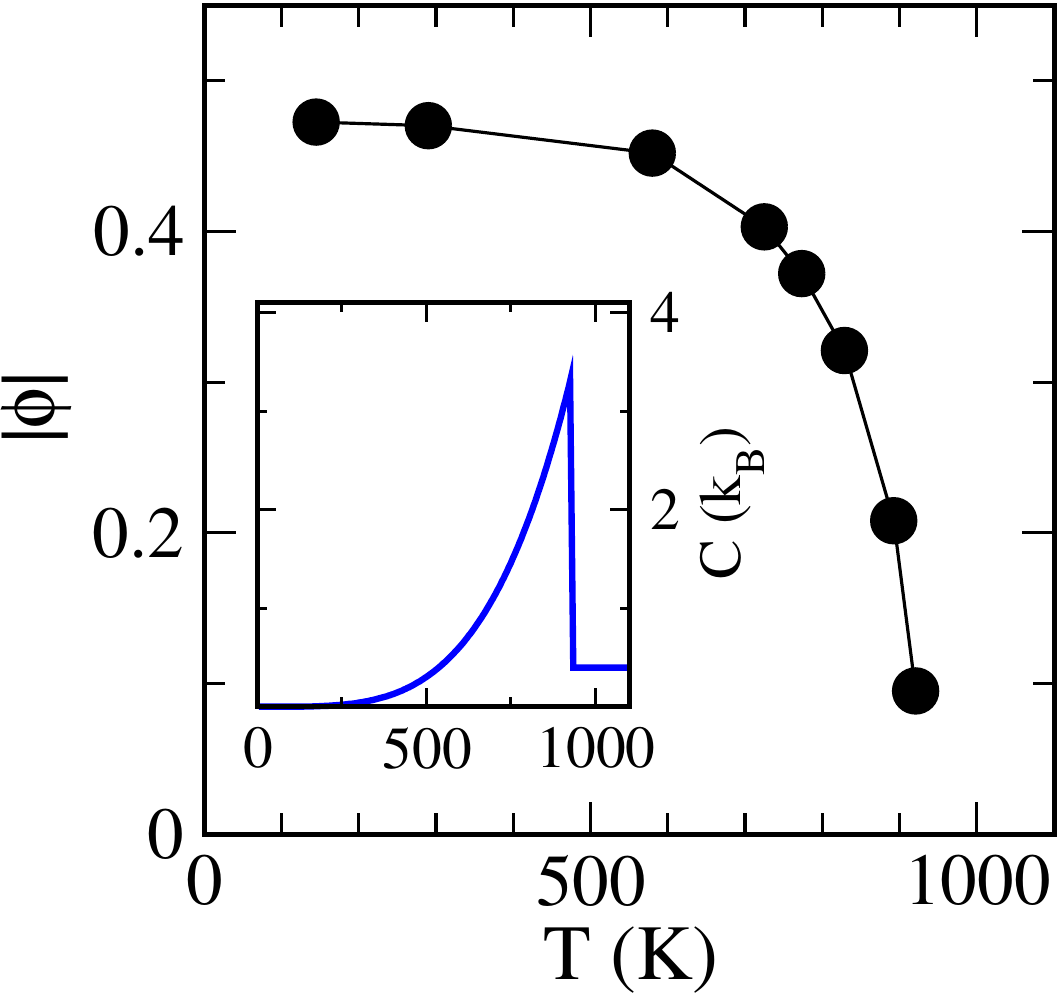}
\caption{\label{fig:ei_inst} Left: The instability temperature of the normal phase for various band asymmetries $t_b/t_a$ and crystal-fields $\Delta$. The squares mark the divergence of the 
orbital susceptibility (instability towards the solid order), the circles correspond to the divergence of the excitonic susceptibility. The sum $t_a^2+t_b^2$ was kept fixed. Data taken from Ref.~\onlinecite{kunes14a}. 
Right: The magnitude of the order parameter $|\phi(T)|$ along the trajectory marked by arrow in the left panel. The specific heat
per site along the same line is shown in the Inset. 
Adapted from Ref.~\onlinecite{kunes14b}. Copyrighted by the American Physical Society.
}
\end{figure}

\begin{figure}
\includegraphics[width=0.36\columnwidth]{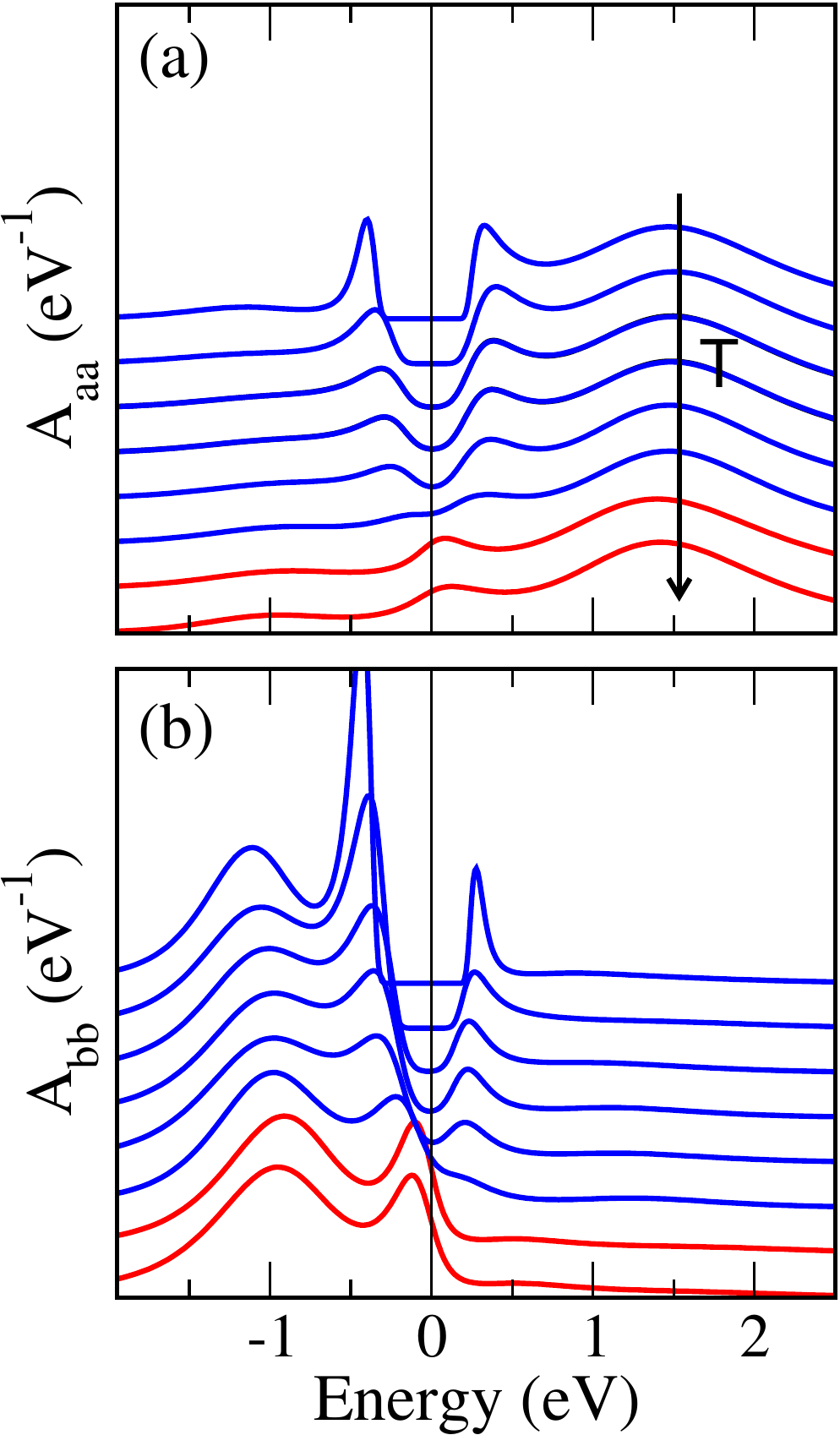}
\includegraphics[width=0.45\columnwidth]{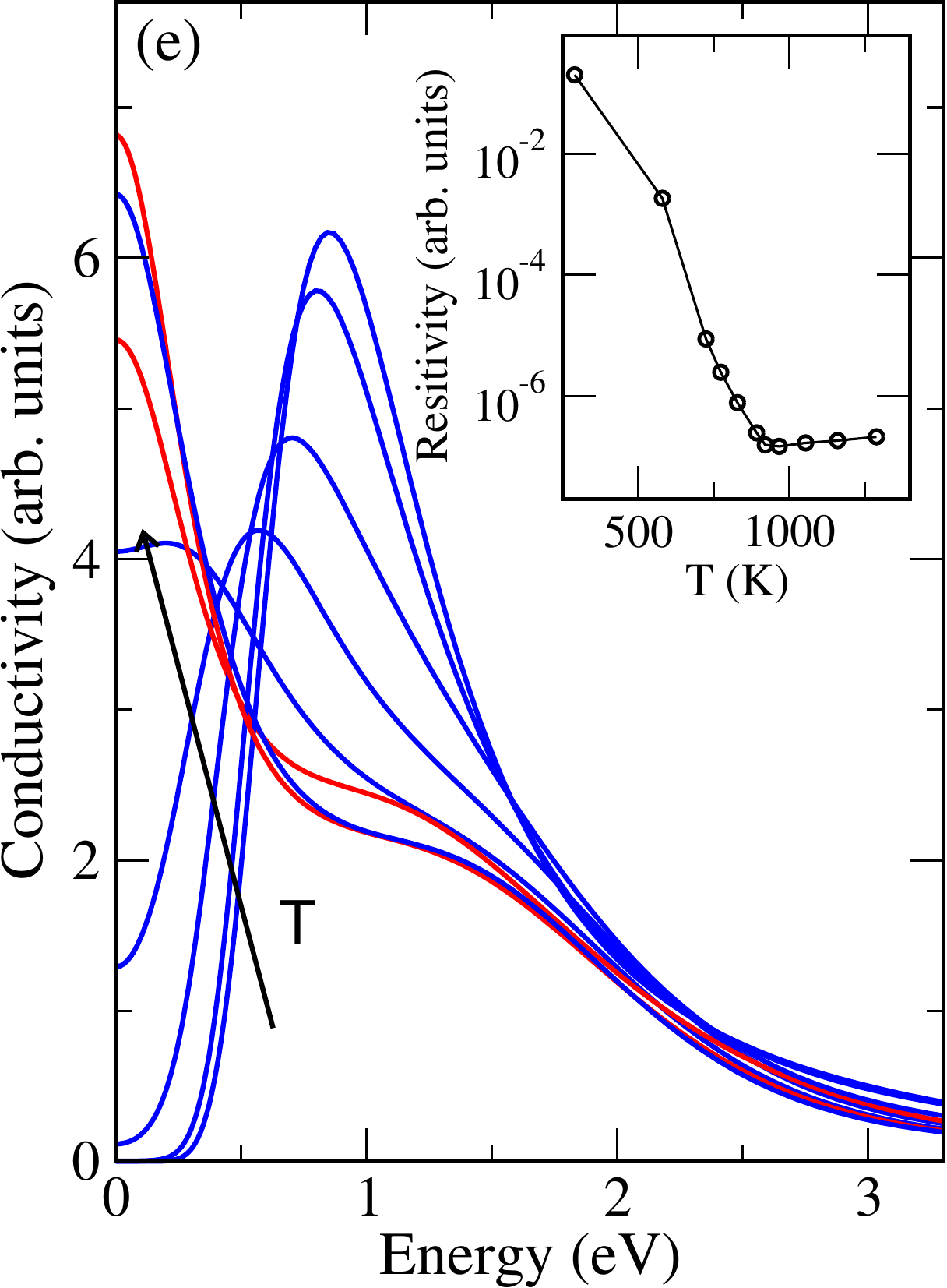}
\caption{\label{fig:orca} Left: The evolution of the one-particle spectral functions (diagonal elements) with temperature along the line marked in Fig.~\ref{fig:ei_inst}a. The spectra
are taken at $T$=1160, 968, 921, 892, 829, 725, 580, and 290 K. Right: The corresponding optical conductivity. The Inset shows the $T$-dependence of the $dc$ resistivity.
Adapted from Ref.~\onlinecite{kunes14b}. Copyrighted by the American Physical Society.
 }
\end{figure}

The basic physical properties of the excitonic phase were studied in Ref.~\onlinecite{kunes14b}. 
In Fig.~\ref{fig:ei_inst}, we show a typical \mbox{$T$-dependence} of the magnitude of the order parameter $|\phi|$. The systems selected a polar EC state, $\bar\bph_i\wedge\bph_i=0$,~\footnote{In Ref.~\onlinecite{kunes14b} the term linear phase was used.} which agrees with the antiferromagnetic nn exchange expected in the the strong-coupling limit (\ref{eq:boson1}). 
The shape of $|\phi(T)|$ is consistent with the mean-field $(1-T/T_c)^{1/2}$ dependence, expected for DMFT method. 
As in the weak-coupling limit, the excitonic condensation leads to opening of a charge gap and appearance of Hebel-Slichter peaks in
the one-particle spectra, see Fig.~\ref{fig:orca}. 
This behaviour reflects the formal analogy to an $s$-wave superconductor discussed in Sec.~\ref{sec:spinless}. 

This analogy however, does not extend to the
electromagnetic properties of the condensate. The neutral exciton condensate does not contribute to the charge transport, which
is facilitated by the quasi-particle excitations.
In Fig.~\ref{fig:orca} we show the optical conductivity and $dc$ resistivity at various temperatures. 
The opening of the charge gap leads to an optical gap and exponential increase of the resistivity below $T_c$. 

\begin{figure}
\includegraphics[width=0.48\columnwidth]{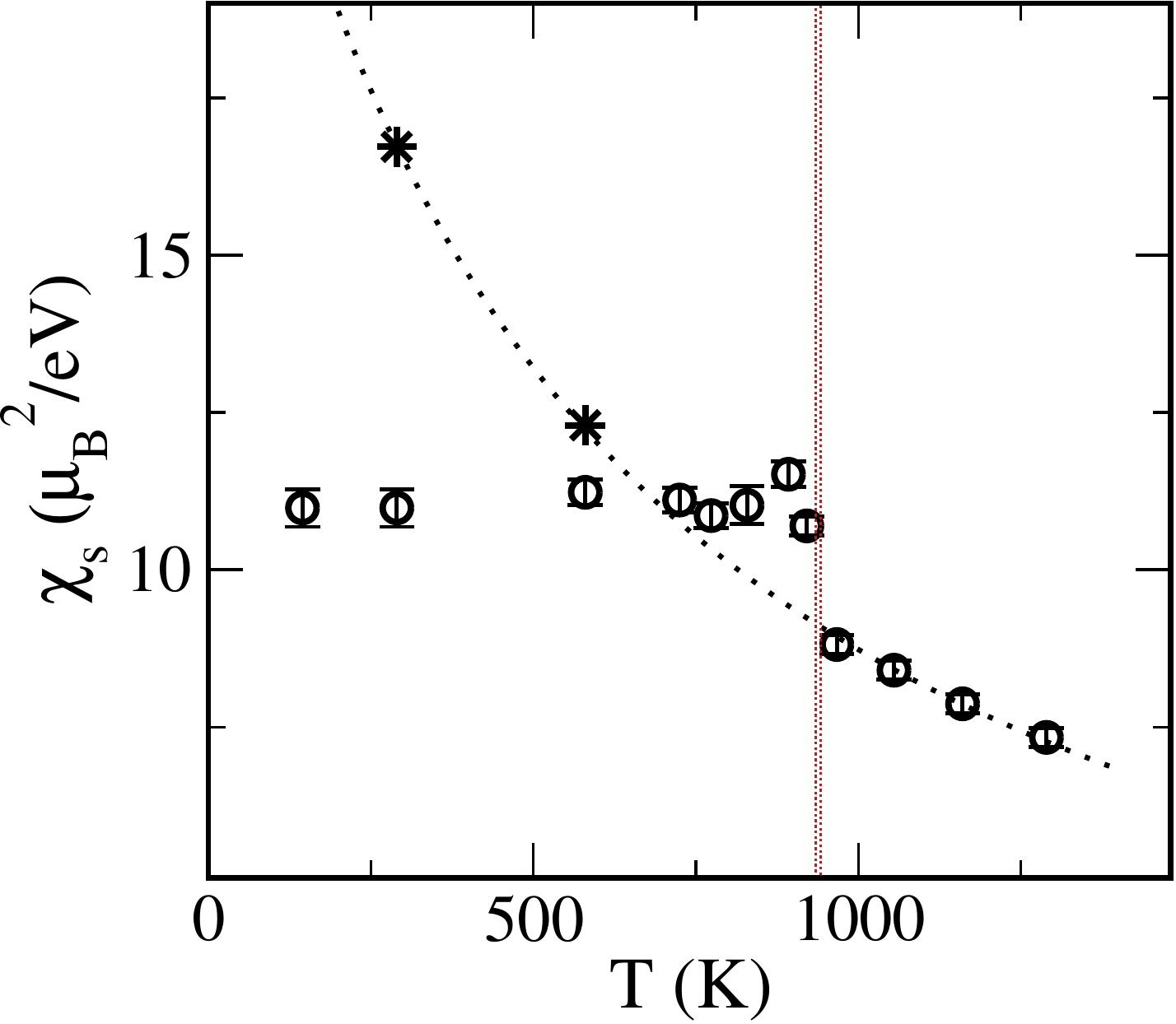}
\includegraphics[width=0.48\columnwidth]{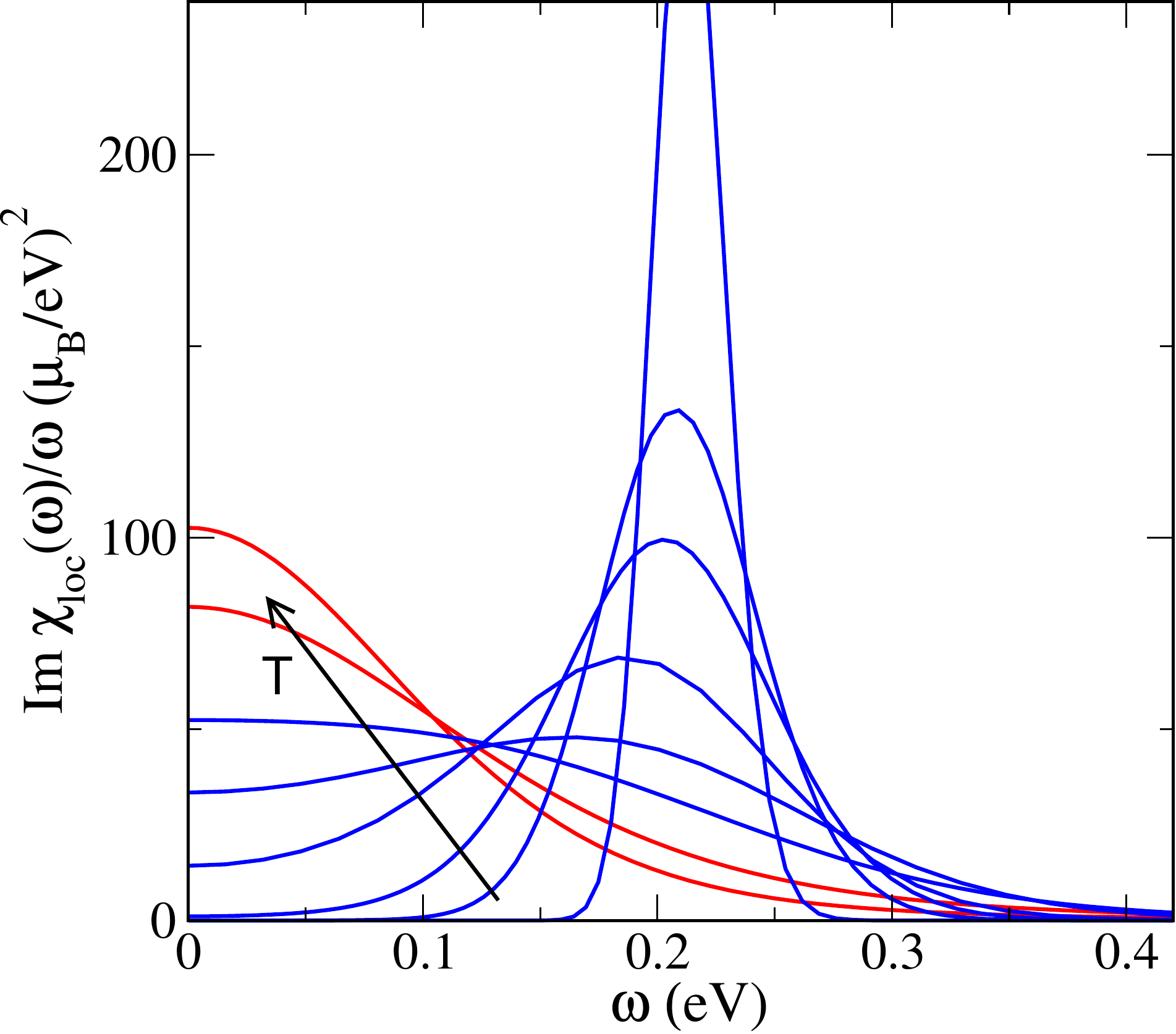}
\caption{\label{fig:bw} Left: The static uniform spin susceptibility $\chi_s$ (circles with error bars) along the line marked in Fig.~\ref{fig:ei_inst}a. The stars and dotted line indicate $\chi_s(T)$ for the system constrained
to the normal state. The red vertical line marks the EC transition. Adapted from Ref.~\onlinecite{kunes14b}. Copyrighted by the American Physical Society. Right: The imaginary part of the local dynamic susceptibility $\chi_{\text{loc}}(\omega)$ at various
temperatures across the EC transition~\cite{kunes_unpub}. }
\end{figure}

Finally, we discuss the spin susceptibility $\chi_s^{zz}$, shown in Fig.~\ref{fig:bw}. While the normal-phase susceptibility exhibits Curie-Weiss behaviour, 
the susceptibility of the polar EC phase appears $T$-independent. 
This observation holds within the numerical accuracy for the uniform susceptibility and approximately also for the local susceptibility. This behaviour can be understood from 
a single atom picture. In the normal phase, the system is locally in a statistical mixture of the LS ground state and thermally populated HS multiplet. This leads
to Curie-Weiss susceptibility and a vanishing of the spin gap. 
In the polar EC phase, the Weiss field mixes the LS and HS states and opens a gap of the order $U'\phi/2\gg T$ between the local ground state and the excited states. The local ground state is a superposition of the form $\alpha |\bvac\rangle+\beta(|-1\rangle + |1\rangle )$. The spin susceptibility has $T$-independent van Vleck character and
a finite spin gap appears, Fig.~\ref{fig:bw}.

\begin{figure}
\includegraphics[width=0.8\columnwidth]{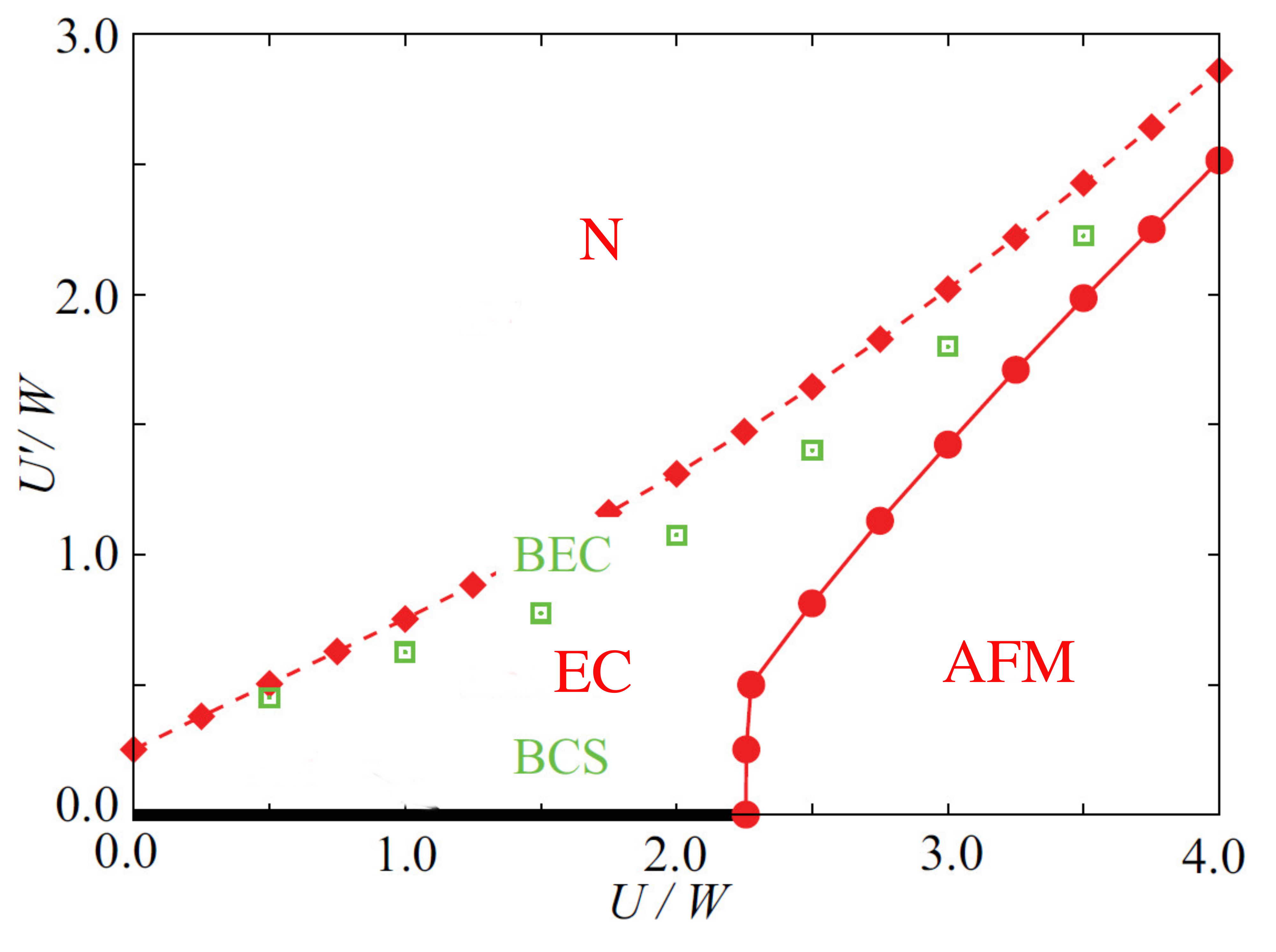}
\caption{\label{fig:kaneko} The $T=0$ phase diagram of 2BHM on a square lattice for $t_a=t_b=1$ ($W=4$), $V_{ab},V_{ba},J,J'=0$, $\Delta=6$ obtained with variational cluster
approximation. The transition between the band insulator (N) and the EC phase is continuous, the EC/AFM phase transition is of the first order. 
The thick black line for $U'=0$ marks a paramagnetic metal.
Adapted with permission of authors from Ref.~\onlinecite{kaneko12}. Copyrighted by the American Physical Society.}
\end{figure}

Kaneko {\it et al.}~\cite{kaneko12} used variational cluster approximation to study the spin-triplet EC in 2BHM without Hund's coupling ($J,J'=0$) and with symmetric bands $t_a=t_b$
for a broad range of interaction parameters.
Similar to the strong-coupling phase diagram Fig.~\ref{fig:btj_mf} the authors of Ref.~\onlinecite{kaneko12} 
found continuous transition between the EC and normal state (band insulator), and a first-order transition
between the EC and AFM phases as shown in Fig.~\ref{fig:kaneko}. In Ref.~\onlinecite{kaneko14}, Kaneko and Ohta extended the study to include the Hund's coupling, which confirmed that $J<0$ favours the spin-singlet
charge-density-wave state, while $J>0$ selects the spin-triplet spin-density-wave state~\cite{halperin68b,burker81}.

\subsubsection{Doping}
The weak-coupling theory of doped excitonic insulator was developed by Volkov and collaborators~\cite{volkov73,volkov75,volkov76} for systems without Hund's coupling. They showed
that such systems tend to develop ferromagnetic order due to simultaneous appearance of the spin-singlet and spin-triplet order. 
In Sec.~\ref{sec:weakc}, we have reviewed the application of the weak-coupling approach by Bascones~{\it et al.}~\cite{bascones02}. Besides finding Volkov's ferromagnetic phase (COL)
the authors of Ref.~\onlinecite{bascones02} found also a spin-triplet phase (NC) induced by an external magnetic field. 
The possibility of the triplet ferromagnetism was discussed also by Balents~\cite{balents00b} who considered  the effect of doping in a strong coupling limit on a qualitative level.

\begin{figure}
\includegraphics[width=0.48\columnwidth]{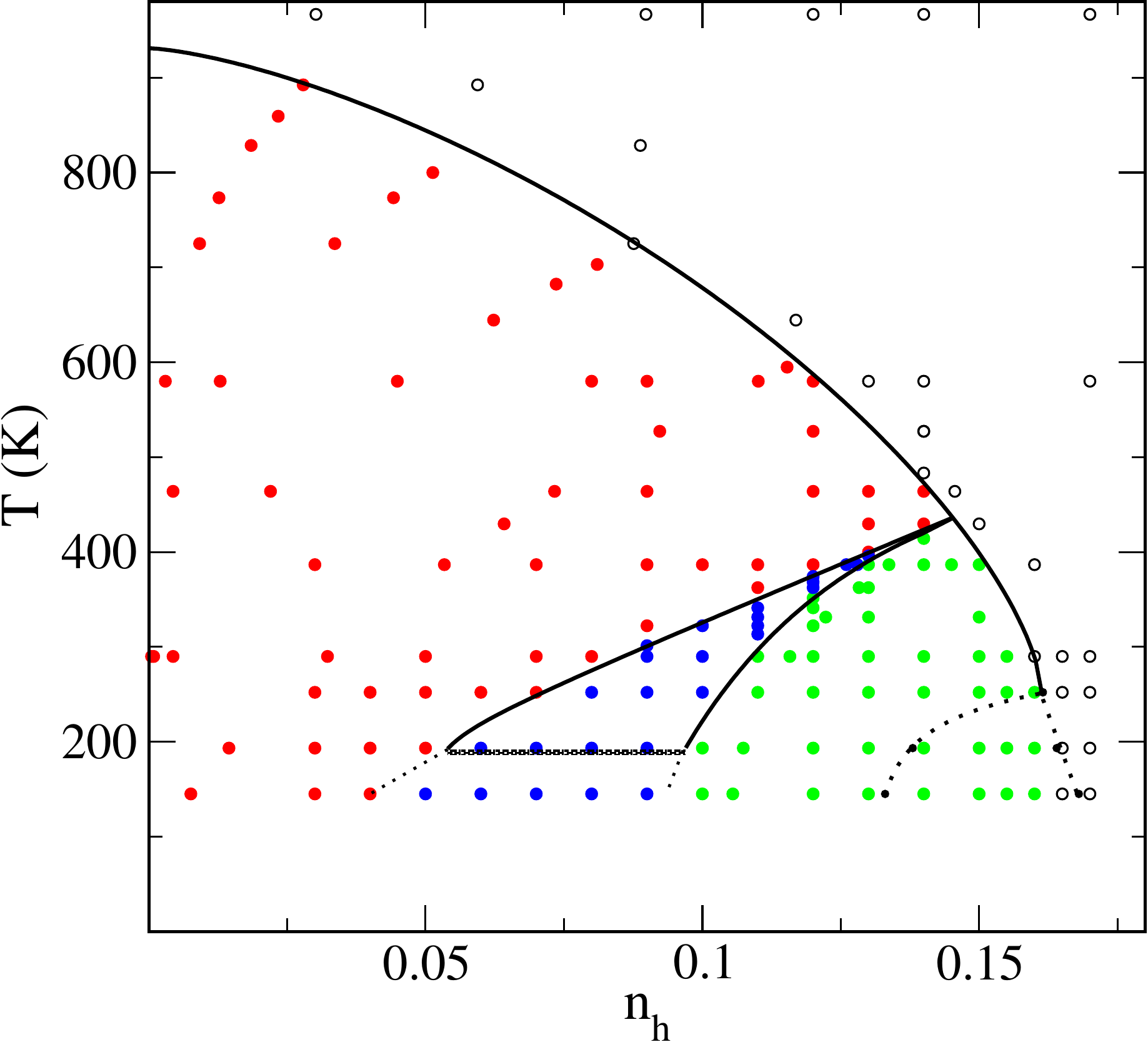}
\includegraphics[width=0.48\columnwidth]{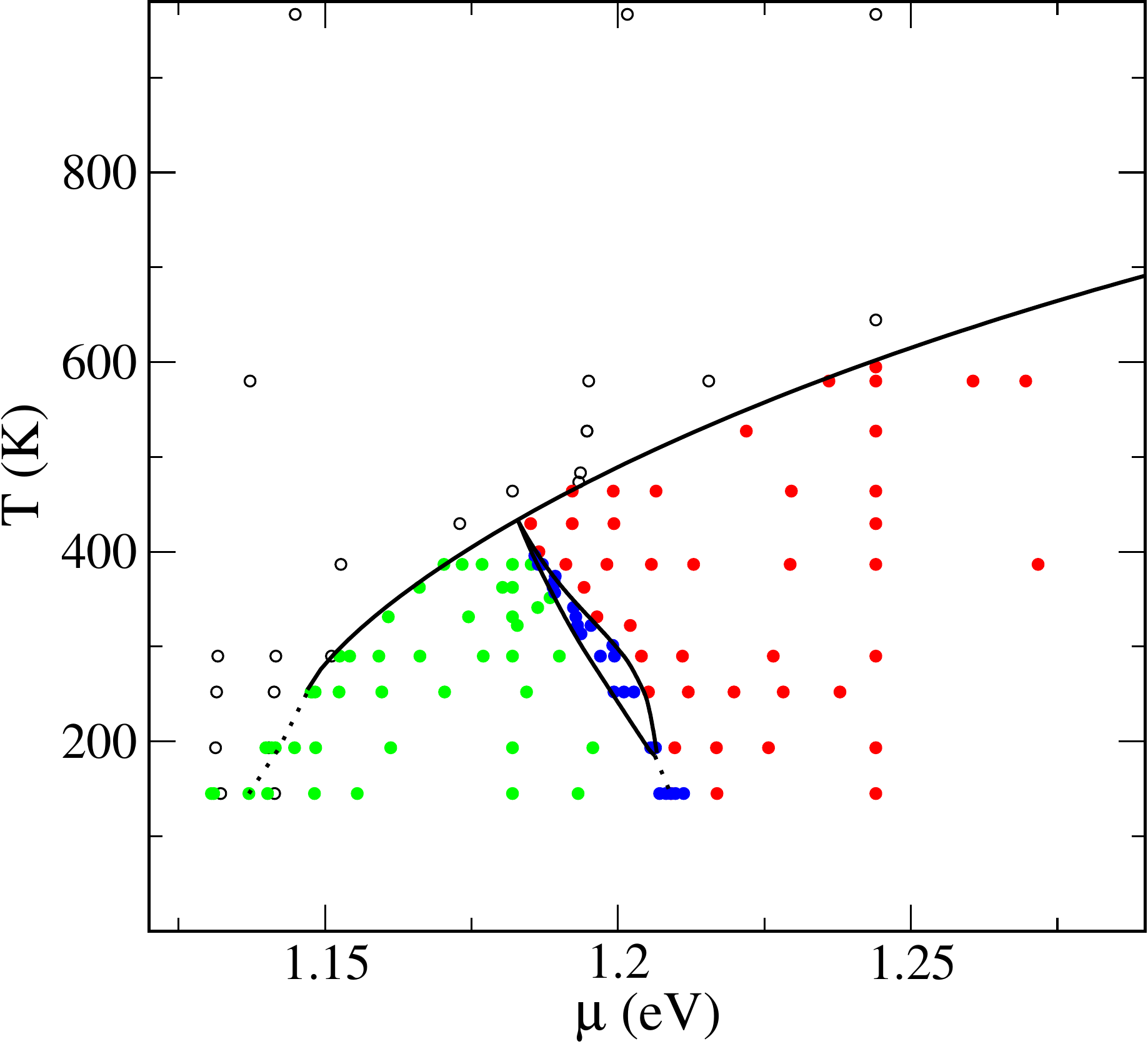}
\caption{\label{fig:phase} Phase diagram of 2BHM with hole-doping (counted from half-filled case) (left) and chemical potential (right) on the horizontal axis. Solid lines
mark continuous transitions, while dashes lines mark the first order transitions (right) or boundaries to the phase separation regime (left). The coloured dots represent the actual
numerical calculations. The red colour marks the polar EC phase, the blue and green colour mark the FMEC phase. The blue and green phases are distinct when the cross-hopping
in (\ref{eq:2bhm}) is strictly zero.  Adapted from Ref.~\onlinecite{kunes14c}. Copyrighted by the American Physical Society.}
\end{figure}

Kune\v{s}~\cite{kunes14c} used DMFT to study the effect of doping in 2BHM with strong Hund's coupling, which limits the possible EC order to spin triplet. 
The phase diagram, shown in Fig.~\ref{fig:phase}, contains the normal phase (open circles) and three excitonic phases. 
The polar phase (red), discussed in preceding section at half-filling, extends to finite doping levels. It is distinguished from the other two phases by absence of ordered spin moment. 
The relation ${\bph^*\wedge\bph=0}$ implies that $\bph$ can be factorised to a real vector and a phase factor. The arbitrariness of the phase factor is connected
to the absence of the cross- and pair-hopping in the studied model. 
The polar phase is equivalent to the EI phase in Fig.~\ref{fig:bascones}. 

At higher doping levels and lower temperatures the system enters the FMEC phase with finite magnetisation ${\mathbf{m}\sim i(\bph^*\wedge\bph)\neq 0}$. 
The transition between polar and ferromagnetic phases proceeds via an intermediate phase (blue), which is distinguished from the FMEC only in absence
of cross- and pair-hopping~\cite{kunes14c}. The FMEC phase is equivalent to the COL phase of Ref.~\onlinecite{bascones02} (Fig.~\ref{fig:bascones}).
Similar to the $T=0$ phase diagram of Ref.~\onlinecite{bascones02}, first-order transitions and phase separation is found at low temperatures. It is not clear 
from the numerical data whether a $T=0$ polar phase exists at finite doping. At the moment we can only speculate that the antiferromagnetic nn coupling
provides a means to stabilise it. A peculiar feature of the FMEC is the $T$-dependence of the magnetisation in the vicinity of the continuous transition
to the normal state, which follows the linear ${1-T/T_c}$ dependence. This behaviour is a consequence of the quadratic dependence of magnetisation on the EC order parameter
${\mathbf{m}\sim i(\bph^*\wedge\bph)}$.

 
 Finally, we briefly discuss the effect of finite cross- and/or pair-hopping, which break the charge conservation per orbital flavour. As a result
 the phase factor in the polar EC phase is fixed and the system selects either the spin-density-wave or spin-charge-density-wave order. It is also possible
 that both types of order are realised in parts of the phase diagram. Another consequence of finite cross-/pair- hopping will be the absence of a 
 continuous normal to FMEC transition. The FMEC state is selected by terms of the order $\phi^4$ in the Ginzburg-Landau functional and a continuous transition
 is only possible if the second order terms do not depend on the phase of $\phi$. Finite cross-/pair- hopping removes this degeneracy. For weak
 cross hopping one can expect a small wedge of polar EC phase separating the normal and FMEC phases. The first-order normal/FMEC
 transitions are still possible.

 %
 
 \section{Multi-orbital Hubbard model}
 Little has been done in generalisation of the physics of Sec.~\ref{sec:2bhm} to systems with more than two orbitals per atom.
 Kune\v{s} and Augustinsk\'y~\cite{kunes14b} used Hatree-Fock (LDA+U) approach to study EC in quasi-cubic perovskite with nominally 6 electrons in the $d$ shell.
 Let us demonstrate the new features on 5-orbital model describing the $d$-orbital atoms on a cubic lattice. 
 The basic setting of such system is similar to half-filled two-orbital model. 
 The crystal-field splits the $d$ orbitals into threefold degenerate $t_{2g}$ and twofold degenerate $e_g$ states. With 6 electrons per atom
 the lower $t_{2g}$ levels are filled and the upper $e_g$ levels are empty. 
 An exciton is formed by moving an electron from a  $t_{2g}$ orbital to an $e_g$ orbital on the same atom.
 Unlike 2BHM with only one possible orbital structure of an exciton, in the $d$-atom there are six possible orbital combinations. 
 The cubic symmetry distinguishes the orbital symmetries of an exciton into two 3-dimensional irreducible representations $T_{1g}$ and $T_{2g}$. 
 Considering the geometry of these two excitons~\cite{kunes14b} one can infer that $T_{1g}$ excitons are more tightly bound and substantially more mobile that the  $T_{2g}$ ones. 
 Therefore only the  $T_{1g}$ excitons need to be considered as candidates for condensation. 
The three-fold orbital degeneracy of the $T_{1g}$-excitons adds to the $S=1$ spin degeneracy making the order parameter a more complex object. 
It can be arranged to a tensorial form of a $3\times3$ matrix $\phi^{\alpha}_{\beta}$, where the $\alpha$ indexes the spin and $\beta$
the orbital components. $\phi^{\alpha}_{\beta}$ thus transforms as a vector under the $SO(3)$ rotations in the spin space and as a pseudo vector ($T_{1g}$ representation) under the discrete symmetry operations of the cubic point group. The explicit expression for  $\phi^{\alpha}_{\beta}$  in terms of the 
on-site occupation matrix can be found in the supplementary material of Ref.~\onlinecite{kunes14b}. 
The structure of the order parameter resembles of superfluid $^3$He~\cite{vollhardt_he} and 
one can use the group theoretical methods applied there to analyse the possible EC phases~\cite{bruder86}. 
More complex structure of the order parameter allows for existence of numerous distinct phases.  

Another important difference to 2BHM concerns the hard-core constraint imposed in the excitons in the strong coupling limit. While in the case of 2BHM
there cannot be more than one exciton on a given atom, in the 5-orbital model the hard-core constraint is less restrictive. 
There cannot be more than one exciton of a given orbital flavour on atom, but it is possible that two excitons with different
orbital flavours meet. 
In the cobaltites terminology a single exciton on atom represents the intermediate spin $S=1$ state, while the $S=2$ 
high-spin state can be viewed as a bi-exciton. The fact that the site energy of the high-spin state is lower than of the 
intermediate-spin state~\cite{tanabe} 
is then expressed as an attraction between excitons of different orbital flavour and parallel spins.
Bosonic models with infinite intra-species, but finite inter-species interaction have been studied for two species (spinless) bosons~\cite{altman03,kuklov03,trousselet14,lv14}
and shown to exhibit phases which are not allowed with inter-species hard-core constraint.

\section{Materials}
\subsection{Bulk materials}
\label{sec:materials}
There have been numerous proposals of materials to exhibit excitonic condensation, but very few realisations of EC were actually documented.
The early candidates for excitonic condensation, which followed the weak-coupling picture of proximity to semimetal--semiconductor transition, included the group V semimetals
(Bi, Sb, As) and divalent metals (Ca, Sr, Yb), possibly under pressure~\cite{jerome67}. However, the search for signatures of EC in these materials was not successful.

Wachter and collaborators~\cite{neuens90,bucher91,wachter01,wachter04} reported observation of EC in TmSe$_{0.45}$Te$_{0.55}$ under pressure. While they did not see
a sharp thermodynamic transition, they argued that the observed anomalies are consistent with a 'transition' in weak (pairing) field. Wakisaka {\it et al.}~\cite{wakisaka09} interpreted their photoemission data on Ta$_2$NiSe$_5$ in terms of exciton condensation~\cite{kaneko12,kaneko13b,seki14}.  Also the lattice distortion and photoemission spectra
of a layer compound 1$T$-TiSe$_2$ have been interpreted in terms of charge-density-wave type exciton condensation and described with the weak-coupling theory~\cite{cercellier07,monney09,monney10,monney11}.

While the previous examples involved cases of spin-singlet condensation,
in the following we will discuss excitonic magnetism.  In 1999 Young {\it et al.}~\cite{young99} reported observation of 600~K ferromagnetism in La$_{0.005}$Ca$_{0.995}$B$_6$. 
Ferromagnetism in a slightly doped semiconductor lead several groups to generalise the weak-coupling theory of excitonic ferromagnetism
pioneered by Volkov and collaborators~\cite{volkov73,volkov75,volkov76} to the case of multiple Fermi surface sheets~\cite{zhitomirsky99,balents00a,veillete00,barzykin00,murakami02}. Subsequent investigations showed, nevertheless, that the band gap in parent compound CaB$_6$ is $\approx 1$~eV~\cite{denlinger02} and thus inconsistent with the EC scenario. It is now generally accepted that the ferromagnetism in La$_{x}$Ca$_{1-x}$B$_6$ arises from defects
rather that doping.

Another group of materials where the EC concept found its use are iron pnictides~\cite{kamihara08, stewart11}. The physics of these materials is governed
by nesting between several Fermi surface sheets formed by bands of different orbital characters~\cite{graser10}. Several groups studied a simplified 
two-orbital model~\cite{han08,chubukov08,brydon09a,brydon09b} and its multi-orbital extensions~\cite{knolle10,eremin10} using weak-coupling approaches,
and observed a spin-density-wave order with a periodicity  given by the nesting vector. Arising from nesting between
bands that mix several orbital characters, the corresponding Weiss field in general couples all possible orbital combinations.
However, since the nested patches of the Femi surface have different dominant orbital characters a large part of the condensation energy comes
from orbital off-diagonal pairing, which produces local magnetic multipoles but no local moments (polar EC state). The generally present, but small,
orbital diagonal contributions than give rise to the apparently small ordered moments. A first principles calculation of ordered state supporting this picture
was done by  Crincchio~{\it et al.}~\cite{crincchio10}. 

Kune\v{s} and Augustinsk\'y~\cite{kunes14a} proposed that a transition observed in some materials of Pr$_x$Ca$_{1-x}$CoO$_3$ (PCCO) family~\cite{tsubouchi02,tsubouchi04,hejtmanek10,hejtmanek13} can be understood as an excitonic condensation.
Materials from this family exhibit a phase transition with $T_c$ as high as 130~K which is characterised by sharp peak in the specific heat, transition from high-$T$ metal 
to a low-$T$ insulator, disappearance of Co local moment response and simultaneous Pr$^{3+}$ $\rightarrow$ Pr$^{4+}$ valence transition. A puzzling feature of the low
temperature phase is the breaking of time reversal symmetry (in absence of ordered moments) evidenced by Schottky anomaly associated with splitting of the Pr$^{4+}$ Kramer's
ground state. The transition to a spin-density-wave EC state provides a comprehensive explanation of these observations and the EC
ground state is obtained with Hartree-Fock-type LDA+U calculations.

Another class of the materials with potential to exhibit the exciton condensation was proposed by Khaliullin~\cite{khaliullin13,akbari14}. He considered a strong-coupling model of a $d$-electron material with cubic crystal field, strong spin-orbit coupling and an average $d$ occupancy of four electrons per atom. A scenario
possibly realised in materials with Re$^{3+}$, Ru$^{4+}$, Os$^{4+}$, or Ir$^{5+}$ ions. The large crystal field restricts the low energy physics to the space spanned by $t_{2g}$ states. Now the spin-orbit coupling plays the role of $\Delta$ in (\ref{eq:2bhm}) competing with the Hund's coupling $J$. Sufficiently strong spin-orbit coupling renders the single-ion ground state a singlet $\tilde{S}=0$~\footnote{This corresponds to the state with the four $J_{\text{eff}}=3/2$ one-particle states filled and the
$J_{\text{eff}}=1/2$ states empty.} and the first excited state a triplet $\tilde{S}=1$. The perturbative treatment of nn hopping results in the model
similar to (\ref{eq:boson2}). There is, however, one important difference between (\ref{eq:boson2}) and the Khaliullin's model formulated in terms of pseudospin 
$\tilde{\bS}$. The real spin $\bS$ is decoupled from the lattice and thus (\ref{eq:boson2}) is invariant under the $SO(3)$ spin spin rotations, in particular the hopping $K_{\perp}$ is spin independent. The pseudospin $\tilde{\bS}$ represents a spin-orbital object, which is coupled to the lattice and thus the model cannot be invariant
under continuous pseudospin rotations. This is reflected in the hopping amplitudes being $\tilde{S}$-dependent, i.e., the hopping containing terms known
from Kitaev model~\cite{kitaev06}.

Finally, the possibility of exciton condensation in layered cuprates~\cite{brinkman05} and oxide heterostructures~\cite{millis10} was discussed
by several authors, but has not been realised so far.

\subsection{Bi-layer structures}
\label{sec:bilayer}
A major direction in the research of exciton condensation are bi-layer structures~\cite{lozovik75}. The basic idea is that a bi-layer with negligible inter-layer tunnelling provides a system
where two orbital flavours are to high accuracy independently conserved. 
Exciton condensation takes place under suitable conditions including small inter-layer distance, so that inter-layer electron-electron
interaction is sufficiently strong, large enough intra-layer electron mobility, so that sufficiently high $T_c$ can be achieved, and a matching doping, such that the electron concentration
in one layer closely matches the hole concentration in the other layer. 
Two types of structures have been studied. 2D quantum well systems in perpendicular magnetic field. The idea here is to use the formation of Landau levels and, in particular, 
the dependence of the number of quantum states per Landau level on the magnetic field as a means to achieve the desired electron and holes concentrations. 
The exciton condensation in these systems was evidenced by enhanced inter-layer tunnelling~\cite{spielman84} or vanishing Hall conductivity~\cite{eisenstein04}.
For a review of experimental challenges probing exciton condensation in bi-layers we refer the reader to specialised literature~\cite{gupta11}.

Another bi-layer system that has been intensely studied is the system of two graphene layers separated with a dielectric where EC was predicted to take place at room temperature~\cite{min08}. 
The important and much debated issue here is the screening of the inter-layer interaction~\cite{kharitonov08}. The system therefore cannot be described with a simple
Hubbard type model for the strongly correlated electrons. For  a review of the physics of electron-electron interaction in graphene structures we refer the reader to a specialised
literature~\cite{kotov12}.

\section{Outlook}
Analogy to superconductivity has been the traditional driving force of the field of exciton condensation. In particular, there was a considerable effort in realisation of supercurrents driven by 
the gradient of the condensate phase -- a hallmark of superfluidity. 
There are two important conditions to realise a supercurrent. First, the phase invariance (conduction/valence charge conservation) in the normal phase.
Second, the ability to contact separately the electron and the hole parts of the exciton. These conditions can be met in bi-layer systems, where the electrons and hole are spatially separated while
maintaining sufficiently strong interaction. 
We find it, however, unlikely that similar situation can be realised in bulk materials. Already the first condition is bound to be violated by various one- and two-particle terms in the Hamiltonian as discussed 
previously. Instead of a superfluid with an arbitrary phase the system selects either charge(spin)-density-wave or charge(spin)-current-density-wave state~\cite{halperin68b} with a fixed
phase. Omitting possible special points in the phase diagram where the symmetry may be enhanced, e.g., close to a second-order boundary between these phases the system will not exhibit
a superfluid behaviour.

The main value of the concept of exciton condensation in bulk materials lies, in our opinion, in providing a comprehensive picture of potentially complex phase diagrams and
understanding long-range orders that are not easy to detect because they do not lead to charge(spin)-density modulations on inter-atomic scale.
Moreover, the geometrical form of the order parameter in EC system may non be as intuitive as for example magnetisation is systems with local moments.
For example, exciton condensation in $s$-$p$ systems leads to formation of electric dipoles~\cite{batista02} or condensation in a system
close to the spin-state transition gives rise to a magnetic multipole order~\cite{kunes14a}. 
There are other examples of orbital-off-diagonal orders, e.g. orbital currents in cuprates~\cite{varma06,weber14,bulut15}, nematic order in iron pnictides~\cite{kasahara12,rosenthal14,fernandes14}, hidden order in URu$_2$Si$_2$~\cite{mydosh11}
where we can only speculate that they may be viewed as a special example of the EC physics. 
It should be clear from this discussion that the concept of exciton condensation extended beyond the strict superfluid state is not sharply defined and, in particular, the weak-coupling
borderline between exciton condensation and  'just a' Fermi surface instability is quite fuzzy.

We see numerous open questions and possible directions of further investigation. In the two-band model, realisation of spin-current-density-wave phase and in general the phase 
diagram in the presence of cross-hopping are to be explored. The Weiss field in the  spin-current-density-wave state can be viewed as a spontaneous spin-orbit coupling - a field
that breaks the spin-rotational symmetry, but preserves the time reversal symmetry. Interaction between the exciton condensate and the lattice has attracted attention recently~\cite{zenker14}.
The multi-band systems with multiple orbital flavours of excitons are completely unexplored to our knowledge. A model that can shed some new light on the long standing problem
of perovskite cobaltites.

\acknowledgements
I would like to thank Anna Kauch for critical reading of the manuscript and numerous suggestions, Pavel Augustinsk\'y for providing his unpublished data and Louk Rademaker, Arno Kampf, Vladislav Pokorn\'y, Dieter Vollhardt, and Pavol Farka\v{s}ovsk\'y for discussions. I would like to acknowledge the hospitality of the Center for Electronic Correlations and Magnetism of the University of Augsburg where part of the work was done and the support of the Research Unit FOR1346 of the Deutsche Forschungsgemainschaft.

\end{document}